\documentclass{article}

\usepackage{PRIMEarxiv}

\usepackage[round]{natbib}
\usepackage{amsmath}

\usepackage[utf8]{inputenc} % allow utf-8 input
\usepackage[T1]{fontenc}    % use 8-bit T1 fonts
\usepackage{hyperref}       % hyperlinks
\usepackage{url}   
\usepackage{algorithm}
\usepackage{booktabs}       % professional-quality tables
\usepackage{amsfonts}       % blackboard math symbols
\usepackage{nicefrac}       % compact symbols for 1/2, etc.
\usepackage{microtype} 
\usepackage{algpseudocode}
\usepackage{soul}
\usepackage{seqsplit}
\usepackage{xcolor}
\usepackage{lipsum}
\usepackage{fancyhdr}       % header
\usepackage{graphicx}       % graphics
\usepackage{bm}

\usepackage{amsmath}
\usepackage{cleveref}
\usepackage{tikz}
\usepackage{listings}
\usepackage{relsize}
\usepackage{multirow}
\usepackage{subcaption}
\usepackage{caption}
\usepackage{tikz}
\usetikzlibrary{positioning}

\def\bSig\mathbf{\Sigma}

\DeclareMathOperator*{\argmax}{arg\,max}

\newcommand*\circled[1]{\tikz[baseline=(char.base)]{
            \node[shape=circle,draw,inner sep=2pt] (char) {#1};}}
            
\newcommand*{\bs}{\boldsymbol}
\newcommand*{\mb}{\mathbf}

\title{Bayesian Scalar-on-network Regression with Applications to Brain Functional Connectivity}

\author{Xiaomeng Ju \thanks{\noindent{Email: jux01@nyu.edu}} \qquad Hyung G. Park \qquad Thaddeus Tarpey \vspace{0.1cm}\\  Division of Biostatistics, Department of Population Health \\ 
New York University School of Medicine \\
New York, NY 10016, U.S.A}

\begin{document}

%  This will produce the submission and review information that appears
%  right after the reference section.  Of course, it will be unknown when
%  you submit your paper, so you can either leave this out or put in 
%  sample dates (these will have no effect on the fate of your paper in the
%  review process!)

%\date{{\it Received October} 2007. {\it Revised February} 2008.  {\it
%Accepted March} 2008.}

%  These options will count the number of pages and provide volume
%  and date information in the upper left hand corner of the top of the 
%  first page as in published papers.  The \pagerange command will only
%  work if you place the command \label{firstpage} near the beginning
%  of the document and \label{lastpage} at the end of the document, as we
%  have done in this template.

%  Again, putting a volume number and date is for your own amusement and
%  has no bearing on what actually happens to your paper!  

%\volume{64}
%\pubyear{2008}
%\artmonth{December}

%  The \doi command is where the DOI for your paper would be placed should it
%  be published.  Again, if you make one up and stick it here, it means 
%  nothing!

%\doi{10.1111/j.1541-0420.2005.00454.x}

%  This label and the label ``lastpage'' are used by the \pagerange
%  command above to give the page range for the article.  You may have 
%  to process the document twice to get this to match up with what you 
%  expect.  When using the referee option, this will not count the pages
%  with tables and figures.  

%\label{firstpage}

%  put the summary for your paper here
\maketitle

\begin{abstract}
This paper presents a Bayesian regression model relating  scalar outcomes to brain functional connectivity represented as symmetric positive definite (SPD) matrices.  Unlike many proposals that simply vectorize the matrix-valued connectivity predictors thereby ignoring their geometric structure, the method presented here respects the Riemannian geometry of SPD matrices by using a tangent space modeling. Dimension reduction is performed in the tangent space, relating the resulting low-dimensional representations to the responses. The dimension reduction matrix is learned in a supervised manner with  a sparsity-inducing prior imposed on a Stiefel manifold to prevent overfitting. Our method yields a parsimonious regression model that allows uncertainty quantification of all model parameters and identification of key brain regions that predict the outcomes.  We demonstrate the performance of our approach in simulation settings and through a case study to predict Picture Vocabulary scores using data from the Human Connectome Project. 
\end{abstract}

%  Please place your key words in alphabetical order, separated
%  by semicolons, with the first letter of the first word capitalized,
%  and a period at the end of the list.
%


%  As usual, the \maketitle command creates the title and author/affiliations
%  display 

%  If you are using the referee option, a new page, numbered page 1, will
%  start after the summary and keywords.  The page numbers thus count the
%  number of pages of your manuscript in the preferred submission style.
%  Remember, ``Normally, regular papers exceeding 25 pages and Reader Reaction 
%  papers exceeding 12 pages in (the preferred style) will be returned to 
%  the authors without review. The page limit includes acknowledgements, 
%  references, and appendices, but not tables and figures. The page count does 
%  not include the title page and abstract. A maximum of six (6) tables or 
%  figures combined is often required.''

%  You may now place the substance of your manuscript here.  Please use
%  the \section, \subsection, etc commands as described in the user guide.
%  Please use \label and \ref commands to cross-reference sections, equations,
%  tables, figures, etc.
%
%  Please DO NOT attempt to reformat the style of equation numbering!
%  For that matter, please do not attempt to redefine anything!

\section{Introduction}
Neuroimaging techniques allow us to gain valuable insights into brain functioning by examining interrelationships among spatially distributed brain regions of interest (ROIs). These interrelationships can be quantified as  ``functional connectivity'' using the covariance (or correlation) between functional magnetic resonance imaging (fMRI) signals measured at multiple ROIs over time. This paper develops a Bayesian regression framework for modeling and predicting 
  cognitive or behaviour outcomes using symmetric positive definite (SPD) matrix-valued measures of functional connectivity,  which  identifies key  brain regions contributing to the predictions. %, which can associate cognitive or behaviour outcomes to region-specific brain functioning.
  %Alternative representations, including partial correlations \citep{ryali2012estimation, weaver2023single}  and graph-based statistics \citep{bullmore2009complex, vogelstein2012graph}, have also been explored. 

 Given the estimated functional connectivity matrices, a common approach to building a regression model takes vectorized predictor matrices as input to conventional regression algorithms.  This approach  disregards the SPD  nature of the features and produces an exceedingly large number of predictors that require substantial regularization or pre-screening through massive univariate tests \citep{craddock2009disease,zeng2012identifying,ming2022flexible}. The resulting estimates can be  difficult to interpret due to the lack of structure in the selected features.  Another modeling approach adopts a two-stage procedure: first applying unsupervised methods to obtain a low dimension representation of the connectivity matrices, then fitting a  regression model with the low-dimensional features \citep{zhang2019tensor,ma2022semi}.  However, features  learnt to minimize the reconstruction error of the matrices may not be highly predictive of the response.

% Advances in neuroimaging techniques and increased accessibility of data have led to a growing interest in studying the relationships of 
 %functional connectivity and cognitive or behaviour outcomes.  For example, functional connectivity has been linked to   intelligence \citep{song2008brain, he2020deep}, language ability \citep{zhang2014resting, tomasi2020network}, cognitive impairment \citep{meskaldji2016prediction, lin2018resting}, and neuropsychiatric disorders \citep{craddock2009disease, venkataraman2012whole,fair2013distinct}. The focus of our study is to build a regression model based on connectivity features that can be used to make predictions of various outcomes of interest while identifying key brain regions that contribute to the predictions. 
%Yet, they require the computation of matrix inversions or selection of representative summary statistics.  Thus, our proposal focuses on correlation/covariance-based connectivity estimates that belong to the class of symmetric positive definite (SPD) matrices. 

Within the frequentist paradigm, a variety of methods have been proposed for regression directly with matrix features, including those  targeting at symmetric matrices that naturally represent connectivity networks and others that  deal with  generic matrices or higher-order tensors, including  symmetric matrices as a special case. Let $\mb{M}_i \in \mathbb{R}^{p \times p}$ denote a SPD predictor matrix of a scalar outcome $y$ with corresponding regression coefficient matrix $\mb{C} \in \mathbb{R}^{p \times p}$.  
 Several proposals impose regularization on $\mb{C}$ to encourage its sparsity, including  penalization on its spectral domain \citep{zhou2014regularized}, edge-wise and node-wise regularization \citep{weaver2023single}, and a combination of  edge-wise and spectral regularization \citep{brzyski2023matrix}. These methods may provide meaningful interpretations, however, they require estimating a large number of parameters and carefully tuning the level of sparsity. On the other hand, tensor regression methods assume $\mb{C}$ to be of low rank  and admit a tensor decomposition, which drastically reduces the number of parameters needed to be estimated.  Previous methods have adopted rank-1 approximations \citep{zhao2014structured}, CANDECOMP/PARAFAC (CP) decomposition \citep{zhou2013tensor}, and Tucker decomposition  \citep{li2018tucker}.  When modeling symmetric matrix predictors, it is natural to impose symmetry on $\mb{C}$.    \cite{wang2019symmetric} and \cite{wang2021learning} assumed such structure by using a  rank-$K$ decomposition: $\mb{C} = \sum_{k=1}^K \lambda_k \bs{\beta}_k \bs{\beta}_k^T$ with some unknown  $\bs{\beta}_k \in \mathbb{R}^p$ and $\lambda_k \in \mathbb{R}$. Our proposal can also be interpreted through this decomposition and it further provides  identifiability and uncertainty quantification of model parameters as  described in \Cref{sec:method}.

  %While some methods specifically model symmetric matrices that naturally represent connectivity networks, other techniques deal with  generic matrices or higher-order tensors  \cite[e.g.,][]{zhou2013tensor}, including  symmetric matrices as a special case. 

  SPD matrices naturally reside on a nonlinear manifold, but the methods introduced above ignore this structure and treat SPD matrices as Euclidean objects,  leading to several shortcomings. For example, the Euclidean distance between SPD matrices may not reflect their geodesic distance along the manifold \citep{you2021re} and may produce distorted results, and the entries of SPD matrices are often inter-correlated due to SPD constraints, forming skewed distributions \citep{schwartzman2016lognormal} that can yield unstable estimates of the corresponding regression coefficients.   Recent studies have explored treating correlation/covariance connectivity matrices as elements in an appropriate tangent space to preserve their Riemannian geometry and showed significant improvements in predictive performance over Euclidean approaches in several benchmark experiments \citep{dadi2019benchmarking,pervaiz2020optimising}. 
Although  providing good prediction results, these modeling attempts solely focus on using vectorized tangent-space representations as inputs to existing  regression models and yield estimates that are difficult to interpret.

Regression with matrix features has been less explored under the Bayesian paradigm.  Some Bayesian methods assume low-rank decompositions of the coefficient matrix $\mb{C}$ along with sparsity-inducing priors on $\mb{C}$, such as in \cite{guhaniyogi2017bayesian} and  \cite{papadogeorgou2021soft}, however, these methods are  not tailored for symmetric matrices. While \cite{guha2021bayesian} assumed a lower diagonal $\mb{C}$ for symmetric matrix predictors with a sparsity-inducing prior to identify important nodes and edges, their method involves a large number of parameters that scales quadratically with $p$.  A joint modeling approach was proposed by \cite{jiang2020bayesian}; however, their method relies on certain assumptions to construct probablistic formulations of the matrix predictors, limiting its flexibility. Furthermore,  all currently available Bayesian methods disregard the Riemannian geometry of SPD  matrices and instead treat these objects as Euclidean.

 The Bayesian regression method developed in this paper, relating a scalar outcome to SPD matrices, provides a novel representation of SPD predictors in a tangent space, thereby leveraging the Riemmanian geometry of SPD matrices and distinguishing our approach from other Bayesian methods.  Unlike previous tangent-space based methods that use vectorized representations, we apply supervised dimension reduction directly to SPD matrices to obtain a parsimonious model. A novel sparsity-inducing prior is developed as a regularization approach on a Stiefel manifold consisting of possible dimension reducing matrices. 
While tangent-space approaches typically lack interpretability, our model enables identification of key brain regions and subnetworks that are related to the response in the context of connectivity matrices.
 
 %This distinguishes our proposal from previous Bayesian methods that view matrices as Euclidean objects. 
 
%\item We conduct extensive numerical experiments comparing our proposal to frequentist and Bayesian competitors, including a simulation study and case study with data from the Human Connectome Project (HCP). The results are analyzed regarding the predictive performance and inference of model parameters.
%\end{itemize}

%The remainder of the paper is organized as follows. \Cref{sec:method} begins by defining the projection to obtain tangent-space representations and introducing the regression model. This is followed by describing the posterior inference procedure and the prior that enables sparsity sampling on the Stiefel manifold.  The interpretation of our model parameters based on  a generative model  is provided in \Cref{sec:interpretation}. \Cref{sec:simulation} evaluates the performance of our proposal through a simulation study, comparing it against existing alternatives.  \Cref{sec:hcp} presents an application to the HCP data for predicting the Picture Vocabulary score. Finally, \Cref{sec:discussion} concludes with a discussion of our approach and results. 

\section{Methodology} \label{sec:method}
We consider functional connectivity estimates derived from fMRI signals where each subject $i$ has fMRI time series $\mb{x}_i(t) \in \mathbb{R}^p$ recorded at $T_i$ time points $t = 1,2,...,T_i$ for $p$ ROIs, producing a size $n$ sample of SPD matrices $\mb{M}_i$'s for $i = 1,..., n$.  As in previous studies (e.g. \cite{pervaiz2020optimising}), we estimate connectivity via sample covariance (or correlation) matrices. 
%With standardized unit variance signals, $\mb{M}_i$ becomes the sample correlation matrix. 
To account for the Riemannian geometry of the space of $p \times p$ SPD matrices (denoted as $\text{Sym}_p^{+}$) we use a tangent-space parameterization as described below.  
 
% Both sample covariance and correlation matrices satisfy the SPD constraint, allowing them to serve as connectivity representations for our model. 

%The space of SPD matrices form a nonlinear Riemannian manifold which we denote as $\text{Sym}_p^{+}$.  Due to the geometry of the Riemannian manifold, the standard practice to model SPD matrices as Euclidean objects has several issues.  As pointed out by \cite{you2021re}, the Euclidean distance between SPD matrices may not reflect their geodesic along the manifold. Additionally, the entries of SPD matrices are often inter-correlated due to SPD constraints, forming skewed distributions \citep{schwartzman2016lognormal, pervaiz2020optimising}, Treating these matrices as Euclidean objects in regression models  can thus yield unstable estimates of regression coefficients and violate assumptions required for statistical inference. 

\subsection{Tangent space projections} \label{subsec:tangent}
A tangent space of the Riemannian manifold $\text{Sym}_p^{+}$ is defined at a reference point $\mb{M}^{\ast} \in \text{Sym}_p^{+}$, containing all tangent vectors passing through $\mb{M}^{\ast}$. We choose $\mb{M}^{\ast}$ to represent the Euclidean average of $\mb{M}_i$'s., i.e. $\mb{M}^{\ast} = \sum_{i=1}^n \mb{M}_i/n$.  Among the estimators examined in previous works \citep{ng2015transport, dadi2019benchmarking, pervaiz2020optimising}, this choice of $\mb{M}^{\ast}$ demonstrated stable performance across various scenarios.  The resulting tangent space consists of $p \times p$ symmetric matrices and provides a locally Euclidean approximation of the manifold.   Working in this tangent space allows us to perform linear operations (e.g. evaluating inner products) without distorting the intrinsic geometry (inter-subject distances) of SPD matrices. 
 
 % in $\text{Sym}_p^{+}$.  It contains all the tangent vectors that are derivatives of the curves passing through that reference point 
  
   %and  provides a locally Euclidean approximation of the manifold \citep{you2021re}. 
   
  %Different from $\text{Sym}_p^{+}$ that do not conform to the Euclidean geometry, its tangent space is a vector space consisting of $p \times p$ symmetric matrices.  Thus,

%Let $\mb{M}^{\ast}$ be a reference point ()in  $\text{Sym}_p^{+}$ 

 %Given its robust performance, we adopt the Euclidean average as our reference point $\mb{M}^{\ast}$. 

 We first normalize each $\mb{M}_i$  by the reference point $\mb{M}^{\ast}$  as ${\mb{M}^{\ast}}^{-1/2}\mb{M}_i{\mb{M}^{\ast}}^{-1/2}$. 
  This normalization step is referred to as the \textit{matrix whitening transport}  \citep{ng2015transport} which creates  ``whitened'' or ``standardized'' SPD objects that are close to the $p \times p$ identity matrix ( $\mb{I}_p$). This step allows us to use the tangent space at $\mb{I}_p$ (denoted as  $\mathcal{T}_{p}$) as a common space for projections via the Log map  % We then map the whitened matrices to the tangent space at $\mb{I}_{p}$, and denote this space as $\mathcal{T}_{p}$.  With a  given reference point $\mb{M}^{\ast}$,  the mapping from $\mb{M}_i$ to $\mathcal{T}_{p}$ is defined as: 
\begin{equation} \label{log:map}
\phi_{\mb{M}^{\ast}}: \mb{M}_i \rightarrow \text{Log}\left({\mb{M}^{\ast}}^{-1/2}\mb{M}_i{\mb{M}^{\ast}}^{-1/2}\right),
\end{equation}
where the  ``Log'' represents the matrix logarithm \citep{varoquaux2010detection}. % The resulting space consist of symmetric matrices no longer linked to the positive definite constraint \citep{pervaiz2020optimising}. 
%Similarly, we can  define the inverse map from $\mb{T}_i  \in \mathcal{T}_p$ to $\text{Sym}_p^{+}$
%\begin{equation} \label{exp:map}
%    \phi_{\mb{M}^{\ast}}^{-1}:\mb{T}_i  \rightarrow {\mb{M}^{\ast}}^{1/2} \text{Exp}\left(\mb{T}_i \right) {\mb{M}^{\ast}}^{1/2},
%\end{equation}
%where the "Exp" represents the matrix exponential \citep{varoquaux2010detection}.

Note that our definition of the tangent space projection differs from the one introduced in \cite{pennec2006riemannian} and \cite{you2021re}, which  maps $\mb{M}_i$ directly to the tangent space defined at $\mb{M}^{\ast}$ without ``whitening''.  Instead, we map $\mb{M}_i$ whitened by $\mb{M}^{\ast}$ to the tangent space at $\mb{I}_p$ ($\mathcal{T}_p$). Within our regression framework, this choice provides meaningful interpretations as will be discussed in \Cref{sec:interpretation}.  The same mapping was adopted by other proposals either in contexts different from ours \citep{varoquaux2010detection,ng2015transport,park2023bayesian} or for regression with vectorized predictors \citep{dadi2019benchmarking,pervaiz2020optimising}.

\subsection{Regression model} 
We propose a regression model relating the tangent-space representation $\phi_{\mb{M}^{\ast}}(\mb{M}_i)$  in \eqref{log:map}  to the response $Y_i$. Directly regressing $Y_i$ against $\phi_{\mb{M}^{\ast}}(\mb{M}_i)$ typically requires a quadratic number of parameters growing with $p$.  For example, the model could have  $p(p+1)/2$ coefficient parameters, one per lower-diagonal element of $\phi_{\mb{M}^{\ast}}(\mb{M}_i) \in \mathbb{R}^{p \times p}$.  For a large $p$, such a model can risk overfitting and in the Bayesian paradigm cause slow mixing and convergence issues of Markov chains when sampling from the posterior distribution.  To this end, we propose a dimension reduced model
\begin{equation} \label{eq:model}
Y_i =  \mu +  \left \langle \bs{\Gamma}^T \phi_{\mb{M}^{\ast}}(\mb{M}_i) \bs{\Gamma}, \mb{B} \right \rangle + \epsilon_i,  \ \epsilon_i \overset{iid}{\sim} N(0, \sigma^2),
\end{equation} 
where $\mu \in \mathbb{R}$ is the intercept and  $\langle \cdot, \cdot \rangle$ is the Frobenius inner product. The matrix $\bs{\Gamma} \in~\mathbb{R}^{p\times d}$, $d < p$, has orthonormal columns  ($\bs{\Gamma}^T \bs{\Gamma} = \mb{I}_d$) and reduces the dimensionality of $\phi_{\mb{M}^{\ast}}(\mb{M}_i)$ from $p \times p$ to $d \times d$.  
 The coefficient matrix $\mathbf{B}$ is a $d \times d$ diagonal matrix where the diagonality assumption is required for model identifiability.   Compared to previous tangent-space approaches \citep{dadi2019benchmarking, pervaiz2020optimising}, the proposed method substantially reduces the number of parameters to scale linearly in $p$. 
%For any matrices $\mb{A}_1$ and $\mb{A}_2$ of the same dimension, $\langle \mb{A}_1,  \mb{A}_2 \rangle = \text{trace}(\mb{A}_1 \mb{A}_2^T) = \text{vec}(\mb{A}_1)^T \text{vec}(\mb{A}_2)$.      
  %From \eqref{eq:model}, our model learns a low-dimensional representation of $\phi_{\mb{M}^{\ast}}(\mb{M}_i)$ in a supervised manner by connecting this representation to the response.
Expressing $\bs{\Gamma}$ in columns as $(\bs{\gamma}_1, \bs{\gamma}_2, ...., \bs{\gamma}_d)$, we can write
$\left \langle \bs{\Gamma}^T \phi_{\mb{M}^{\ast}}(\mb{M}_i) \bs{\Gamma}, \mb{B} \right \rangle = \sum_{j=1}^d b_j 
\bs{\gamma}_j^T \phi_{\mb{M}^{\ast}} \left(\mb{M}_i \right) \bs{\gamma}_j$,  a linear combination of $d$ projections of $\phi_{\mb{M}^{\ast}}(\mb{M}_i)$, where each projection extracts a one-dimensional component~$\bs{\gamma}_j^T \phi_{\mb{M}^{\ast}} \left(\mb{M}_i \right) \bs{\gamma}_j$.

 Due to $\bs{\Gamma}$ having 
 orthonormal columns,  model \eqref{eq:model} can be equivalently expressed as 
\begin{equation} \label{eq:model:version2}
Y_i =  \mu + \left \langle  \phi_{\mb{M}^{\ast}}\left( \mb{M}_i \right),  \bs{\Gamma}\mb{B} \bs{\Gamma}^T \right \rangle + \epsilon_i, \ \epsilon_i \overset{iid}{\sim} N(0, \sigma^2),
\end{equation}
which provides an alternative view of our model as a tensor regression approach.  With $\mb{B} = \text{Diag}(b_1,..., b_d)$, 
$\bs{\Gamma}^T \mb{B}\bs{\Gamma} = \sum_{j=1}^d b_j \bs{\gamma}_j\bs{\gamma}_j^T$
 defines a  decomposition  with  $d$ rank-1 components.  Under this decomposition, \eqref{eq:model:version2} fits a rank-d coefficient matrix to the order-2 symmetric tensor $\phi_{\mb{M}^{\ast}}\left( \mb{M}_i \right)$.  Similar decompositions have been explored under the frequentist paradigm by \cite{wang2019symmetric} and \cite{wang2021learning}, but using $\mb{M}_i$ instead of $\phi_{\mb{M}^{\ast}}\left( \mb{M}_i \right)$ and without orthonormality constraints on $\bs{\Gamma}$. Their model parameters are unidentifiable without penalization added to $\bs{\Gamma}$. In contrast, our proposed  model is identifiable up to column sign and permutation of $\mb{\Gamma}$ (or equivalently that of $b_1$,..., $b_d$), enabling convenient Bayesian inference and interpretation.  A proof for identifiability  is provided in Supporting Information A.

\subsection{Bayesian posterior inference}
Let  $\mathcal{M} = \{\mathbf{M}_i\}_{i=1}^n$  denote the SPD  matrix features and $\mathcal{Y} = \{Y_i\}_{i=1}^n$ the responses.  The reference matrix is $\mb{M}^{\ast} = \frac{1}{n}\sum_{i=1}^n  \mb{M}_i$. The posterior distribution of the parameters $(\boldsymbol{\Gamma}, \mathbf{B}, \mu, \sigma)$ of \eqref{eq:model} can be expressed as $p(\boldsymbol{\Gamma}, \mathbf{B}, \mu, \sigma|\mathcal{M}, \mathcal{Y}) 
 \propto p(\mathcal{Y}|\boldsymbol{\Gamma}, \mathbf{B}, \mu, \sigma, \mathcal{M}) p(\boldsymbol{\Gamma}, \mathbf{B},  \mu, \sigma)$, where the likelihood 
$p(\mathcal{Y} | \boldsymbol{\Gamma}, \mathbf{B},  \mu, \sigma,\mathcal{M}) =  \prod_{i=1}^n p(Y_i| \boldsymbol{\Gamma}, \mathbf{B}, \mu, \sigma, \mb{M}_i)$ is specified based on a Gaussian assumption of the errors $\epsilon_i$ in  \eqref{eq:model}. 
%& \propto \frac{1}{\sigma^n} \prod_{i=1}^n \text{exp} \left(- \frac{1}{2\sigma^2} \left(Y_i - \mu -  \langle \boldsymbol{\Gamma}^T \phi_{\mb{M}^{\ast}}(\mb{M}_i)\boldsymbol{\Gamma}, \mathbf{B} \rangle \right)^2  \right)  
%\end{align*}
%based on the assumption in \eqref{eq:model} that the errors $\epsilon_i$'s follow i.i.d. $N(0, \sigma^2)$.  
We use independent priors for the parameters:
\begin{align}\label{eq:prior}
p(\boldsymbol{\Gamma}, \mathbf{B},  \mu, \sigma)  =    p(\boldsymbol{\Gamma}) p(\mu) p(\sigma) p(\mb{B}), \ \text{and} \ \ p(\mb{B})= \prod_{j=1}^d p(b_j). 
\end{align} 
%where we place separate priors on each parameter $\mathbf{\Gamma}, \mu, \sigma$, and $\mathbf{B}$, with the prior on $\mathbf{B}$ further split into independent priors on each $b_j$.
%In the absence of prior knowledge, we use mean-zero (weakly informative) Gaussian priors on each $b_j$ and $\mu$. 
 Without loss of generality, we can center the response (at zero) and use mean-zero (and weekly informative) Gaussian priors on each $b_j$ and $\mu$ in the absence of prior knowledge and  enforce $b_1<...<b_d$ to address identifiability up to column permutation of $\bs{\Gamma}$.  For $\sigma$, we use a weakly informative prior  or alternatively can use a prior based on a preliminary regression fit, such as  the one used in \Cref{sec:simulation}.  Defining the prior on $\bs{\Gamma}$ is delicate due to its orthonormal constraints. As in \cite{pourzanjani2021bayesian}, we use  Givens rotations for sampling $\bs{\Gamma}$  with a shrinkage prior imposed to encourage sparsity. The details are provided next. 

\subsection{Sparse sampling of $\mathbf{\Gamma}$ via Givens rotations} \label{sec:gamma}
Givens rotations \citep{givens1958computation} provide an algorithm to find the QR decomposition of any $p \times d$  matrix $\mb{A} = \mb{Q}\mb{R}$, factorizing it into a $p \times p$ orthonormal matrix  $\mb{Q}$  and a $p \times d$ upper triangular matrix $\mb{R}$.  By sequentially applying a series of Givens rotation matrices (whose form is defined in \eqref{eq:givens}), we can ``zero out'' the lower diagonal entries of any given $\mb{A}$.  If $\mb{A}$ is orthonormal,  $\mb{R}$ equals to $\mb{I}_{p \times d}$: a $p \times d$ matrix with ones along the diagonal and zeros elsewhere.  Under this QR decomposition, Givens rotations sequentially transform an orthonormal $\bs{\Gamma} \in \mathbb{R}^{p \times d}$  to $\mb{I}_{p \times d}$.  Reversing these rotations, we derive a sampling procedure to generate sparse $\bs{\Gamma}$ from $\mb{I}_{p \times d}$, based on the proposal of \cite{pourzanjani2021bayesian}.

We define a Givens rotation matrix 
 $\mb{G}_{i,j}(\theta)$ to  perform a clockwise rotation in the $(i,j)$-plane
 of angle  $\theta$ 
  \begin{equation} \label{eq:givens}
\mb{G}_{i,j}(\theta) = \begin{pmatrix}
1 & \cdots & 0 & \cdots & 0 & \cdots & 0 \\
\vdots & \ddots & \vdots & & \vdots & & \vdots \\
0 & \cdots & \cos(\theta) & \cdots & \sin(\theta) & \cdots & 0 \\
\vdots & & \vdots & \ddots & \vdots & & \vdots \\
0 & \cdots & -\sin(\theta) & \cdots & \cos(\theta) & \cdots & 0 \\
\vdots & & \vdots & & \vdots & \ddots & \vdots \\
0 & \cdots &  0 & \cdots & 0   &  \cdots & 1 
\end{pmatrix},
\end{equation}
which embeds within the $\mb{I}_p$ a rotational operation at the intersections of the $i$-th and the $j$-th rows and columns, characterized by angle $\theta$.  Likewise, $\mb{G}_{i,j}^T(\theta)$  performs a counter-clockwise rotation of angle $\theta$ in the $(i,j)$-plane.  Successively applying a series of $\mb{G}_{i,j}(\theta)$'s can introduce zeros to the lower diagonals of $\bs{\Gamma}$, one zero at a time.  
  The order to eliminate these entries is not unique and we choose to proceed from top to bottom and left to right. An illustrative example of this procedure is provided in Supporting Information B. 
%For example, for $d = 5$ and $d=3$, the elimination order of each entry is shown in \eqref{eq:order} in circled numbers, where $X$'s denote entries that do not require elimination:
%\begin{equation}\label{eq:order}
%\begin{pmatrix}
%X & X & X  \\
%\circled{1} & X & X  \\
%\circled{2} & \circled{5} & X \\
%\circled{3} & \circled{6} & \circled{8} \\
%\circled{4} & \circled{7}& \circled{9} \\
%\end{pmatrix}. 
%\end{equation}
Sequentially applying Givens rotation matrices of form  \eqref{eq:givens} to an orthonormal $\bs{\Gamma}$ yields
\begin{equation} 
\mathbf{G}_{d,p}(\theta_{d,p})
...\mathbf{G}_{d,d+1}(\theta_{d,d+1}) ...\mathbf{G}_{2,p}(\theta_{2,p}) ...\mathbf{G}_{2,3}(\theta_{2,3})
\mathbf{G}_{1,p}(\theta_{1,p}) ...\mathbf{G}_{1,2}(\theta_{1,2})\boldsymbol{\Gamma} = \mathbf{I}_{p\times d},
\end{equation}
where the rotation angles $\theta_{i,j}$'s are selected to ``zero out'' the lower-diagonal elements of $\bs{\Gamma}$ in the aforementioned order. Reversing these rotations 
 constructs an orthonormal  $\bs{\Gamma}$ from $\mb{I}_{p \times d}$ 
\begin{equation} \label{eq:givens:generate}
\mathbf{G}_{1,2}^T(\theta_{1,2})...\mathbf{G}_{1,p}^T(\theta_{1,p})\mathbf{G}_{2,3}^T(\theta_{2,3}) ...\mathbf{G}_{2,p}^T(\theta_{2,p}) ...
 \mathbf{G}_{d,d+1}^T(\theta_{d,d+1}) ... \mathbf{G}_{d,p}^T(\theta_{d,p}) 
 \mathbf{I}_{p\times d} = \boldsymbol{\Gamma}.  
 \end{equation}
 The number of parameters (i.e. $\theta_{i,j}$'s) in this formulation is $pd - d^2/2 - d/2$, substantially lower (i.e. linear in $p$) compared to previous studies \citep{dadi2019benchmarking, pervaiz2020optimising}.

 We introduce sparsity to $\bs{\Gamma}$ by 
 imposing sparsity on the rotation angles $\theta_{i,j}$.  A zero angle  $\theta_{i,j}=0$ skips the corresponding rotation, preserving the sparsity pattern in the matrix it immediately applies to. As a result, the sparsity pattern in $\mb{I}_{p\times d}$ propagates to the generated $\bs{\Gamma}$ in \eqref{eq:givens:generate} as detailed in \cite{cheon2003sparse}. We use the horseshoe prior \citep{carvalho2010horseshoe}  that encourages the sparsity of  $\theta_{i,j}$ at global and local levels
\begin{equation}  \label{eq:horseshoe}
\theta_{i,j} \sim \text{Truncated\ Normal} (0, \tau^2 \lambda_{i,j}^2), \ 
\lambda_{i,j} \sim \text{Half\ Cauchy}(0, 1), \end{equation}
where $\tau$ controls the global sparsity across $\theta_{i,j}$'s and $\lambda_{ij}$ allows local shrinkage for each $\theta_{i,j}$. 

Of note, our sparsity prior differs from the one used by \cite{pourzanjani2021bayesian} due to the configuration of our regression model. In   \cite{pourzanjani2021bayesian},  $\theta_{1,2}$, $\theta_{2,3}$, ..., $\theta_{d,d+1}$ range from  $-\pi$ to $\pi$ while the rest range from $-\pi/2$ to $\pi/2$. As sign flips of the columns of $\bs{\Gamma}$ do not impact the responses for our model, we let all $\theta_{i,j} \in [-\pi/2, \pi/2]$ to reduce the multi-modality of the posterior distribution.  Further explanation is provided in Supporting Information B. Additionally, unlike \cite{pourzanjani2021bayesian} which adopted the regularized horseshoe prior \citep{piironen2017sparsity} to prevent $\theta_{i,j}$ from approaching its boundary, we use the  original horseshoe prior \citep{carvalho2010horseshoe} to provide increased flexibility. Our boundary angles of $\pi/2$ and $-\pi/2$ correspond to row swap operations, while the angle $-\pi/2$ also introduces sign flips. The row swaps exchange the sparsity patterns of two rows in the matrix that $\mb{G}_{i,j}(\theta_{i,j})$  immediately applies to in \eqref{eq:givens:generate}, and thus permit $\bs{\Gamma}$ to  have a variety of sparsity patterns inherited from $\mb{I}_{p\times d}$ as well as row permuted variants of $\mb{I}_{p\times d}$. 
   %Hence, we adopt the original horseshoe prior \citep{carvalho2010horseshoe}, providing this increased flexibility.

\section{Model interpretation} \label{sec:interpretation}
While the tangent space representation captures the Riemannian geometry of SPD matrices, the resulting models often lack interpretability.  With existing tangent-space approaches \citep{dadi2019benchmarking, pervaiz2020optimising}
it is not straightforward to localize which ROIs greatly influence the response. 
 In contrast, our formulation in \eqref{eq:model} enables meaningful interpretations based on a generative model of the fMRI signals.  
 The approach of using generative models for interpretation has been adopted for other tangent-space regression methods \citep{sabbagh2019manifold,kobler2021interpretation}, but they rely on  more stringent assumptions compared to ours.  
%Our generative model is based on the whitened signals ${\mb{M}^{\ast}}^{-1/2}\mb{x}_i(t)$ and has more relaxed assumptions .

 %The approach based on a generative model for interpretation has also been adopted by \cite{sabbagh2019manifold}, and \cite{kobler2021interpretation}  for other tangent-space regression models.  Their generative model is based on the original signals $\mb{x}_i(t)$ whereas ours is based on the whitened signals ${\mb{M}^{\ast}}^{-1/2}\mb{x}_i(t)$ and has more relaxed assumptions.

%Assume $\mb{M}^{*-1/2}\mb{x}_i(t) \in \mathbb{R}^p$ are linear mixtures of source signals $\mb{s}_i(t) \in \mathbb{R}^d$ (those relate to the response) and noise signals $\bs{\eta}_i(t) \in \mathbb{R}^{p-d}$. 

We consider a generative model for $\mb{x}_{i} \in \mathbb{R}^p$: $\mb{x}_{i}(t)  =  {\mb{M}^{\ast}}^{1/2} \bs{\Gamma} \mb{s}_{i}(t) +  {\mb{M}^{\ast}}^{1/2} \mb{V}_i \bs{\eta}_{i}(t)$, where ${\mb{M}^{\ast}}^{1/2} \bs{\Gamma} \in \mathbb{R}^{p\times d}$ is a mixing matrix for the ``signal'' sources $\mb{s}_i(t) \in \mathbb{R}^d$ and ${\mb{M}^{\ast}}^{1/2} \mb{V}_i$ is a subject-specific mixing matrix for the ``noise'' sources $\bs{\eta}_i(t) \in \mathbb{R}^{p-d}$, with  $\bs{\Gamma}^T\bs{\Gamma} = \mb{I}_d$, $\mb{V}_i^T\mb{V}_i = \mb{I}_{p-d}$, and $\bs{\Gamma}^T\mb{V}_i = \mb{0}_{d \times (p-d)}$ for $i = 1,..., n$. This model can be equivalently written as 
\begin{equation} \label{eq:signal}
{\mb{M}^{\ast}}^{-1/2}\mb{x}_{i}(t) =  \bs{\Gamma} \mb{s}_{i}(t) +  \mb{V}_i \bs{\eta}_{i}(t). 
\end{equation}
Assume $\mb{C}_{s,i} = \text{Cov}(\mb{s}_i(t))$ and $\mb{C}_{\eta,i} = \text{Cov}(\bs{\eta}_i(t))$ are time-invariant. As in \Cref{subsec:tangent}, let $\mb{M}_i = \text{Cov}(\mb{x}_i(t))$ and $\tilde{\mb{M}}_i = {\mb{M}^{\ast}}^{-1/2}\mb{M}_i{\mb{M}^{\ast}}^{-1/2}$. Evaluating the covariance of ${\mb{M}^{\ast}}^{-1/2}\mb{x}_{i}(t)$ in \eqref{eq:signal} gives
$\tilde{\mb{M}}_i = \bs{\Gamma} \mb{C}_{s,i}  \bs{\Gamma}^T + \mb{V}_i \mb{C}_{\eta,i}  \mb{V}_i^T.$
Under this generative model, our regression \eqref{eq:model}  can be interpreted as  relating $\mb{C}_{s,i}$ to $Y_i$ via the model $Y_i = \mu + \langle \text{Log}(\mb{C}_{s,i}) ,  \mb{B}  \rangle + \epsilon_i$, 
where the ``$\text{Log}$'' represents the matrix logarithm and the assumptions on the parameters $\mu$, $\mb{B}$, and $\epsilon$ are identical to those used in \eqref{eq:model}.   This equivalency follows since $\langle \text{Log}(\mb{C}_{s,i} ) ,  \mb{B}  \rangle =   \langle \bs{\Gamma}^T\phi_{\mb{M}^{\ast}}(\mb{M}_i) \bs{\Gamma}, \mb{B} \rangle$ (see Supporting Information C for derivation).  Thus, by pre-multiplying $\bs{\Gamma}^T$ on both sides of \eqref{eq:signal}, $\bs{\Gamma}^T {\mb{M}^{\ast}}^{-1/2} \in \mathbb{R}^{d\times p}$ is  key to interpretation as it acts as a ``spatial filter'' for extracting the relevant signals $\mb{s}_i(t)$, whose covariance ($\mb{C}_{s,i}$) is related to the response.     Each of its rows  $\gamma_j^T{\mb{M}^{\ast}}^{-1/2} \in \mathbb{R}^p$ forms a subnetwork that produces the $j$-th ``signal'' source.

%Thus,  model 
%\eqref{eq:model} essentially relates the covariance of the ``signal'' sources ($\mb{s}_i(t)$) to the response.  From \eqref{eq:signal}, the ``signal''
% sources are computed as $\mb{s}_i(t) = \bs{\Gamma}^T {\mb{M}^{\ast}}^{-1/2} \mb{x}_i(t)$. The matrix $\bs{\Gamma}^T {\mb{M}^{\ast}}^{-1/2}$ is  key to interpretation as it acts as a spatial filter for extracting the relevant signals $\mb{s}_i(t)$, whose covariance is related to the response. 

Finally, we  define the conditional expectation $E_{\mb{M}^{\ast}}(Y|\mb{M})  =  \mu + \langle \bs{\Gamma}^T \phi_{\mb{M}^{\ast}}(\mb{M}) \bs{\Gamma}, \mb{B} \rangle$. 
 The intercept $\mu$ can be interpreted as the expected response for an individual with the ``average'' connectivity $\mb{M}^{\ast}$, since $E_{\mb{M}^{\ast}}(Y|\mb{M}^{\ast})  = \mu$.  The interpretation of $\mb{B}$ based on the deviation of $\mb{M}$ from $\mb{M}^{\ast}$ along $\bs{\Gamma}$ in the tangent space is provided in Supporting Information C.

\section{Simulation study} \label{sec:simulation}
To assess the performance of our proposal, extensive simulations were conducted to provide a comparison to several alternative frequentist and Bayesian approaches, with respect to their predictive accuracy and  model parameter inference. 

\subsection{Setup} \label{subsec:setup}
We generated $N$ subject-level covariance matrices $\mb{M}_i$ based on the  eigen-decomposition
$\mb{M}_i = \mb{A}_{i} \mb{E}_{i} \mb{A}_{i}^T, i = 1,..., N,$
where $\mb{A}_i \in \mathbb{R}^{p \times p}$ is the matrix of eigenvectors and $\mb{E}_i = \text{Diag}(e_{i,1},..., e_{i,p})$ contains the  eigenvalues.  We generated the eigenvectors $\mb{A}_i$  by the method of \cite{stewart1980efficient} implemented in the \texttt{pracma} package \citep{pracma}, independently for each subject $i$, and the 
eigenvalues by $e_{i,j} = e^{u_{i,j}}$ where $u_{i,j}$  were sampled i.i.d. from $\mathcal{U}(-2,2)$ for $i = 1,..., N$ and $j = 1,..., p$.  Out of the total $N$ observations, the first $n$ observations were designated as the training set and the remaining $N-n$ were set aside as a test set fixed at size 1000.

%started by simulating $N$ independent covariance matrices $\mb{M}_i$
%using the approach of , implemented in the \texttt{pracma} R package \citep{pracma}, based on the  eigen-decomposition
%$\mb{M}_i = \mb{A}_{i} \mb{E}_{i} \mb{A}_{i}^T, i = 1,..., N, $
%where $\mb{A}_i \in \mathbb{R}^{p \times p}$ contains the eigenvectors and $\mb{E}_i = \text{Diag}(e_{i,1},..., e_{i,p})$ is a diagonal matrix of eigenvalues. The eigenvectors were generated independently for each subject $i$ by the method of \cite{stewart1980efficient}
%\st{We generated the \st{eigenvectors} $\mathbf{A}_i$ independently for each subject $i$. This was done} using the approach by \cite{stewart1980efficient}, implemented in the \texttt{randortho} function from the \texttt{pracma} R package \citep{pracma}.  
%The 

We computed the tangent space representation $\phi_{\mb{M}^{\ast}}(\mb{M}_i)$ of $\mb{M}_i$  based on \eqref{log:map} and generated a sparse $\bs{\Gamma}$ using Givens rotations in \Cref{sec:gamma}, which requires $pd - d^2/2 - d/2$ angles ($\theta_{i,j}$'s) to obtain  $\bs{\Gamma}$. 
   We randomly sampled 50\% (rounding down to the nearest integer) of the angles to be zero, and the rest from $\mathcal{U}(-\pi/2, \pi/2)$.

The responses ($Y_i$'s) were simulated via two processes -- one aligns with our regression model and the other one allows model misspecification. For the 
``correctly specified'' regression case, the response $Y_i$ was  generated based on model \eqref{eq:model} with $\mu = 0$.  For the ``misspecified'' regression case, we define $\tilde{ \bs{\Gamma}}_{i} = \bs{\Gamma} + \mb{O}_i,$ where the elements of the perturbation  matrix $\mb{O}_i \in \mathbb{R}^{p \times d}$ are sampled i.i.d. from $N(0, \nu^2)$, and used $\tilde{ \bs{\Gamma}}_{i}$ in place of $\bs{\Gamma}$. Different levels of perturbation were considered, corresponding to $\nu \in \{0.05, 0.1, 0.2\}$.

 % Our experiment is thus structured in two parts, Part~(1) and Part~(2), corresponding to using responses generated from separate processes.  
  
  %For Part~(1) of our experiment, the : and for Part~(2)
%\begin{equation} \label{eq:sim:model:p}
%Y_i = \langle \tilde{ \bs{\Gamma}}_{i}^T \phi_{\mb{M}^{\ast}}(\mb{M}_i)\tilde{ \bs{\Gamma}}_{i}, \mb{B}\rangle  + \epsilon_i,  \ \epsilon_i \overset{iid}{\sim} N(0, \sigma^2),
% \end{equation}
%where  $\tilde{ \bs{\Gamma}}_{i}$ represent $\bs{\Gamma}$ with added perturbation for each subject. We define 
%where the perturbation   $\mb{O}_i \in \mathbb{R}^{p \times d}$ has elements sampled from i.i.d. $N(0, \nu^2)$.   

We compared the performance of different methods across settings of varying complexity based on combinations of the training sample size ($n$), dimensionality of $\bs{\Gamma}$ ($p, d$), and the signal-to-noise ratio (SNR).   Let $g_i = \langle \bs{\Gamma}^T \phi_{\mb{M}^{\ast}}(\mb{M}_i)\bs{\Gamma}, \mb{B}\rangle$ represent the expected response for the $i$th subject,  and $\bar{g} = \sum_{i=1}^n g_i/n$.  The value of $\sigma$ in \eqref{eq:model} was computed based on SNR: $\sigma = \sqrt{ \sum_{i=1}^n (g_i - \bar{g})^2/((n-1)\text{SNR})}$.  
For the ``correctly specified'' case, we considered combinations of $n = \{200, 400\}$, SNR $= \{1, 5\}$, and $(p, d) \in \{(5,2), (15,2), (15,4) \}$, yielding $2 \times 2 \times 3 = 12$ settings.  For the ``misspecified'' case, we focused on the more challenging $n = 200$ and $\text{SNR} = 1$ scenario for the same ($p, d$) combinations as the  ``correctly specified'' case  under three  perturbation levels ($\nu$), giving $3 \times 3 = 9$ settings.  For  $d = 2$ settings, $\mb{B} = \text{Diag}(1, -1)$, and for $d = 4$ settings, $\mb{B} = \text{Diag}(2,1,-1,-2)$. The experiment was repeated over 100 independent runs for each setting, with $\bs{\Gamma}$ independently generated for each run.

\subsection{Implementation details}
We compared our proposal with available alternatives introduced below:
\begin{itemize}
\item \textbf{LS}: least squares linear regression \citep{legendre1806nouvelles}; 
\item \textbf{LASSO}: linear regression with  LASSO penalty \citep{tibshirani1996regression};
\item  \textbf{BTR}:  Bayesian tensor regression \citep{guhaniyogi2017bayesian};
\item  \textbf{SBL}: symmetric bilinear regression \citep{wang2019symmetric},
\item  \textbf{FCR}:  functional connectivity regression \citep{weaver2023single}. 
\end{itemize}

\textbf{LS} and \textbf{LASSO} use the vectorized 
upper triangular portion of each $\mb{M}_i$, including diagonal elements, as features. The regularization parameter in \textbf{LASSO} was selected through 10-fold cross-validation using functions from the \texttt{glmnet} package \citep{friedman2010regularization}.

\textbf{BTR} and \textbf{SBL} assume a low-rank decomposition of the regression coefficient $\mb{C} \in \mathbb{R}^{p \times p}$ in the model $E(Y_i|\mb{M}_i) =  \langle \mb{M}_i , \mb{C} \rangle $.  \textbf{BTR} considers the typical CP decomposition \citep{kolda2009tensor} whereas \textbf{SBL}   constrains $\mb{C}$ to be symmetric and adopts bilinear decomposition. For each method, we ran the algorithm with decomposition ranks of $2, 3, 4$, and report results for the rank that yields the lowest average test error across the 100 runs of the experiment.  %The remaining algorithm parameters were set to default values. 

\textbf{FCR} 
uses partial correlation matrices as predictors and a single index model $E(Y_i|\mb{P}_i) = h \left(  \langle \mb{P}_i , \mb{C} \rangle \right)$, where $\mb{P}_i$ is the partial correlation matrix derived from $\mb{M}_i$, and $h$ is an unknown link function.  We used an outer loop of 30 iterations to estimate $h$, with 10 inner iterations within each outer loop to update the estimate of $\mb{C}$ in \textbf{FCR}.

We also evaluated the four variants of our proposal to demonstrate 
the benefit of the two key features in our approach: tangent ({\bf T}) space representation and sparsity ({\bf S}) sampling for $\bs{\Gamma}$. These proposals are referred to as {\em Bayesian Scalar-On-Network} (\textbf{BSN}) regression: (i) the na\"ive approach  (\textbf{BSN-N}) with the model $Y_i = \mu + \langle \bs{\Gamma}^T\mb{M}_i  \bs{\Gamma}, \mb{B}^T \rangle + \epsilon_i$, that does not use the tangent space mapping for $\mb{M}_i$, (ii) the na\"ive approach with sparsity sampling (\textbf{BSN-NS}). (iii)  the tangent space mapping model (\textbf{BSN-T}) (\ref{eq:model}), and  (iv) the \textbf{BSN-T} approach with  sparsity sampling (\textbf{BSN-TS}). \textbf{BSN-TS} is the recommended approach.

We implemented \textbf{BSN} regression in the R software \citep{citeR} with the Bayesian posterior sampling performed in Stan \citep{carpenter2017stan} through the \texttt{cmdstanr} package \citep{cmdstanr}. For all \textbf{BSN} methods, we set priors $b_j \sim N(0, 10^2)$ for $j = 1,..., d$,  and $\mu \sim N(0,1)$. The residual scale ($\sigma$ in \eqref{eq:model})  followed an exponential prior with its median set to the standard deviation of the  residuals from the \textbf{LASSO} fit. For comparibility,  the dimension $d$ (either 2 or 4) of \textbf{BSN} methods was set to match with the data generating process. %though in practice, selecting $d$ based on information criteria such as the Watanabe-Akaike information criterion (WAIC) \citep{watanabe2010asymptotic} is more recommended.  
 We used four Markov chain Monte Carlo (MCMC) chains with each chain having 1500 warm-up iterations and 500 sampling iterations.

For \textbf{BSN-NS} and \textbf{BSN-TS}, the horseshoe shrinkage hyperparameter ($\tau$ in \eqref{eq:horseshoe}) was set to 0.1, 0.2, and 0.3.  We found that  $\tau=0.1$ and 0.3 performed best for  $p=5$ and 15 settings respectively, indicating  less shrinkage may be preferred for higher dimensions.   The choice of the best $\tau$ is insensitive to the other factors. The reported results of \textbf{BSN-NS} and \textbf{BSN-TS} are based on these optimized $\tau$ values: $0.1$ for $p = 5$, and $0.3$ for $p = 15$. The hierarchical structure of the horseshoe prior can produce funnel-shaped posterior distributions, sometimes causing  poor mixing and convergence for MCMC chains \citep{piironen2017sparsity}. A noncentered parameterization \citep{papaspiliopoulos2007general} was used to address this.

\subsection{Evaluation metrics}
We evaluated the predictive performance for all methods by calculating the mean-squared-prediction-error (MSPE) on the test data: $\text{MSPE} = \sum_{i \in \mathcal{I}}(\hat{Y}_i - Y_i)^2/|\mathcal{I}|$, 	where $\mathcal{I} = \{n+1,..., N\}$ denotes the subject indices for the test set, and $|\cdot|$ denotes the cardinality. For the Bayesian methods (\textbf{BTR} and \textbf{BSN}), $\hat{Y}_i$ is defined as the posterior median of the predicted response for the $i$-th subject. For the frequentist methods, $\hat{Y}_i$ was computed using the point estimates of the model parameters.

For \textbf{BTR} and \textbf{BSN}, we evaluated the response coverage  (RC) of the predicted responses' 90\% credible intervals denoted as $[\hat{Y}_i^{L}, \hat{Y}_i^{U}]$, with ``$L$'' and ``$U$'' indicating the lower and upper bounds, respectively. The RC is defined as $\text{RC} = \sum_{i \in \mathcal{I}}I(g_i \in [\hat{Y}_i^{L}, \hat{Y}_i^{U}])/|\mathcal{I}|$,  which gives the proportion of test samples where the predictive intervals covered the true signal ($g_i$).  To further quantify the uncertainty of the predictions, we computed the length of the credible intervals as $\text{Len}(\hat{Y}_i) = \hat{Y}_i^U - \hat{Y}_i^L$.

 For the proposed tangent-space methods (\textbf{BSN-T}  and \textbf{BSN-TS}), we examined their sampled posterior distributions of $\bs{\Gamma}$ and $\mb{B}$ as follows. 
  Let $S$ be the total number of MCMC samples (pooled across all chains). Denote the $s$-th MCMC draw of $\bs{\Gamma}$ and  $b_j$ as  $\bs{\Gamma}^{(s)} =  (\bs{\gamma}_1^{(s)}$, ...,  $\bs{\gamma}_d^{(s)})$ and  $b_j^{(s)}$ for $j = 1,...,d$ respectively.   The column($j$)-wise similarity between  $\bs{\Gamma}^{(s)}$ and the true $\bs{\Gamma}$ was assessed using the absolute cosine similarity (ACS): $\text{ACS}\left(\bs{r}_j, \bs{r}_j^{(s)}\right) = \left|\bs{r}_j^T\bs{r}_j^{(s)}\right|$.   The coverage (denoted as 		``Cover'') of the credible intervals was computed for each entry in $\bs{\Gamma}$ and $\mb{B}$.  For parameter $z$ with 90\% credible interval $[z^L, z^U]$,  we define $\text{Cover}\left( z\right)  =  I \left(z \in [ z^L , z^U ] \right)$.  For each entry in $\bs{\Gamma}$ and $\mb{B}$,   we computed $\text{Cover}\left( \gamma_{j,k}\right)$ and $\text{Cover}\left( b_j \right)$. In computing  $\text{Cover}\left( \gamma_{j,k}\right)$, 
   the sign of $\bs{\gamma}_j^{(s)}$ was matched with that of  $\bs{\gamma}_j$ considering model identifiability. Let $l = \argmax_k (|\gamma_{j,k}|)$.    For each $s$ and $j$, we compared the sign of  $\gamma_{j,l}^{(s)}$ and the true $\gamma_{j,l}$, and if the signs disagreed, we multiplied  $\bs{\gamma}_j^{(s)}$  by -1.  The aligned $\bs{\gamma}_{j}^{(s)}$'s were used to construct the credible intervals of $\bs{\gamma}_j$.

%  Using this notation, we consider the following metrics for evaluation. 

 % to assess if the 
%inference results align with our expectations. The other competing methods use models that do not match our data generation process, and thus their parameter () values  are not comparable to the true values .  

 %The absolute value is used in \eqref{eq:acs} due to the identifiability properties in \Cref{thm:identifiability}: the response is invariant to sign-flips of $\bs{r}_j^{(s)}$. Thus, $\bs{r}_j^{(s)}$ and $-\bs{r}_j^{(s)}$ should be treated as equivalent directions by taking the absolute values. 

 % Additionally, we observed that the signs of $\gamma_{j,l}^{(s)}$ tend to stay consistent within individual Markov chains, yet could differ across chains due to varied initialization converging to approximations of $\bs{\Gamma}$ with distinct column signs. As an alternative, we could adjust the sign per each chain instead of per MCMC draw. 

\subsection{Results} \label{sec:sim:results}
We focus on presenting the simulation results for the ``correctly specified'' case with $p = 15$, $d = 4$, and SNR = 1, and
summarize the ``misspecified'' case.  The dimension $p = 15$ matches the HCP data in \Cref{sec:hcp}, and $d = 4$ corresponds to the largest \textbf{BSN} model considered in that section.  The complete set of results is provided in Supporting Information D. 

\begin{figure}
\hspace{-0.5cm}
\begin{subfigure}{.5\textwidth}
  \centering
\includegraphics[scale = 0.45]{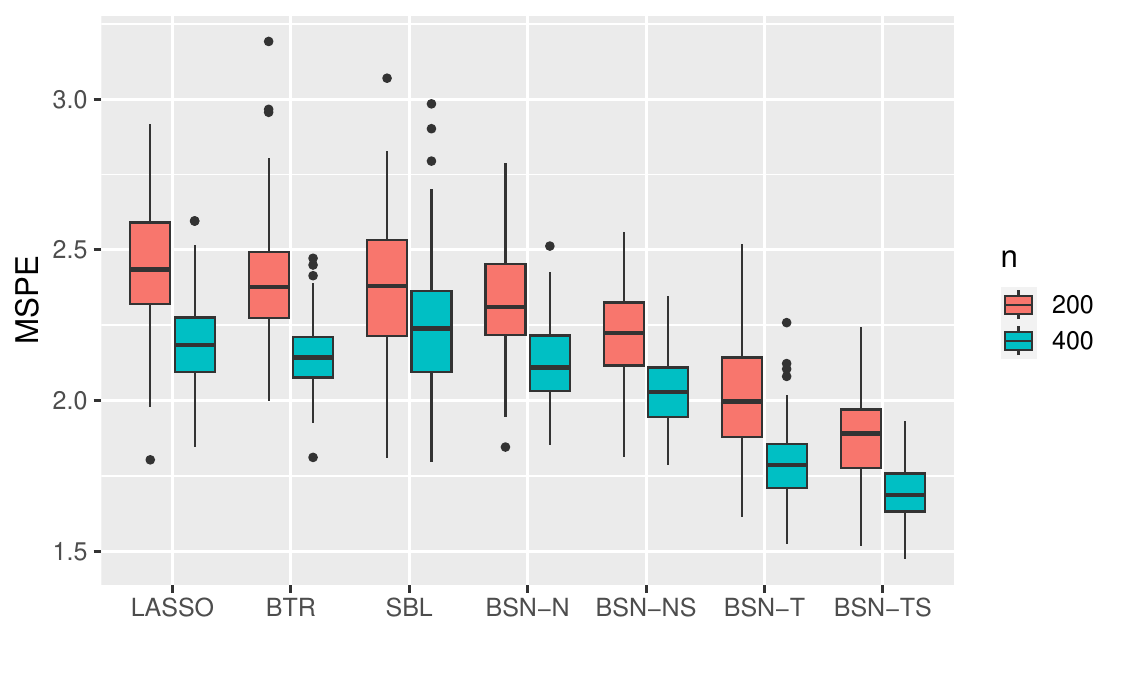}
\subcaption{MSPE}
\end{subfigure}
\begin{subfigure}{.5\textwidth}
\hspace{0.1cm}
\includegraphics[scale = 0.45]{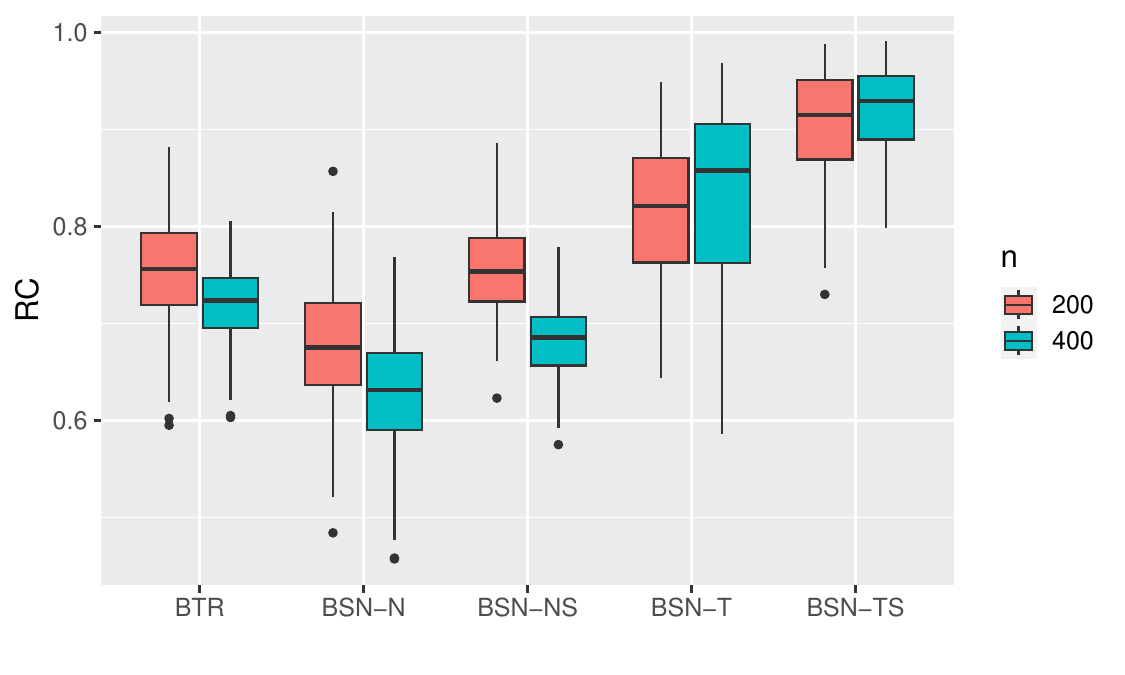}
\subcaption{RC}
\end{subfigure}
\caption{Mean-squared prediction error (MSPE) and response coverage (RC) on test sets for settings with $p= 15$ and $d = 4$ from 100 runs of the experiment in the ``correctly specified'' case, displayed in Panel~(a) and (b) respectively.}
 \label{fig:mspe}
\end{figure}

Panel~(a) of  \Cref{fig:mspe} shows the MSPE on the test sets for each method, obtained from 100 simulation runs.  Results from \textbf{LS}  and \textbf{FCR} are not reported here due to their poor performance across all settings  (\textbf{LS} severely overfits the training data and \textbf{FCR} disregards variance information by only using partial correlations as features).    As $n$ increases from $200$ to $400$, the predictive performance improves for all methods, with lower average MSPEs and reduced variability of the MSPEs.   The proposed \textbf{BSN-TS} performs best (lowest MSPE accompanied by lowest variance) among all techniques.   Within \textbf{BSN} approaches, methods based on the tangent space parametrization (\textbf{BSN-T} and \textbf{BSN-TS}) outperform the ones without such  parametrization (\textbf{BSN-N} and \textbf{BSN-NS}).  By comparing \textbf{BSN-T} with \textbf{BSN-TS}, we observe the performance gain from incorporating sparse sampling to $\bs{\Gamma}$.

  Panel~(b) of  \Cref{fig:mspe} compares the response coverage (RC) for the Bayesian approaches (\textbf{BTR} and \textbf{BSN}) with notably higher RC produced by  \textbf{BST-T} and \textbf{BST-TS} compared to the other methods. In particular, \textbf{BST-TS} achieves the highest RC at roughly 90\%, aligning with the specified level (90\%) to construct the credible intervals.  More training examples slightly improve the RC for \textbf{BSN-T} and \textbf{BSN-TS} but not for the other approaches. This resulted from tightened credible intervals with greater $n$, making it harder for less accurate intervals from the other methods to cover the true values.  In the ``misspecified'' case presented in  Supporting Information D,  \textbf{BSN-TS} maintains the lowest MSPEs at all perturbation levels ($\nu$)  though the  MSPEs deteriorate for all methods as $\nu$ increases.  \textbf{BSN-TS} also produces the highest RCs at around $90\%$ across perturbations.

 %the predictive accuracy declines for all methods.  \textbf{BSN-TS} continues to achieve the lowest MSPEs at all perturbation levels. \textbf{BSN-TS} produces the highest RSs at around ~90\%.  It is worth noting that the RSs improve  with greater noise ($\nu$)  for \textbf{BSN-N} and \textbf{BSN-NS}. This results from higher $\nu$ introducing more uncertainty, leading to wider credible intervals that are more likely to capture the true values, though the estimates (e.g. median of the samples) are potentially less accurate.

   %Panel~(b) of \Cref{fig:mspe} compares model performance when perturbation was introduced into the data generation process for the case with $n = 200$ and SNR = 1.   

%\Cref{fig:mspe} shows MSPE on the test sets for each method, obtained from 100 runs of the experiment.  To better visualize the differences among the better performing  methods, we exclude \textbf{LS}  and \textbf{FCR} from the subsequent analysis due to their poor performance across all settings.  \textbf{LS} severely overfits the training data.  \textbf{FCR} uses partial correlations as features which disregards the predictive information contained in the within-ROI variances. 

%Panel~(a) of \Cref{fig:mspe} presents results from Part~(1)  across training sample sizes ($n$). As $n$ increases from $200$ to $400$, performance improves for all methods, with lower MSPEs on average and reduced variability of the MSPEs.  The proposed \textbf{BSN-TS} shows the best performance, achieving the lowest average MSPE among all techniques  accompanied by a low  variance.  

\begin{figure}[htp] 
\centering
\includegraphics[scale=0.6]{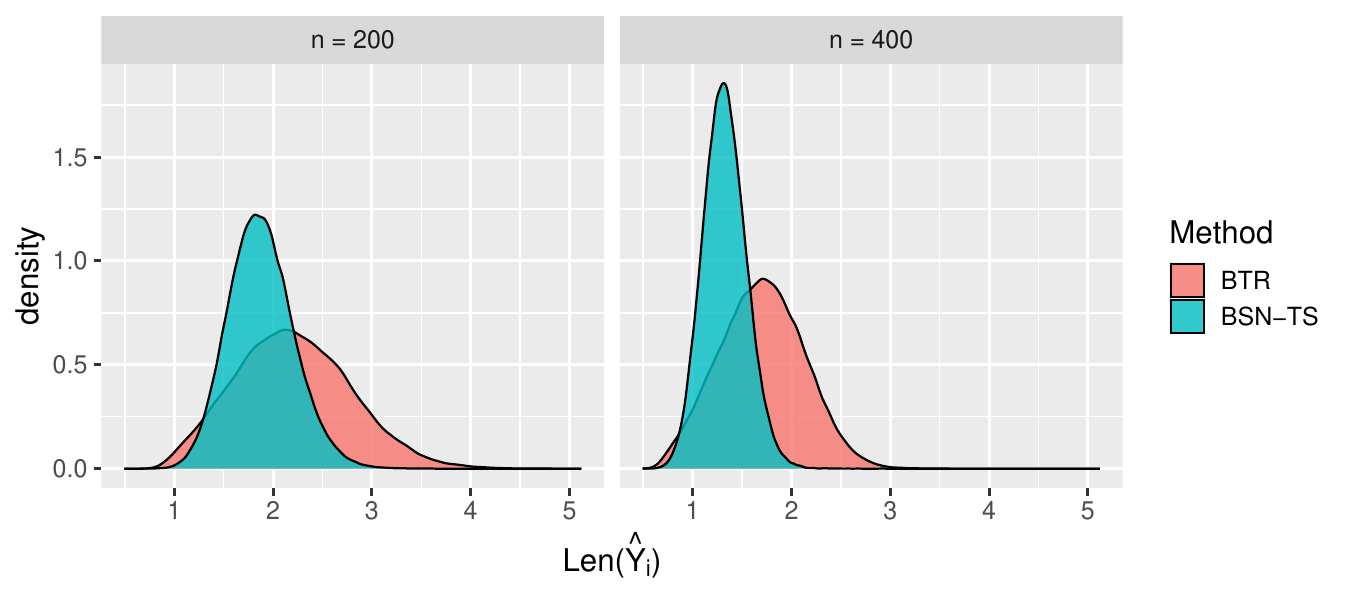}
\caption{Distribution of $\text{Len}(\hat{Y}_i)$ produced by BSN-TS and BTR for settings with $p = 15$ and $d = 4$ from 100 runs of the experiment in the ``correctly specified'' case. }
\label{fig:len}
\end{figure}

We further compared the lengths of 90\% credible intervals ($\text{Len}(\hat{Y}_i)$) obtained from \textbf{BSN-TS} to those from \textbf{BTR}.  For each run, we computed $\text{Len}(\hat{Y}_i)$ from \textbf{BSN} and \textbf{BTR} for each of the 1000 test data points, which were then concatenated across the 100 simulation runs. The distributions of  $\text{Len}(\hat{Y}_i)$ displayed in \Cref{fig:len} show that \textbf{BSN-TS} tends to produce narrower intervals compared to \textbf{BTR}, while still achieving better prediction accuracy and coverage as shown in \Cref{fig:mspe}. Also, \textbf{BSN-TS} generated more consistent interval lengths across different runs.  Similarly, in the ``misspecified'' case provided in  Supporting Information~D, \textbf{BSN-TS} produces more stable interval lengths versus \textbf{BTR} across perturbations. At $\nu = 0.05$ and 0.1, the credible intervals from \textbf{BSN-TS} are narrower than those from \textbf{BTR}, while at $\nu = 0.2$ the methods produce similar interval lengths but  \textbf{BSN-TS} achieves better prediction accuracy.

For the inference of parameters $\bs{\Gamma}$ and $\mb{B}$. we focus our analysis on \textbf{BSN-T} and \textbf{BSN-TS} since their model aligns with the the data generating model in the ``correctly specified'' case, allowing comparison with the true  parameter values.  In the ``misspecified'' case, the $\tilde{\bs{\Gamma}}_i$ differs across subjects but comparison with $\bs{\Gamma}$ is possible since $E(\tilde{\bs{\Gamma}}_i) = \bs{\Gamma}$. The absolute cosine similarity (ACS) compares the estimation accuracy for  each column of $\bs{\Gamma}$.   On average, \textbf{BSN-TS} produces higher ACS  than \textbf{BSN-T}, demonstrating a better estimation performance for $\bs{\Gamma}$. In the $d = 4$ setting, lower ACS is observed for  $\bs{\gamma}_2$ and $\bs{\gamma}_3$  compared to $\bs{\gamma}_1$ and $\bs{\gamma}_4$.  This is due to the higher scale of the coefficients $b_1$ and $b_4$ (versus $b_2$ and $b_3$) that enables easier recovery of the corresponding $\bs{\Gamma}$ columns  in our simulation settings.  Increasing $n$ from 200 to 400 improves the ACS for all $\bs{\gamma}_j$'s, especially for $\bs{\gamma}_2$ and $\bs{\gamma}_3$. In the ``misspecified'' case,  the same conclusion holds that  \textbf{BSN-TS} consistently achieves higher ACS  than \textbf{BSN-T}.

  %In the $d = 4$ setting, we observe lower ACS for $\bs{\gamma}_2$ and $\bs{\gamma}_3$  compared to $\bs{\gamma}_1$ and $\bs{\gamma}_4$. This is due to the higher scale of the coefficients $b_1$ and $b_4$ (versus $b_2$ and $b_3$)  enables easier recovery of the corresponding $\bs{\Gamma}$ columns,
%since they contribute more signal to construct the responses.   Increasing $n$ from 200 to 400 improves the ACS for all $\bs{\gamma}_j$'s, especially for $\bs{\gamma}_2$ and $\bs{\gamma}_3$.  

The coverage (Cover) was computed for $\gamma_{i,j}$'s and $b_j$'s,  and averaged across 100 simulation runs. For most entries, \textbf{BSN-TS} yields a higher average coverage than \textbf{BSN-T}, with values close to 90\%, aligning with our specified credible level.  For the other entries, \textbf{BSN-TS} and \textbf{BSN-T} have similar average coverages. \textbf{BSN-T} produces outliers with average coverage  below 50\% for some entries.  In the ``misspecified'' case, \textbf{BSN-TS}'s average coverages remain near 90\% with values for a few entries slightly below the ``correctly specified'' case, whereas \textbf{BSN-T} produces unstable results with low average coverage for many $\bs{\Gamma}$ entries.

 %\textbf{BSN-TS} yields average coverage values close to 90\% for all $b_j$'s, overall higher than \textbf{BSN-T}.   The coverage patterns are similar for $n = 200$ and $n = 400$, with \textbf{BSN-TS} consistently achieving higher and more stable average coverage of the true $\bs{\Gamma}$.  

%From the other settings covered in the Appendix,  we find that \textbf{BSN-TS} consistently achieves higher ACS  than \textbf{BSN-T} in Part~(1) and (2).  Similar to our observations from \Cref{fig:cover:g} and \Cref{fig:cover:b},  
%the average coverage of $\bs{\Gamma}$ and $\bf{B}$ produced by   \textbf{BSN-TS} stays close to 90\% across all Part~(1) cases, outperforming  \textbf{BSN-T}.  Under perturbations in Part~(2),      

Extending our analysis across various $p$, $d$, and SNR combinations (reported in Supporting Information D), we found \textbf{BSN-TS} consistently outperforms in the predictive performance measured by MSPE and RC.    %Despite sharing the same CP decomposition ($\bs{\Gamma}\mb{B}\bs{\Gamma}^T$) as our model, \textbf{SBL} shows unstable performance, producing outlying MSPEs in some cases. 
 Regarding $\text{Len}(\hat{Y}_i)$,  \textbf{BSN-TS} and \textbf{BTR} are comparable for $p=5$. With $p=15$, \textbf{BSN-TS} consistently generates narrower credible intervals than \textbf{BTR} in the ``correctly specified'' case.  In the ``misspecified'' case across different $p$ and $d$ combinations, $\text{Len}(\hat{Y}_i)$  are similar for \textbf{BSN-TS} and \textbf{BTR} at $\nu = 0.2$; \textbf{BSN-TS} produces narrower intervals at $\nu = 0.05$ and 0.1. For inference, \textbf{BSN-TS} consistently achieves higher ACS than \textbf{BSN-T} in all cases. The average coverage of $\bs{\Gamma}$ and $\bf{B}$ produced by  \textbf{BSN-TS} stays close to 90\% across all ``correctly specified'' settings, outperforming  \textbf{BSN-T}.  Under perturbations in ``misspecified'' case, the \textbf{BSN-TS}'s average coverages remain higher than \textbf{BSN-T} in general.

\section{Human Connectome Project} \label{sec:hcp}
The \textbf{BSN} methods were applied to the Human Connectome Project (HCP) data and compared with the alternative approaches considered in \Cref{sec:simulation}. The results are presented for the predictive performance and identification of the key brain regions associated with the Picture Vocabulary (Pic Vocab) score.

\subsection{Data description}
We studied the resting-state functional magnetic resonance imaging (rs-fMRI) data from the HCP S1200 release \citep{van2013wu}, which consists of behavioural and imaging data collected from healthy young adults. For each subject, the rs-fMRI data were collected over four complete 15-minute sessions, with 1200 timepoints per session. %Each session of the rs-fMRI data was preprocessed according to \cite{smith2013functional} prior to analysis.
To obtain region-mapped time series (signals) from the rs-fMRI data, we adopted a data-driven parcellation based on group spatial independent component analysis (ICA) \citep{calhoun2001method}  with $p = 15$ components described as ICA maps \citep{filippini2009distinct} and denoted them as Nodes 1-15. The output from the parcellation  includes data for $N=1003$ subjects from the S1200 release \citep{van2013wu}.  Considering the temporal dependency in the time series, we performed thinning of the signals based on the effective sample size (ESS) with details provided in Supporting Information E. The thinned signals for  subject $i$ were then used to compute the covariance matrix $\mb{M}_i$.  The Pic Vocab score, collected from the NIH Toolbox Picture Vocabulary Test and adjusted for age  \citep{gershon2013nih} was used as the continuous response variable.

 \subsection{Implementation details}
 We compared the same methods from \Cref{sec:simulation}, including \textbf{LS}, \textbf{LASSO}, \textbf{FCR}, \textbf{BTR}, \textbf{SBL}, and  \textbf{BSN}. The implementation details for \textbf{LS}, \textbf{LASSO}, and \textbf{FCR} are identical to the simulation study.  We randomly split data into a training set with $n = 300$ or 800 subjects, and used the remaining $(N-n)$ as the test set, reporting results separately for each $n$.  The experiment was repeated for 50 random data splits. \textbf{BTR} and \textbf{SBL}  were fit with rank 2, 3, and 4. \textbf{BSN} used  $d = 2, 3,$ and 4.   For each of \textbf{BTR}, \textbf{SBL}, and four  \textbf{BSN} variants, we report the results using the rank or $d$ that achieved the lowest MSPE averaged across 50 runs, with each run fitted with a random split of the data. The selected parameters are rank  3 ($n=300$) and rank  4  ($n=800$) for  \textbf{BTR}; rank 2 for \textbf{SBL}  and $d=4$ for all \textbf{BSN} methods at both $n$ values.

The original scale of the covariance matrices ($\mb{M}_i$'s) in the HCP data are on a higher order of magnitude compared to their tangent space representations ($\phi_{\mb{M}^{\ast}}(\mb{M}_i)$). 
To make the scales comparable across different parameterizations, we divided all covariance entries by $10^4$, which allows a fairer comparison between \textbf{BSN} with the na\"ive parameterization (\textbf{BSN-N} and \textbf{BSN-NS}) and that with the tangent representation (\textbf{BSN-T} and \textbf{BSN-TS}) fitted with the same prior on $\mb{B}$. The Pic Vocab response was standardized to have mean zero and unit variance before fitting the models. The priors we used for \textbf{BSN} remain unchanged from \Cref{sec:simulation}. For each \textbf{BSN} variant, we ran two MCMC chains with 1500 warm-up iterations and 500 sampling iterations for each chain. We tested  $\tau = 0.1, 0.2,$ and 0.3 for \textbf{BSN-NS} and \textbf{BSN-TS}, and report the results for $\tau = 0.3$ that yielded the best predictive performance.

We assessed the performance of the different methods by  MSPE and  compared the Bayesian methods by the length of the 90\% credible intervals ($\text{Len}(\hat{Y}_i)$). Following \Cref{sec:interpretation},  inference was conducted on $\bs{\Gamma}^T {\mb{M}^{\ast}}^{-1/2}$ to identify key regions predictive of the response.

\subsection{Results}
\begin{figure}
\centering 
\includegraphics[scale= 0.6]{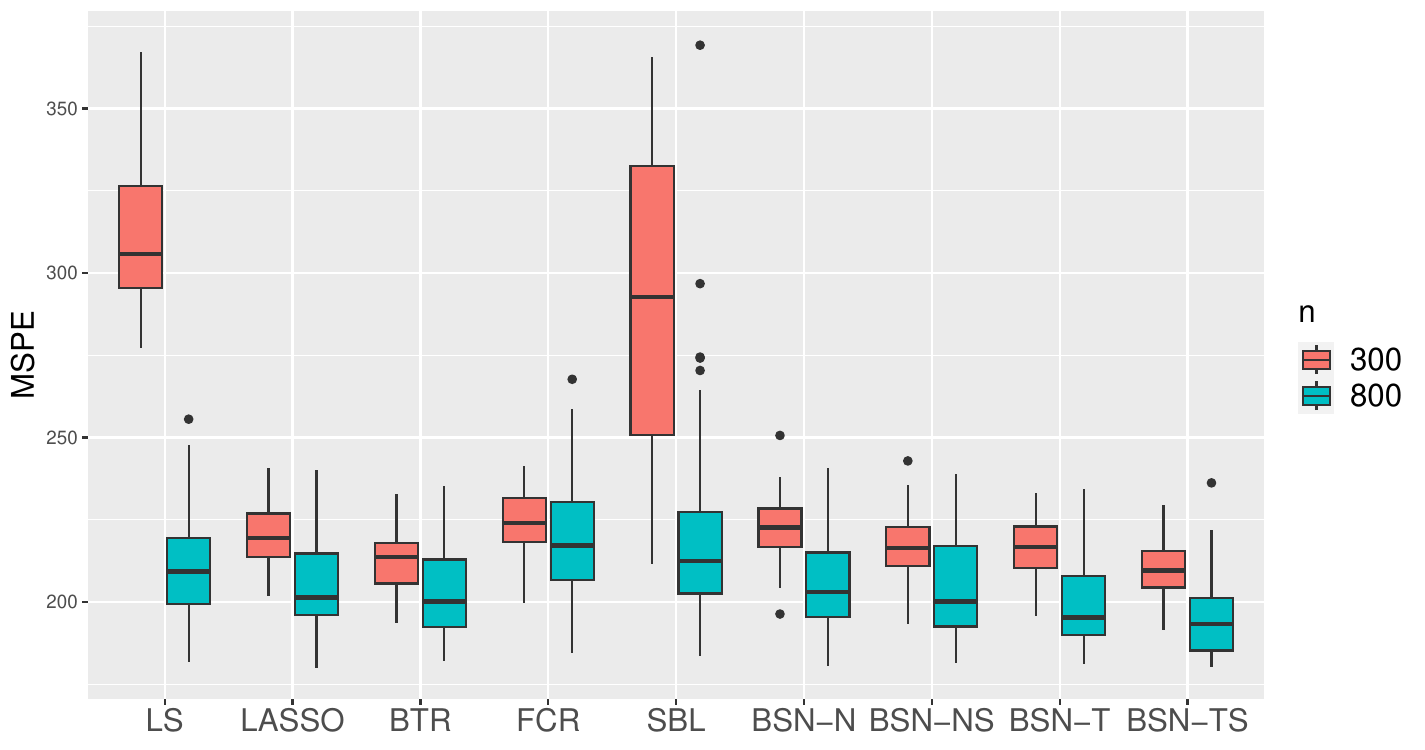}	
\caption{Mean-squared prediction error (MSPE) on test sets obtained from 50 random data splits of the HCP data.}
\label{fig:hcp:mspe}
\end{figure}

\Cref{fig:hcp:mspe} shows the MSPE on test sets over 50 random data splits with $n = 300$ and 800.  The y-axis is truncated to exclude outliers from \textbf{LS} and \textbf{SBL}.  On average, \textbf{BSN-TS} achieves the best performance at both sample sizes.  With $n = 300$, \textbf{LS} and \textbf{SBL} performs poorly, likely due to insufficient regularization.  Increasing $n$ improves predictions for all methods. 
  
  Among the \textbf{BSN} approaches,  tangent space models (\textbf{BSN-T} and \textbf{BSN-TS}) outperform na\"ive models (\textbf{BSN-N} and \textbf{BSN-NS}),  suggesting advantages in accounting for the nonlinear geometry of SPD matrices. Additionally, sparse sampling methods (\textbf{BSN-NS} and \textbf{BSN-TS}) improve over their non-sparse counterparts (\textbf{BSN-N} and \textbf{BSN-T}), demonstrating the performance gains  from regularized estimation of $\bs{\Gamma}$.

   Comparing the  distributions of $\text{Len}(\hat{Y}_i)$,  the average lengths are similar for \textbf{BSN-TS} and \textbf{BSN-S}, whereas \textbf{BSN-TS} produces  more consistent intervals lengths, especially for $n = 800$. (See Supporting Information E for more details.) 

\begin{figure}
\centering
\begin{tikzpicture}[>=latex,node distance=2em]
 \node(a){\includegraphics[width=10cm]{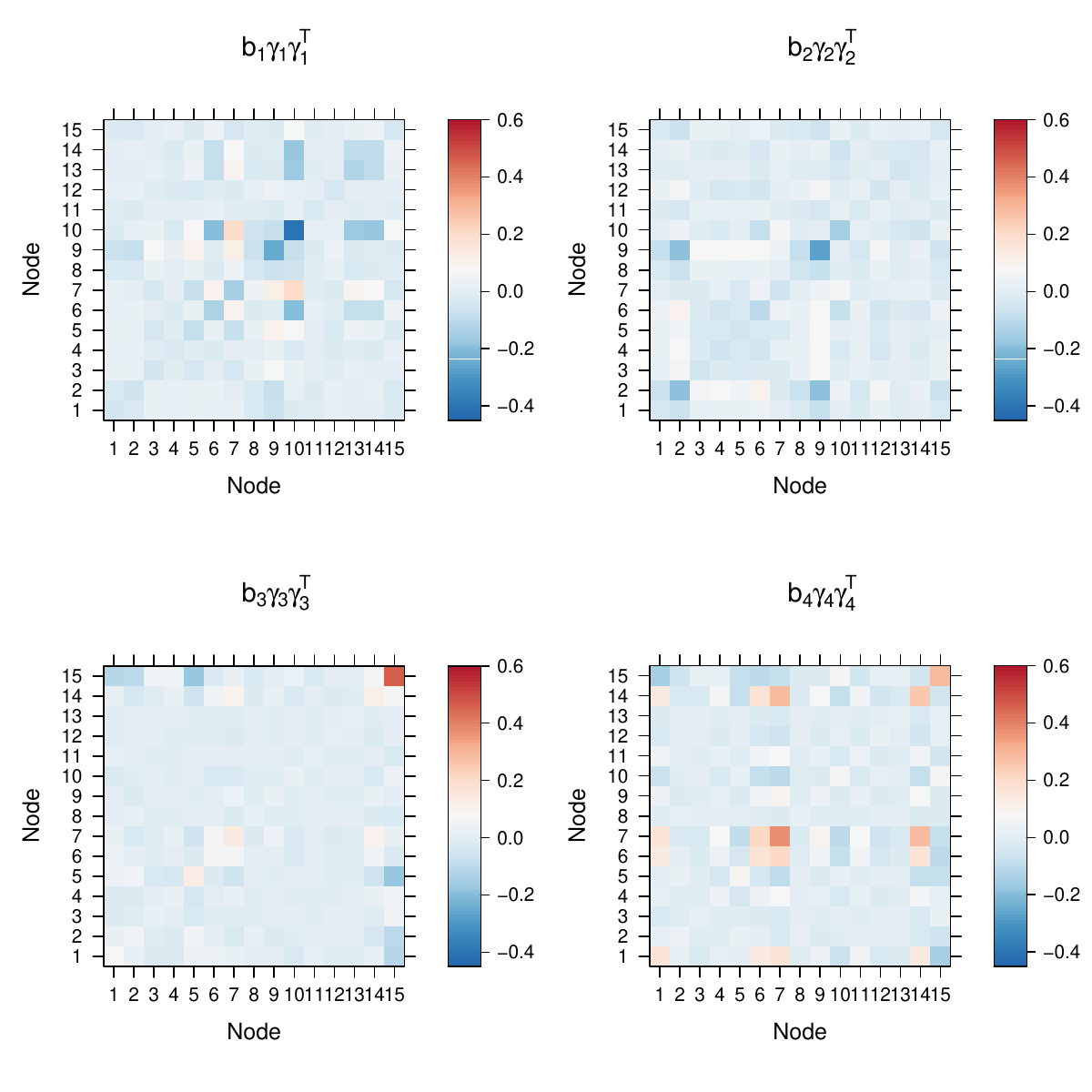}};
 \node[right=of a](b){\includegraphics[width=5.5cm]{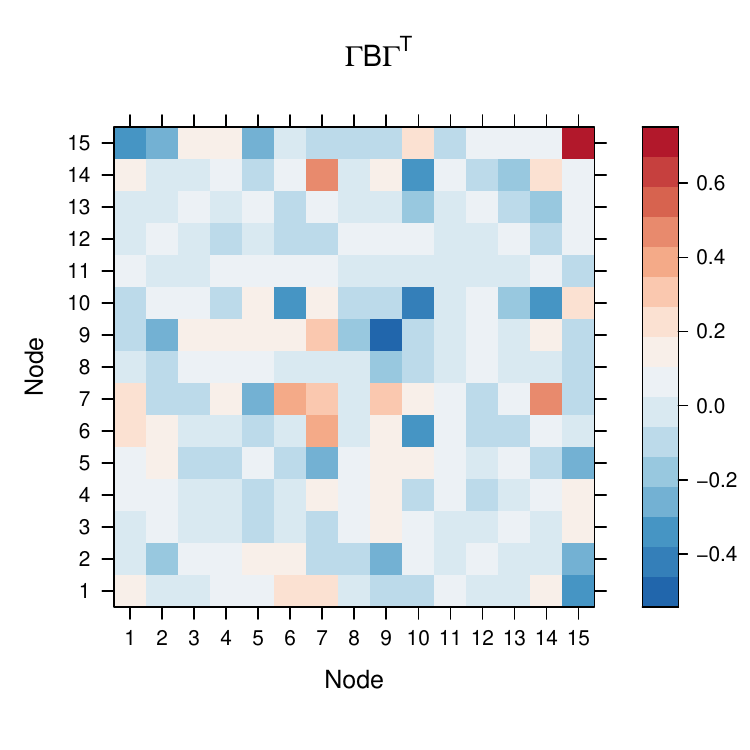}};
 \draw[->] (a) -- (b);
\end{tikzpicture}
\caption{Posterior mean of $b_1\gamma_1\gamma_1^T$,  $b_2\gamma_2\gamma_2^T$,  $b_3\gamma_3\gamma_3^T$,  $b_4\gamma_4\gamma_4^T$ , and  $\bs{\Gamma}\mb{B} \bs{\Gamma}^T$ computed from MCMC samples across the 50 runs, each with a different random data split of the HCP data.}
\label{fig:hcp:coefs}
\end{figure}

\Cref{fig:hcp:coefs} shows the  posterior mean estimates of $\bs{\Gamma}\mb{B}\bs{\Gamma}^T$  for the $n  = 800$ case, along with the rank-1 components $b_j \bs{\gamma}_j\bs{\gamma}_j^T$ for $j = 1,..., 4$, that constructs this matrix. The averages were taken over MCMC samples across 50 runs. Viewing $\bs{\Gamma}\mb{B}\bs{\Gamma}^T$ as the coefficient for $\phi_{\mb{M}^{\ast}}(\mb{M}_i)$, individual (rank-1) components are combined into a rank-4 coefficient, providing flexibility to model the potentially complex mapping from $\phi_{\mb{M}^{\ast}}(\mb{M}_i)$ to $Y_i$ based on multiple projections. 

%As illustrated in \Cref{sec:interpretation}, the matrix $\bs{\Gamma}^T {\mb{M}^{\ast}}^{-1/2}$ 
%extracts ``signal'' sources from the original times series.

\begin{figure}
\centering 
\includegraphics[scale= 0.27]{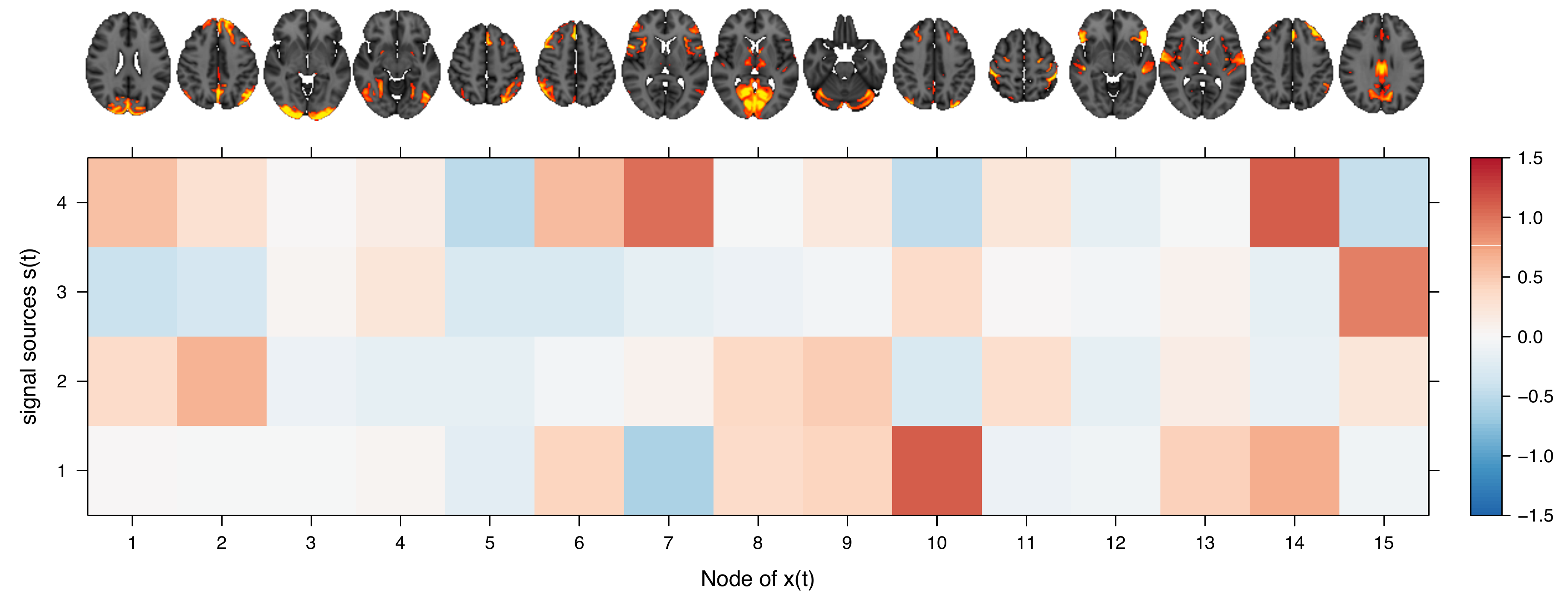}	
\caption{Posterior mean of $\bs{\Gamma}^T {\mb{M}^{\ast}}^{-1/2} \in \mathbb{R}^{d \times p} (=\mathbb{R}^{4 \times 15}$), with each component $\bs{\gamma}_j^T {\mb{M}^{\ast}}^{-1/2}~\in~\mathbb{R}^p$ ($=\mathbb{R}^{15}$) for $j \in  \{1, 2, 3, 4\}$ representing a ``subnetwork'' of the 15 network nodes, computed from aggregated MCMC samples across the 50 runs, each with a random data split of the HCP data. On the top of the figure, the axial slices most relevant to each node are displayed as brain images on the horizontal axis.}
\label{fig:hcp:network}
\end{figure}

For each $\bs{\Gamma}$ column, we aligned the signs of its MCMC samples for identifiability so that 
 the element with the largest absolute value averaged across samples is positive. The posterior mean of $\bs{\Gamma}^T {\mb{M}^{\ast}}^{-1/2}$ for $n = 800$ with these sign-adjusted samples is shown in \Cref{fig:hcp:network}.  The regions identified as important are Nodes 2, 10, 14, and 15, corresponding to entries with large scales in \Cref{fig:hcp:coefs} and positive lower bound of the 90\% credible intervals (see Supporting Information E and note that Node 7 has a large posterior median but a negative lower hound).  Referencing the study by \cite{smith2011network}, these regions relate to  cognition-language networks (Nodes 2, 10 and 14) and the default network (Node 15). Given that we are predicting language skills  measured as Pic Vocab, the significance of these cognition-language nodes aligns with our expectations. 

\section{Discussion}  \label{sec:discussion}
We developed a Bayesian regression method built on tangent space representations to account for the Riemannian geometry of SPD matrix predictors. Dimension reduction in the tangent space and sparse sampling on the Stiefel manifold result in a parsimonious model that effectively prevents overfitting.  Numerical experiments demonstrated that our method  achieves better performance than the alternatives in terms of prediction and inference.  While existing tangent space approaches are difficult to interpret, our model allows identifying important brain regions related to the response.

There are several ways we could enhance our proposal.  Our  implementation uses a fixed  $\tau$ in the horseshoe prior, which controls the global shrinkage of the rotation angles.  We found a higher dimensionality $p$ tends to prefer larger values for the global shrinkage parameter ($\tau$). This is likely due to more trigonometric terms in Givens rotations in~\eqref{eq:givens:generate} for larger $p$, thus requiring less shrinkage per term to maintain the  overall sparsity of $\bs{\Gamma}$.  Placing a hyperprior on $\tau$, as studied in \cite{piironen2017hyperprior}, would automatically control the level of sparsity.  Though we have focused on predicting continuous responses, our model  extends to classification and ordinal regression by modifying the likelihood function for generalized responses.  Exploring  nonlinear model formulations in the reduced dimensional space is also of interest for future work.

\section*{Acknowledgements}
 This work was supported by the National Institute of Health (NIH Grant No. 5 R01 MH099003). Data were provided by the Human Connectome Project, WU-Minn Consortium (Principal Investigators: David Van Essen and Kamil Ugurbil; 1U54MH091657) funded by the 16 NIH Institutes and Centers that support the NIH Blueprint for Neuroscience Research; and by the McDonnell Center for Systems Neuroscience at Washington University.

\section*{Data availability statement}
The Human Connectome Project (HCP) data that support the findings in this paper are openly available at  \url{https://www.humanconnectome.org/study/hcp-young-adult.}

\section*{Supporting Information}
Web Appendices, Tables, and Figures referenced in Sections 2-5 are available with this paper at the Biometrics website on Wiley Online Library.  
R code for the competing methods can be found in their authors' Github repositories: https://github.com/clbwvr/FC-SIM for \textbf{FCR}, https://github.com/wangronglu/Symmetric-Bilinear-Regression for \textbf{SBL}, and \seqsplit{https://github.com/rajguhaniyogi/Bayesian-Tensor-Regression} for \textbf{BTR}. The code implementing our proposal is available at  https://github.com/xmengju/BSN. 

\bibliography{references}

\newpage
\begin{center}
	{\Large{\textbf{Supporting Information}}}
\end{center}

\setcounter{figure}{0}

\hspace{1cm}
\section*{Appendix A: Identifiability}
\textbf{Theorem 1}:  Let $\boldsymbol{\Gamma}_1 = \left(\boldsymbol{\gamma}_1^{(1)},..., \boldsymbol{\gamma}_d^{(1)} \right) \in \mathbb{R}^{p \times d}$ and  $\boldsymbol{\Gamma}_2 = \left(\boldsymbol{\gamma}_1^{(2)},..., \boldsymbol{\gamma}_d^{(2)} \right) \in \mathbb{R}^{p \times d}$  satisfy $\boldsymbol{\Gamma}_1^T\boldsymbol{\Gamma}_1 = \boldsymbol{\Gamma}_2^T\boldsymbol{\Gamma}_2   = \mathbf{I}_{d}$; 
$\mathbf{B}_1 = \text{Diag}(b_1^{(1)},..., b_d^{(1)})$  and  $\mathbf{B}_2 = \text{Diag}(b_1^{(2)},..., b_d^{(2)})$;  and $\mu_1, \mu_2 \in \mathbb{R}$. If 
\begin{equation*}\label{eq:index}
 \mu_1 + \langle \phi_{\mb{M}^{\ast}}(\mathbf{M}), \boldsymbol{\Gamma}_1 \mathbf{B}_1 \boldsymbol{\Gamma}_1^T\rangle  = \mu_2 + \langle \phi_{\mb{M}^{\ast}}(\mathbf{M}),\boldsymbol{\Gamma}_2 \mathbf{B}_2 \boldsymbol{\Gamma}_2^T\rangle  \ \text{for any} \  \mb{M}  \in \text{Sym}_p^{+},  
\end{equation*}
then there exist a permutation $\pi(1), ..., \pi(d)$ of $1,..., d$, and $l_j \in \{0, 1\}$  such that for $j = 1,..., d$, 
$$\boldsymbol{\gamma}_j^{(1)} = (-1)^{l_j} \boldsymbol{\gamma}_{\pi(j)}^{(2)}, \  b_j^{(1)} =   b_{\pi(j)}^{(2)}, \ \text{and} \ \mu_1 = \mu_2.$$

\textbf{Proof}:  $\phi_{\mb{M}^{\ast}}(\mathbf{M})$ spans the space of symmetric matrices for $\mb{M} \in \text{Sym}_p^{+}$ and a fixed $\mb{M}^{\ast}$. We first show that $\mu_1 = \mu_2$ by letting  $\phi_{\mb{M}^{\ast}}(\mathbf{M}) =\mb{0}_p$. Under this condition if 
$$ \mu_1 + \langle \phi_{\mb{M}^{\ast}}(\mathbf{M}), \boldsymbol{\Gamma}_1 \mathbf{B}_1 \boldsymbol{\Gamma}_1^T\rangle  = \mu_2 + \langle \phi_{\mb{M}^{\ast}}(\mathbf{M}),\boldsymbol{\Gamma}_2 \mathbf{B}_2 \boldsymbol{\Gamma}_2^T\rangle, $$
then $\mu_1 = \mu_2$. 

Next, for any symmetric matrix  $\phi_{\mb{M}^{\ast}}(\mathbf{M})$, if $$\langle \phi_{\mb{M}^{\ast}}(\mathbf{M}), \boldsymbol{\Gamma}_1 \mathbf{B}_1 \boldsymbol{\Gamma}_1^T\rangle  =  \langle \phi_{\mb{M}^{\ast}}(\mathbf{M}),\boldsymbol{\Gamma}_2 \mathbf{B}_2 \boldsymbol{\Gamma}_2^T\rangle, $$
we have $\boldsymbol{\Gamma}_1 \mathbf{B}_1 \boldsymbol{\Gamma} = \boldsymbol{\Gamma}_2 \mathbf{B}_2 \boldsymbol{\Gamma}_2^T$. Since $\boldsymbol{\Gamma}_1^T\boldsymbol{\Gamma}_1 = \boldsymbol{\Gamma}_2^T\boldsymbol{\Gamma}_2   = \mathbf{I}_{d}$ and $\mb{B}_1$ and $\mb{B}_2$ are diagonal matrixes, $\boldsymbol{\Gamma}_1 \mathbf{B}_1 \boldsymbol{\Gamma}$ and $\boldsymbol{\Gamma}_2 \mathbf{B}_2 \boldsymbol{\Gamma}$   represent eigenvalue decompositions. The eigenvectors are the columns of $\mb{\Gamma}_1$ and $\mb{\Gamma}_2$, and the eigenvalues are the diagonal elements of $\mb{B}_1$ and $\mb{B}_2$.  Such decompositions are unique up to permutations of eigenvectors together with the corresponding eigenvalues, and sign flips of each eigenvector, and thus the proof is complete.

\section*{Appendix B: Givens rotations}

\subsection*{An illustrative example}
For $d = 5$ and $d=3$, the elimination order of each entry is  illustrated below in circled numbers, where $X$'s denote entries that do not require elimination:
\begin{equation*}
\begin{pmatrix}
X & X & X  \\
\circled{1} & X & X  \\
\circled{2} & \circled{5} & X \\
\circled{3} & \circled{6} & \circled{8} \\
\circled{4} & \circled{7}& \circled{9} \\
\end{pmatrix}. 
\end{equation*}

Applying Givens rotation matrices to an orthonormal $\bs{\Gamma} \in \mathbb{R}^{5 \times 3}$ yields
\begin{equation*} 
\mathbf{G}_{3,5}(\theta_{3,5}) \mathbf{G}_{3,4}(\theta_{3,4})  \mathbf{G}_{2,5}(\theta_{2,5})  \mathbf{G}_{2,4}(\theta_{2,4})  \mathbf{G}_{2,3}(\theta_{2,3})  \mathbf{G}_{1,5}(\theta_{1,5})  \mathbf{G}_{1,4}(\theta_{1,4})  \mathbf{G}_{1,3}(\theta_{1,3}) \mathbf{G}_{1,2}(\theta_{1,2})\boldsymbol{\Gamma} = \mathbf{I}_{5\times 3},
\end{equation*}

The generation of $\bs{\Gamma}$ from $\mb{I}_{5\times 3}$ is given by 
\begin{equation*}
  \mathbf{G}_{1,2}^T(\theta_{1,2})  \mathbf{G}_{1,3}^T(\theta_{1,3})  \mathbf{G}_{1,4}^T(\theta_{1,4})   \mathbf{G}_{1,5}^T(\theta_{1,5})  \mathbf{G}_{2,3}^T(\theta_{2,3})  \mathbf{G}_{2,4}^T(\theta_{2,4})  \mathbf{G}_{2,5}^T(\theta_{2,5}) \mathbf{G}_{3,4}^T(\theta_{3,4}) \mathbf{G}_{3,5}^T(\theta_{3,5}) 
 \mathbf{I}_{5\times 3} = \boldsymbol{\Gamma}.  
 \end{equation*}

\subsection*{Remark: range of  $\theta_{i,j}$'s} 
 
Let $\bs{\Gamma}_0$ be a $p \times d$ orthonormal matrix generated with the rotation angles $\theta_{i,j}^{(0)} \in [-\pi/2, \pi/2]$ for $i = 1,..., d$, and $j = i+1,..., p$.  Define $\mathcal{S}_{0}$ as the set containing all matrices generated by column sign flips of $\bs{\Gamma}_{0}$:  $\mathcal{S}_{0}= \{ \mb{F}_{\mb{l}} \bs{\Gamma}_0 | \mb{l} \in { \{0, 1 \}}^d \}$, where $\mb{F}_{\mb{l}} = \text{Diag}( (-1)^{l_1}, ..., (-1)^{l_d})$  represents a column sign flip matrix based on the vector $\mb{l} = (l_1,...,l_d)$ with $l_j \in \{-1,0\}$. By identifiability results stated in Theorem 1, 
all $\bs{\Gamma} \in \mathcal{S}_0$ yield the same response  under the regression model.  Our restricted angle range $[-\pi/2, \pi/2]$ for all $\theta_{i,j}$'s would only generate $\bs{\Gamma}_0$ from the set $\mathcal{S}_0$, whereas the wider range used by \cite{pourzanjani2021bayesian} can generate all matrices in $\mathcal{S}_0$, creating posterior multi-modality. To see this, let  $\mathcal{I} = \{i, | l_i = -1,  i = 1,..., d\}$ index sign-flipped columns. For a non-empty $\mathcal{I}$, the matrix  $\mb{F}_{
\mb{l}} \bs{\Gamma}_0$ is generated from rotation angles $\tilde{\theta}_{i,j}$'s where    $\tilde{\theta}_{i,i+1} = \theta_{i,i+1} - \text{Sign}(\theta_{i,i+1})\pi$ for $i \in \mathcal{I}$ and $\tilde{\theta}_{i,j}  = \theta_{i,j}^{(0)}$ otherwise. The angles $\tilde{\theta}_{i,j}$'s lie in \cite{pourzanjani2021bayesian}'s range but not ours.  

\section*{Appendix C: Model interpretation}

Denote the eigenvalue decomposition of $\mb{C}_{s,i} = \mb{U}_i \bs{\Lambda}_{s_i}\mb{U}_i^T$, where $\mb{U}_i = (\mb{u}_{i,1},..., \mb{u}_{i,d})$ contains the eigenvectors and  $\bs{\Lambda}_{s_i} = \text{Diag}(\lambda_{s_i,1}, ..., \lambda_{s_i,d})$ contains the eigenvalues.  The equivalence of $Y_i = \mu + \langle  \text{Log}(\mb{C}_{s,i}), \mb{B} \rangle + \epsilon_i$  and the regression model  
 \begin{equation} \label{eq:reg}
 	Y_i = \mu +  \langle  \bs{\Gamma}^T \phi_{\mb{M}^{\ast}}(\mb{M}_{i}) \bs{\Gamma}, \mb{B} \rangle
 \end{equation}
 under the generative model 
 \begin{equation} \label{eq:gen}
 	{\mb{M}^{\ast}}^{-1/2}\mb{x}_{i}(t) =  \bs{\Gamma} \mb{s}_{i}(t) +  \mb{V}_i \bs{\eta}_{i}(t)
 \end{equation}

 can be derived by, based on $\bs{\Gamma}$ satisfying $\bs{\Gamma}^T\bs{\Gamma} = \mb{I}_d$ and  $\bs{\Gamma}^T\mb{V}_i = \mb{0}_{d,(p-d)}$, 
\begin{align*}
\langle \text{Log}(\mb{C}_{s,i} ) ,  \mb{B}  \rangle &= \langle \mb{U}_i \text{Log}(\bs{\Lambda}_{s,i})\mb{U}_i^T, \mb{B} \rangle \\
&= \langle \bs{\Gamma}^T\bs{\Gamma} \mb{U}_i \text{Log}(\bs{\Lambda}_{s,i})\mb{U}_i^T\bs{\Gamma}^T\bs{\Gamma} + \bs{\Gamma}^T \mb{V}_i \text{Log}(\mb{C}_{\eta,i})\mb{V}_i^T \bs{\Gamma}, \mb{B} \rangle \\
&= \langle \bs{\Gamma}^T (\bs{\Gamma} \mb{U}_i \text{Log}(\bs{\Lambda}_{s,i})\mb{U}_i^T \bs{\Gamma}^T  + \mb{V}_i \text{Log}(\mb{C}_{\eta,i})\mb{V}_i^T) \bs{\Gamma}, \mb{B} \rangle \\
&=  \langle \bs{\Gamma}^T \text{Log}(\tilde{\mb{M}}_i) \bs{\Gamma}, \mb{B} \rangle \\
&=  \langle \bs{\Gamma}^T\phi_{\mb{M}^{\ast}}(\mb{M}_i) \bs{\Gamma}, \mb{B} \rangle,
\end{align*}
where the second last equality follows from $\tilde{\mb{M}}_i = \bs{\Gamma} \mb{C}_{s,i}  \bs{\Gamma}^T + \mb{V}_i \mb{C}_{\eta,i}  \mb{V}_i^T$ implied by \eqref{eq:gen}. This shows that  the model~\eqref{eq:reg} essentially relates  the   (logarithm of the)  covariance of the``signal'' sources ($\mb{s}_i(t)$) to the response.  
From \eqref{eq:gen}, the source signals are computed as $\mb{s}_i(t) = \bs{\Gamma}^T {\mb{M}^{\ast}}^{-1/2} \mb{x}_i(t)$. Thus, the matrix $\bs{\Gamma}^T {\mb{M}^{\ast}}^{-1/2}$ is  key for interpretation as it identifies important ROIs that construct $\mb{s}_i(t)$, the covariance of which relates to the response.   
 
To interpret $\mb{B}$, we consider an individual with covariance $\mb{M}$ and its tangent space representation $\mb{T} = \phi_{\mb{M}^{\ast}}(\mb{M})$. 
 Let the orthogonal complement of $\bs{\Gamma}$ be $\bs{\Gamma}^{\perp}$, and  forms a basis in $\mathbb{R}^p$.    The interpretation of $\mb{B}$ is linked to the deviation  of   $\mb{M}$  from $\mb{M}^{\ast}$ along  $\bs{\Gamma}$ in the tangent space.   
 We can write 
 $\mb{T} = \mb{T} - \phi_{\mb{M}^{\ast}}(\mb{M}^{\ast})$ since $\phi_{\mb{M}^{\ast}}(\mb{M}^{\ast})=0$ and thus interpret $\mb{T}$ as the deviation of $\mb{M}$ from $\mb{M}^{\ast}$ in the tangent space. We consider deviations along directions of $\mb{Q}$ by letting $\mb{T} = \phi_{\mb{M}^{\ast}}(\mb{M}) = \mb{Q} \bs{\Delta}\mb{Q}^T$, where $\bs{\Delta} = \text{Diag}(\delta_1,..., \delta_p)$ with $\delta_j \in \mathbb{R}$ for $j = 1, ..., p$.  Based on model \eqref{eq:reg}, it can be shown that $$E_{\mb{M}^{\ast}}(Y|\mb{M}) - E_{\mb{M}^{\ast}}(Y|\mb{M}^{\ast})  = \sum_{j=1}^d \delta_j b_j.$$ 
 Therefore, the contribution to the response from $\mathbf{M}$ deviating from $\mathbf{M}^{\ast}$ in the $\boldsymbol{\Gamma}$ directions in the tangent space is additive in $b_j$'s. Each $b_j$ corresponds to the contribution of the deviation of $\mb{M}$ from $\mb{M}^{\ast}$ in the $\bs{\gamma}_j$ direction. Deviations in the  $\bs{\Gamma}^{\perp}$ directions do not impact the response. 

\section*{Appendix D: Simulation results}

\textbf{Mean-squared prediction error (MSPE)}

\begin{figure}[H]
\begin{subfigure}{.49\textwidth}
  \centering
\includegraphics[width = 9cm, height = 6cm]{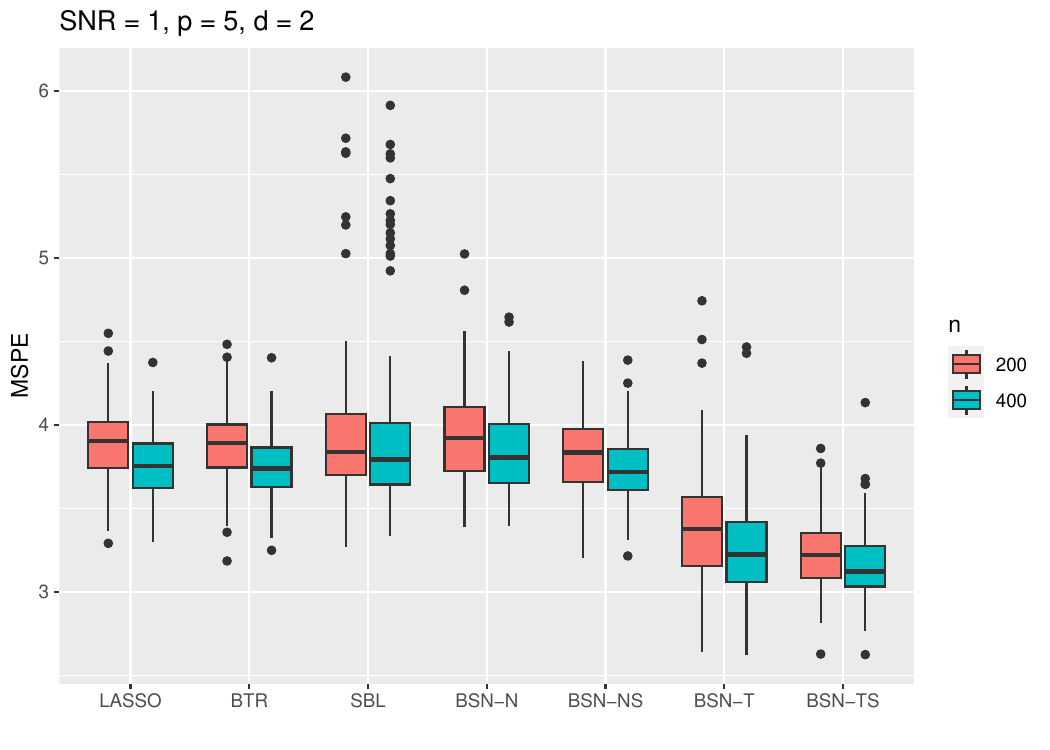}
\end{subfigure}
\begin{subfigure}{.49\textwidth}
\hspace{1cm}
\includegraphics[width = 9cm, height = 6cm]{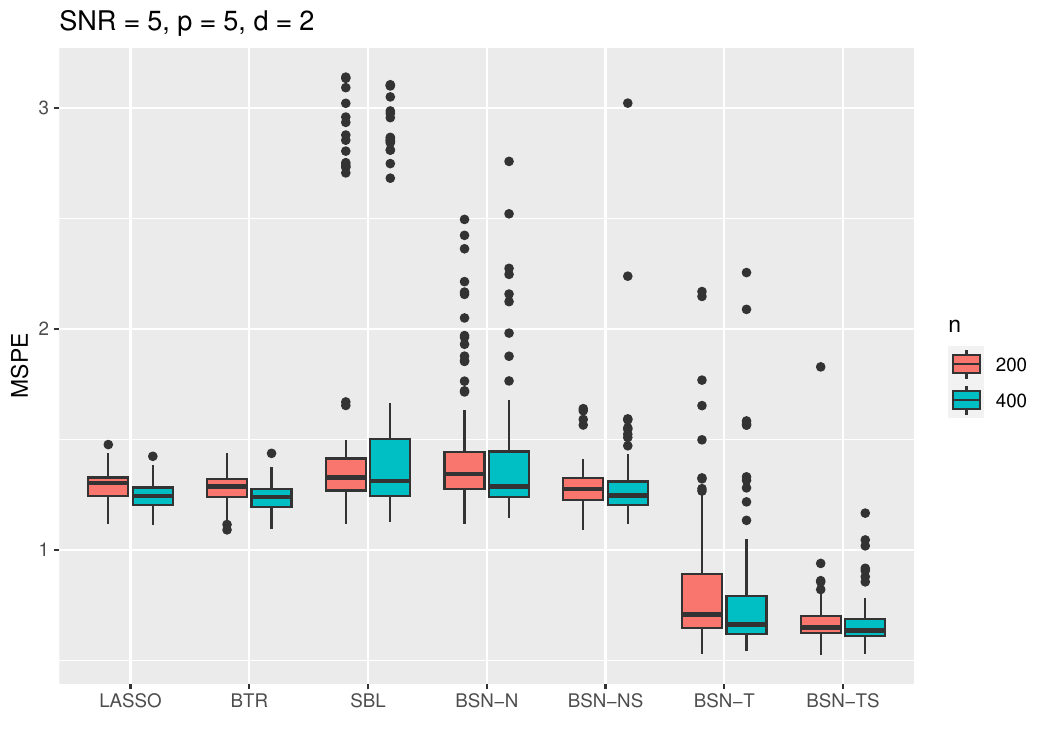}
\end{subfigure}
\caption{Mean-squared prediction error (MSPE) on test sets for settings with $p= 5$ and $d = 2$ under SNR = 1 (left panel) and SNR = 5 (right panel), from 100 runs in the ``correctly specified'' case.}
\end{figure}

\begin{figure}[H]
\begin{subfigure}{.49\textwidth}
  \centering
\includegraphics[width = 9cm, height = 6cm]{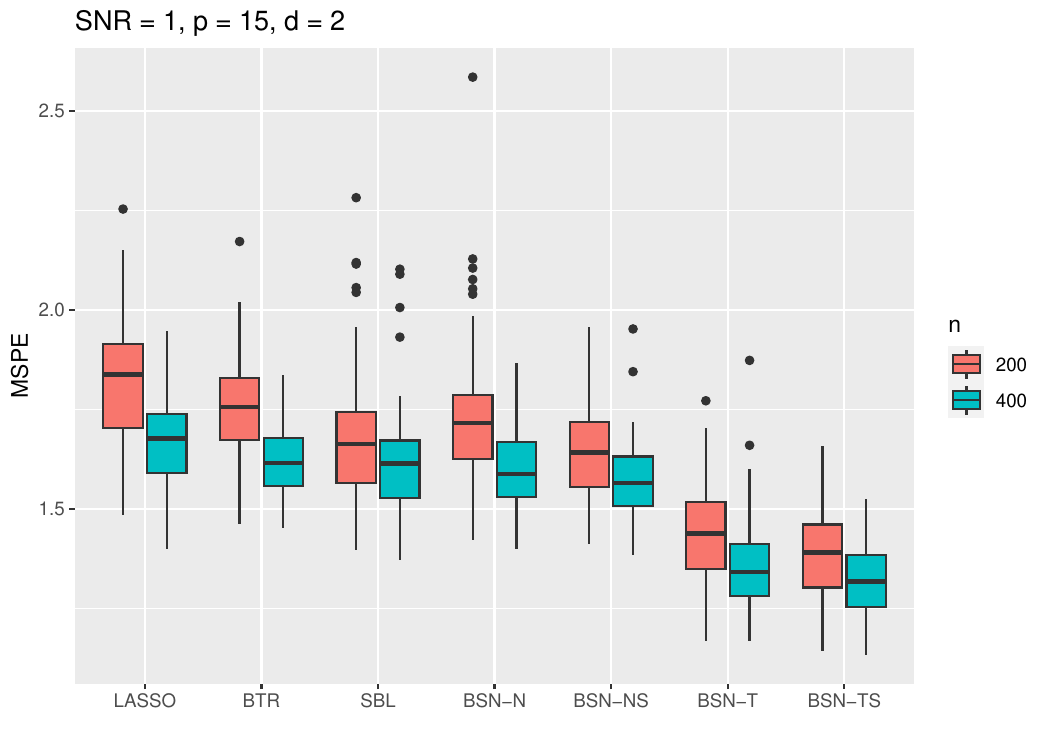}
\end{subfigure}
\begin{subfigure}{.49\textwidth}
\hspace{1cm}
\includegraphics[width = 9cm, height = 6cm]{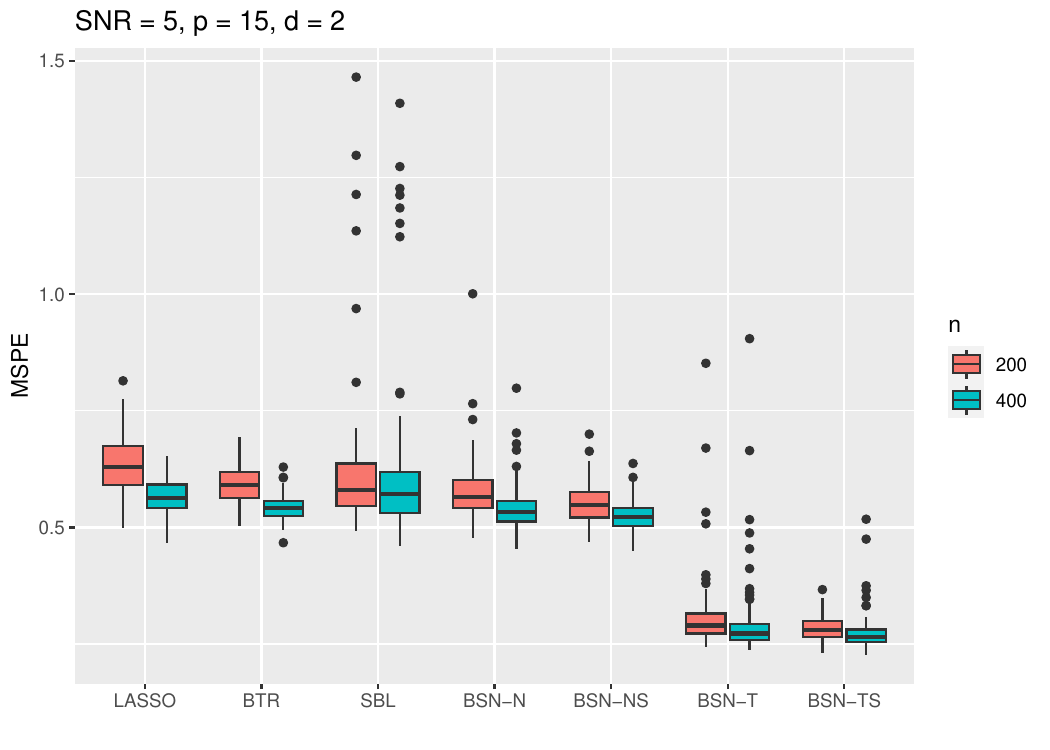}
\end{subfigure}
\caption{Mean-squared prediction error (MSPE) on test sets for settings with $p= 15$ and $d = 2$ under SNR = 1 (left panel) and SNR = 5 (right panel), from 100 runs in the ``correctly specified'' case.}
\end{figure}

\begin{figure}[H]
\begin{subfigure}{.49\textwidth}
  \centering
\includegraphics[width = 9cm, height = 6cm]{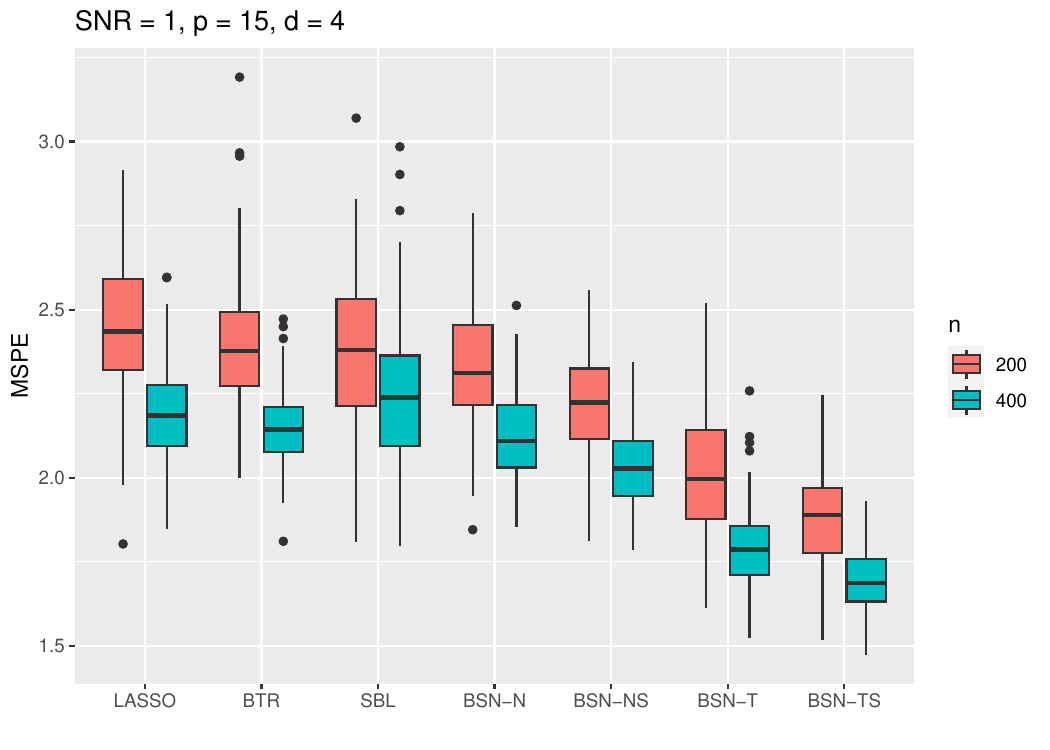}
\end{subfigure}
\begin{subfigure}{.49\textwidth}
\hspace{1cm}
\includegraphics[width = 9cm, height = 6cm]{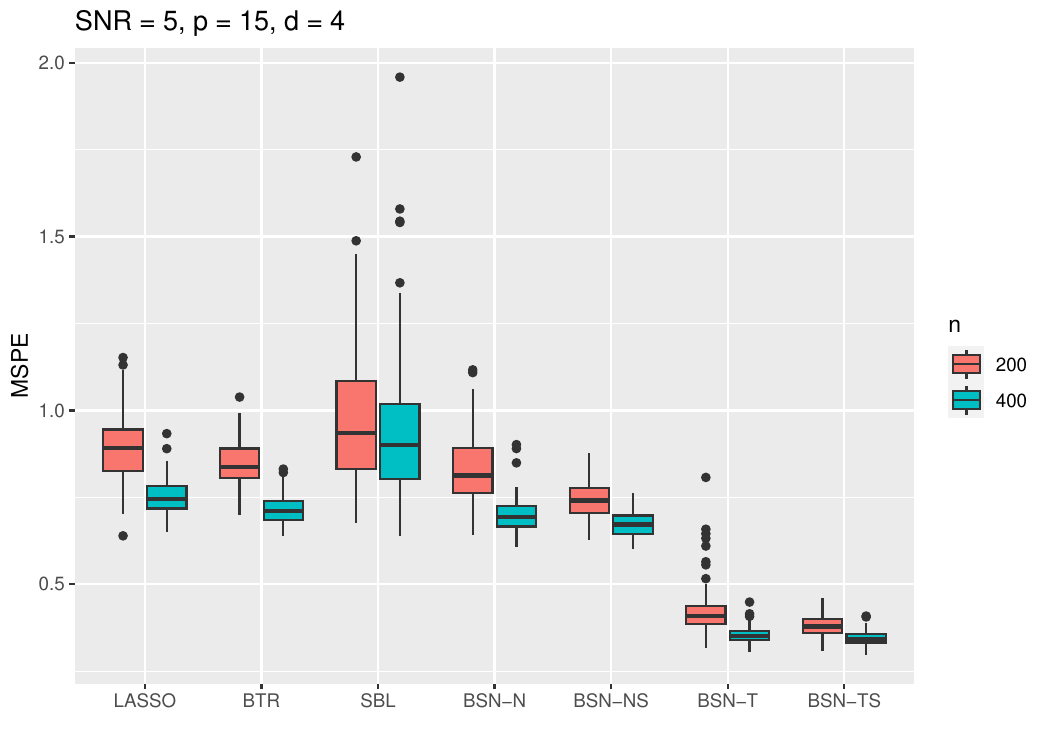}
\end{subfigure}
\caption{Mean-squared prediction error (MSPE) on test sets for settings with $p= 15$ and $d = 4$ under SNR = 1 (left panel) and SNR = 5 (right panel), from 100 runs in the ``correctly specified'' case.}
\end{figure}

\begin{figure}[H]
  \centering
\includegraphics[width = 13cm, height = 8.5cm]{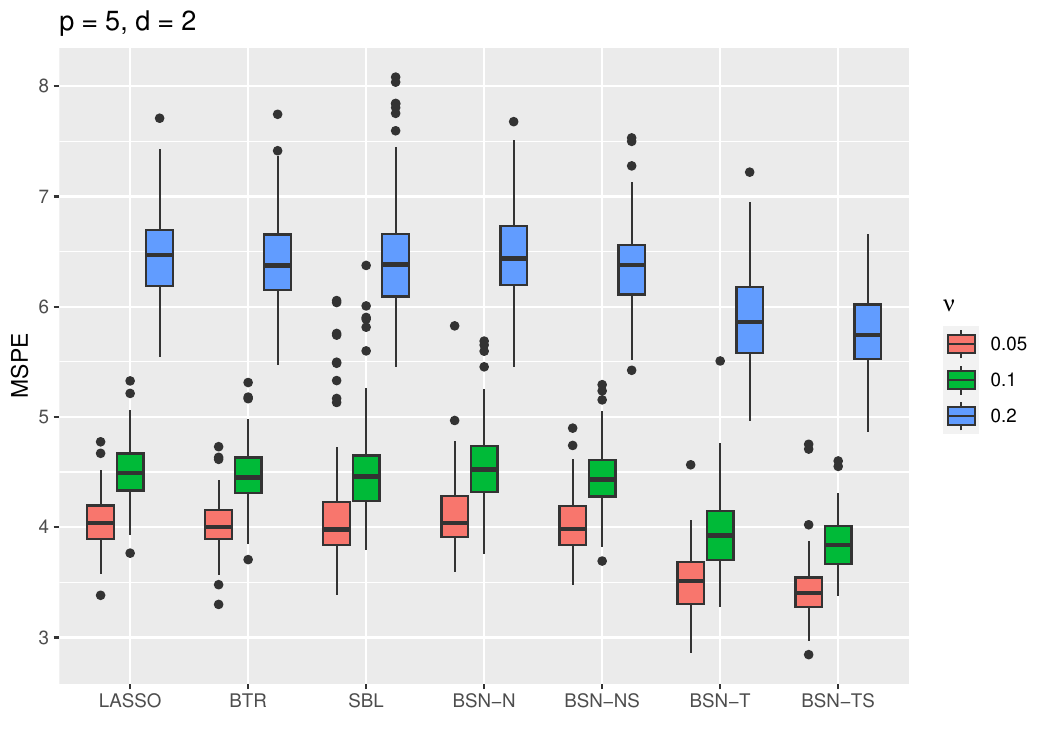}
\caption{Mean-squared prediction error (MSPE) on test sets for settings with $p= 5$ and $d = 2$ under SNR = 1 and $n = 200$ from 100 runs in the ``misspecified'' case.}
\end{figure}

\begin{figure}[H]
  \centering
\includegraphics[width = 13cm, height = 8.5cm]{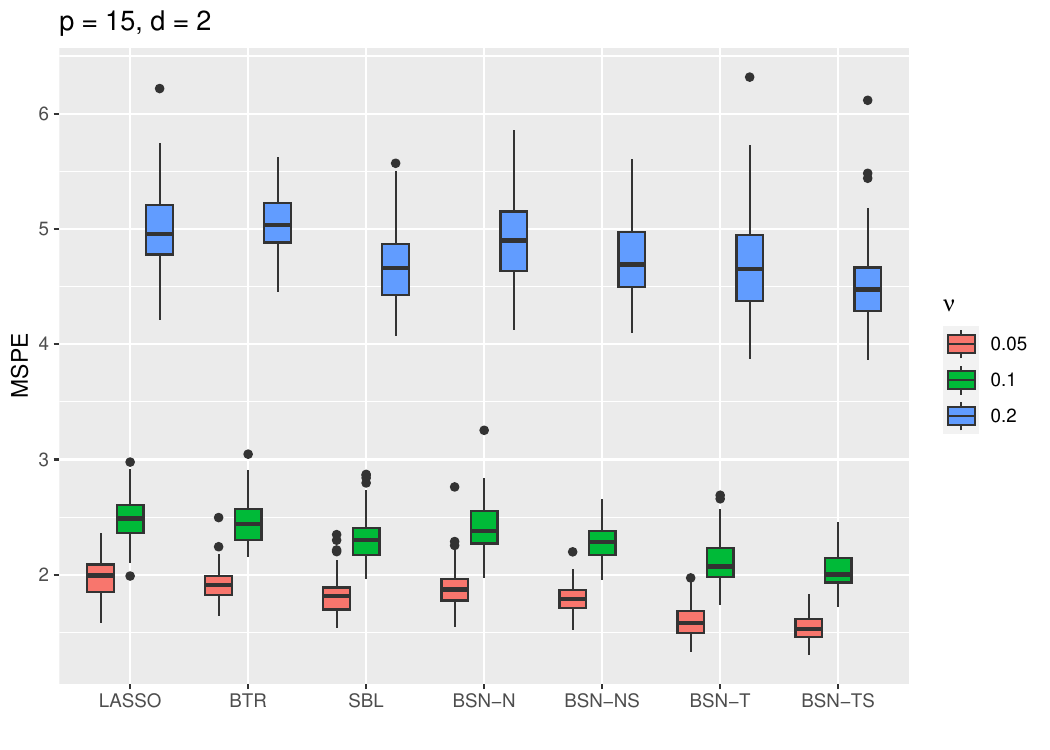}
\caption{Mean-squared prediction error (MSPE) on test sets for settings with $p= 15$ and $d = 2$ under SNR = 1 and $n = 200$ from 100 runs in the ``misspecified'' case.}
\end{figure}

\begin{figure}[H]
  \centering
\includegraphics[ width = 13cm, height = 8.5cm]{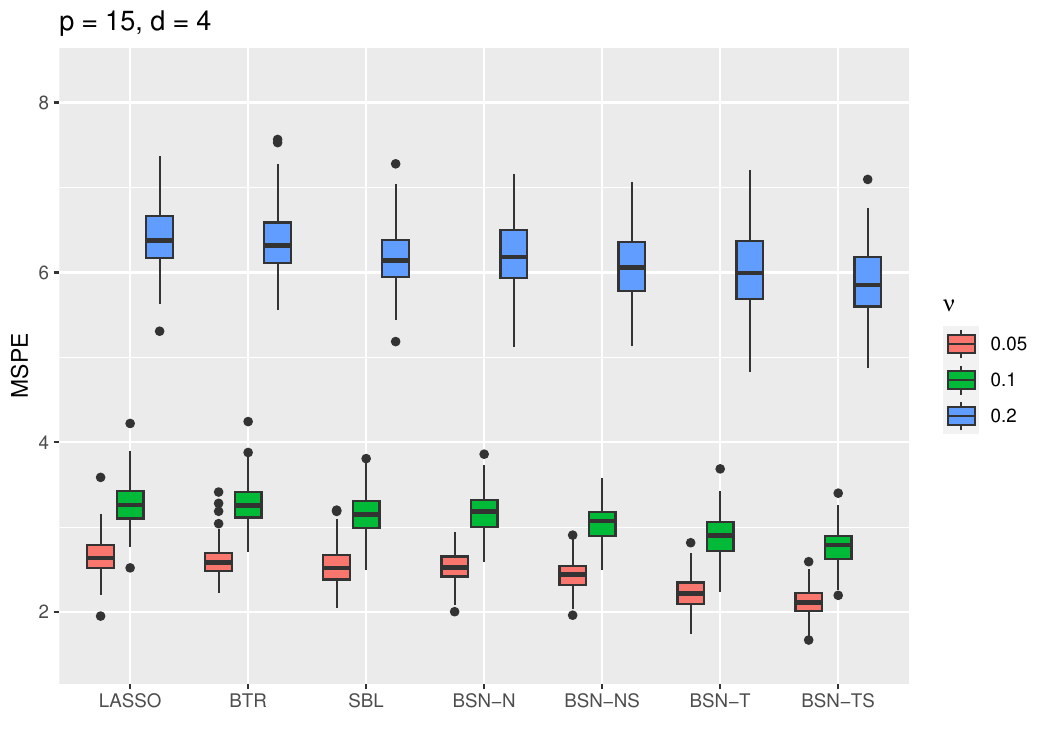}
\caption{Mean-squared prediction error (MSPE) on test sets for settings with $p= 15$ and $d = 4$ under SNR = 1 and $n = 200$ from 100 runs in the ``misspecified'' case.}
\end{figure}

\subsection*{Response coverage (RC)}

\begin{figure}[H]
\begin{subfigure}{.49\textwidth}
  \centering
\includegraphics[width = 9cm, height = 6cm]{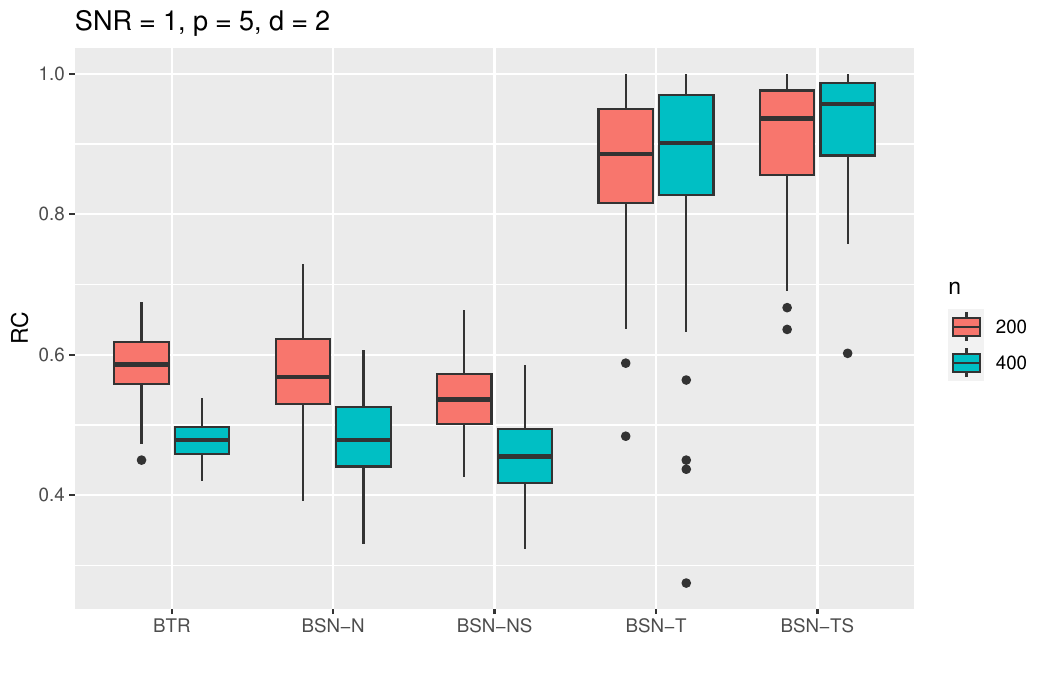}
\end{subfigure}
\begin{subfigure}{.49\textwidth}
\hspace{1cm}
\includegraphics[width = 9cm, height = 6cm]{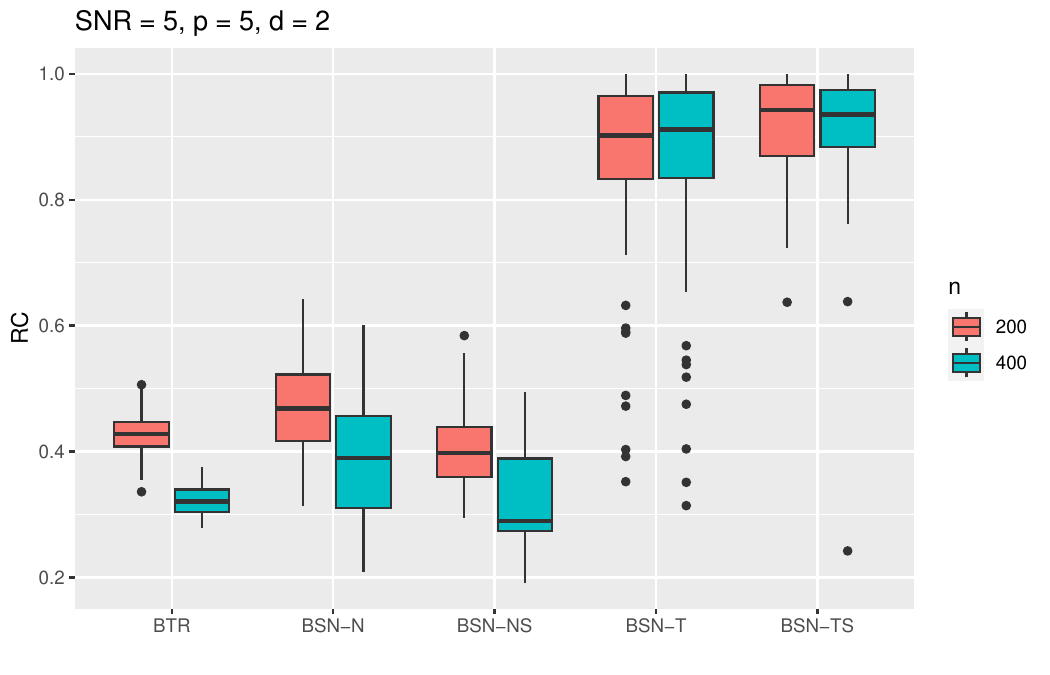}
\end{subfigure}
\caption{Response coverage (RC) on test sets for settings with $p= 5$ and $d = 2$ under SNR = 1 (left panel) and SNR = 5 (right panel), from 100 runs in the ``correctly specified'' case. }
\end{figure}

\begin{figure}[H]
\begin{subfigure}{.49\textwidth}
  \centering
\includegraphics[width = 9cm, height = 6cm]{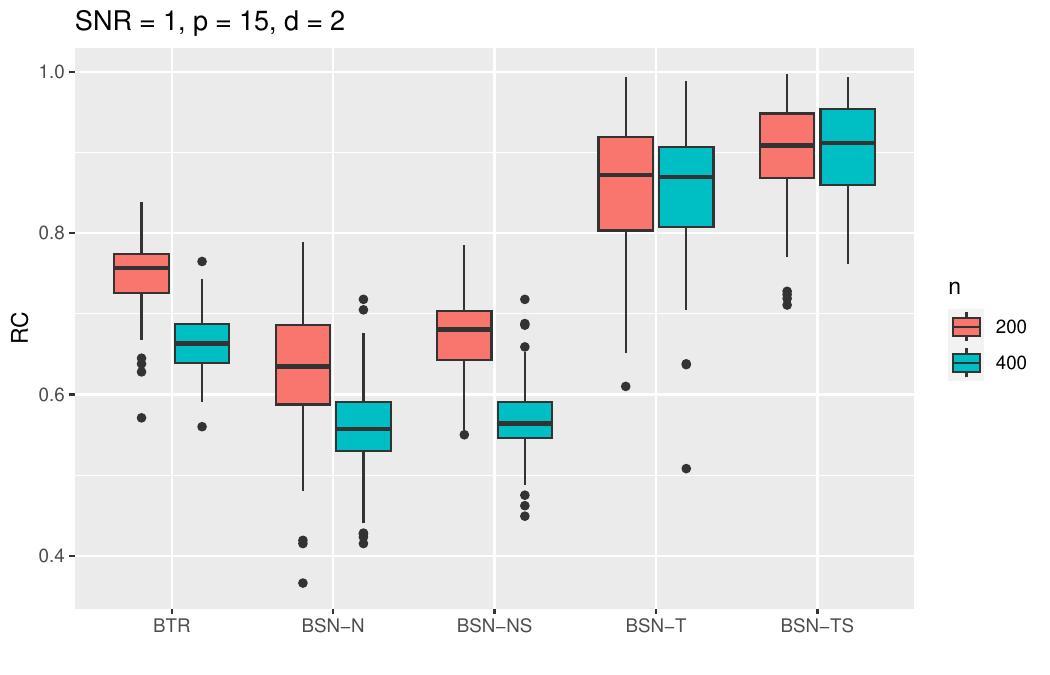}
\end{subfigure}
\begin{subfigure}{.49\textwidth}
\hspace{1cm}
\includegraphics[width = 9cm, height = 6cm]{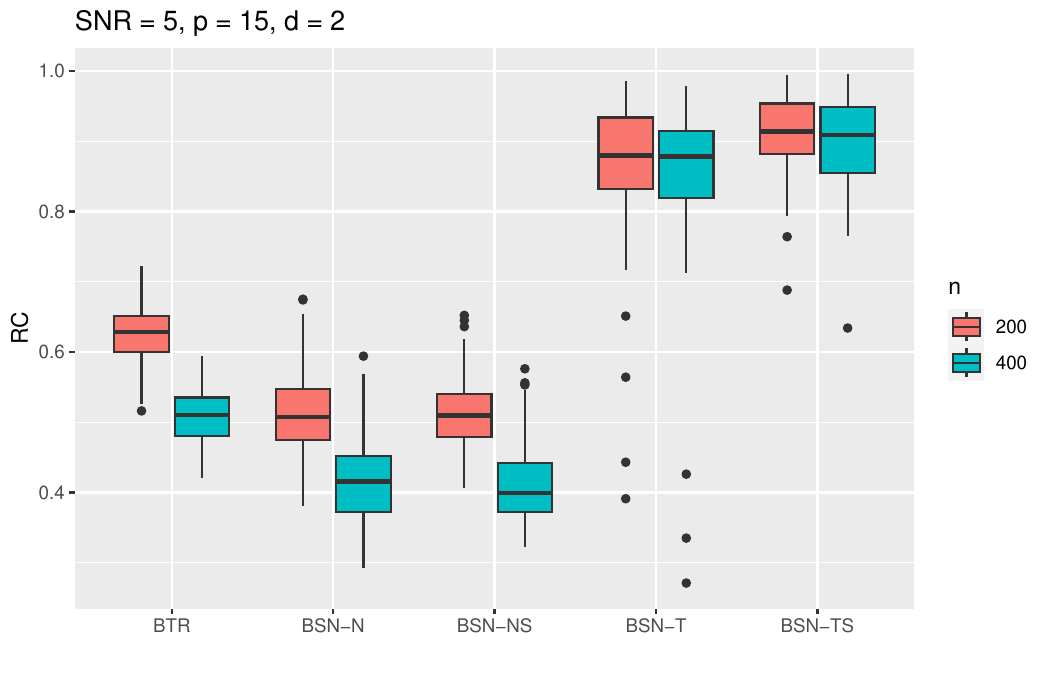}
\end{subfigure}
\caption{Response coverage (RC) on test sets for settings with $p= 15$ and $d = 2$ under SNR = 1 (left panel) and SNR = 5 (right panel), from 100 runs in the ``correctly specified'' case.}
\end{figure}

\begin{figure}[H]
\begin{subfigure}{.49\textwidth}
  \centering
\includegraphics[width = 9cm, height = 6cm]{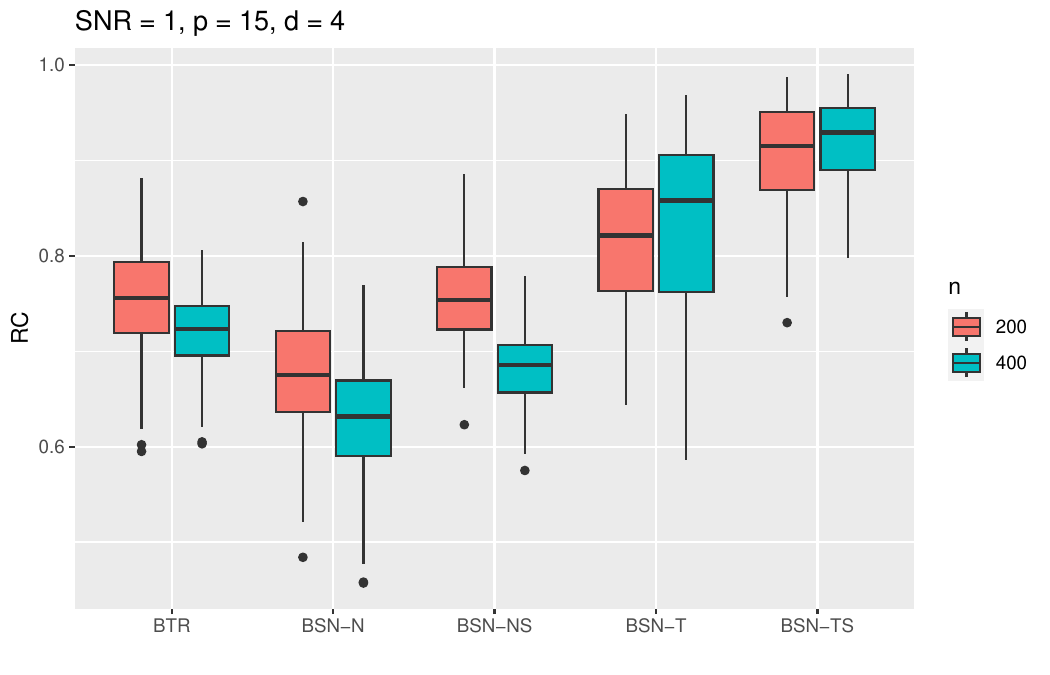}
\end{subfigure}
\begin{subfigure}{.49\textwidth}
\hspace{1cm}
\includegraphics[width = 9cm, height = 6cm]{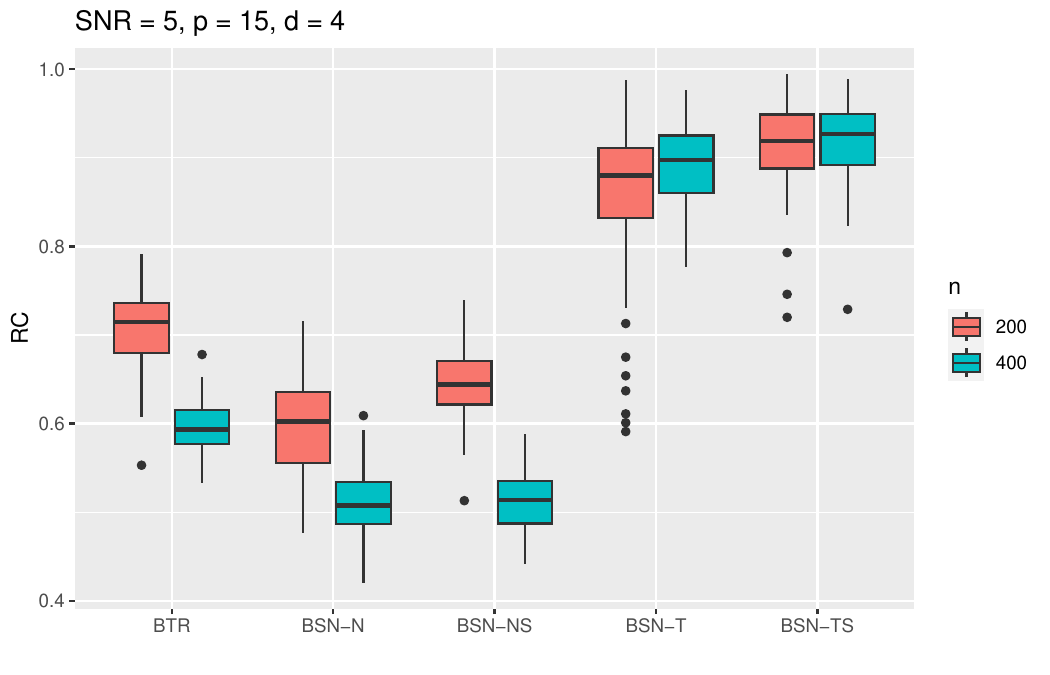}
\end{subfigure}
\caption{Response coverage (RC) on test sets for settings with $p= 15$ and $d = 4$ under SNR = 1 (left panel) and SNR = 5 (right panel), from 100 runs in the ``correctly specified'' case.}
\end{figure}

\begin{figure}[H]
  \centering
\includegraphics[width = 13cm, height = 8cm]{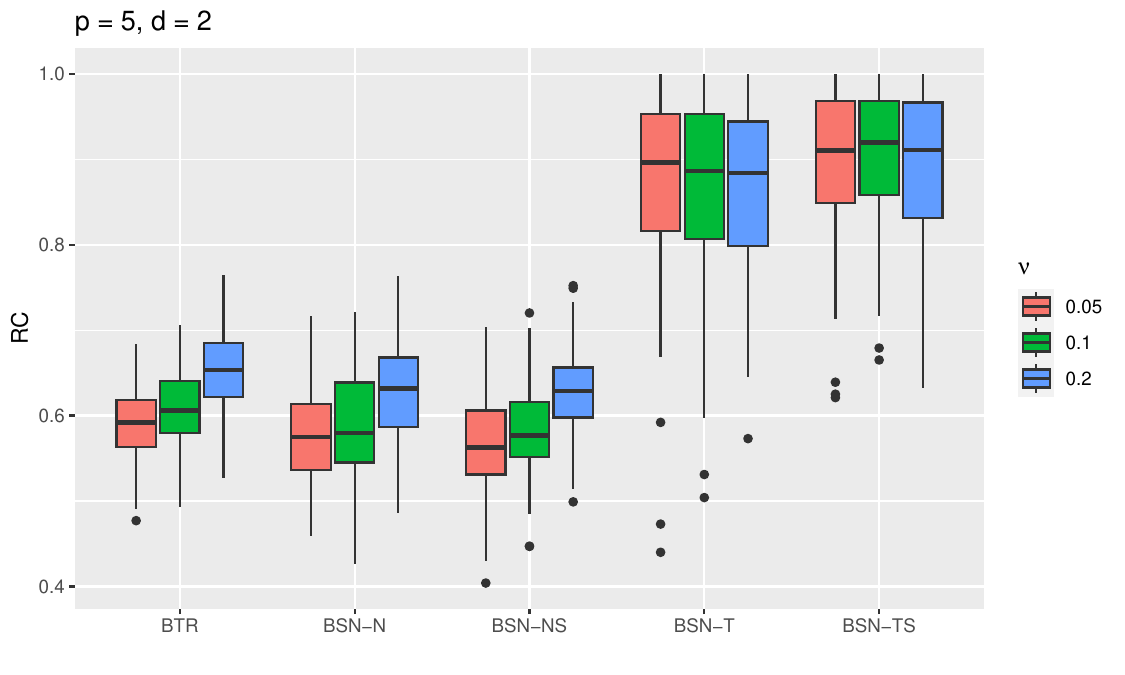}
\caption{Response coverage (RC) on test sets for settings with $p= 5$, $d = 2$, $n = 200$, and SNR = 1 from 100 runs in the ``misspecified'' case. }
\end{figure}

\begin{figure}[H]
  \centering
\includegraphics[width = 13cm, height = 8cm]{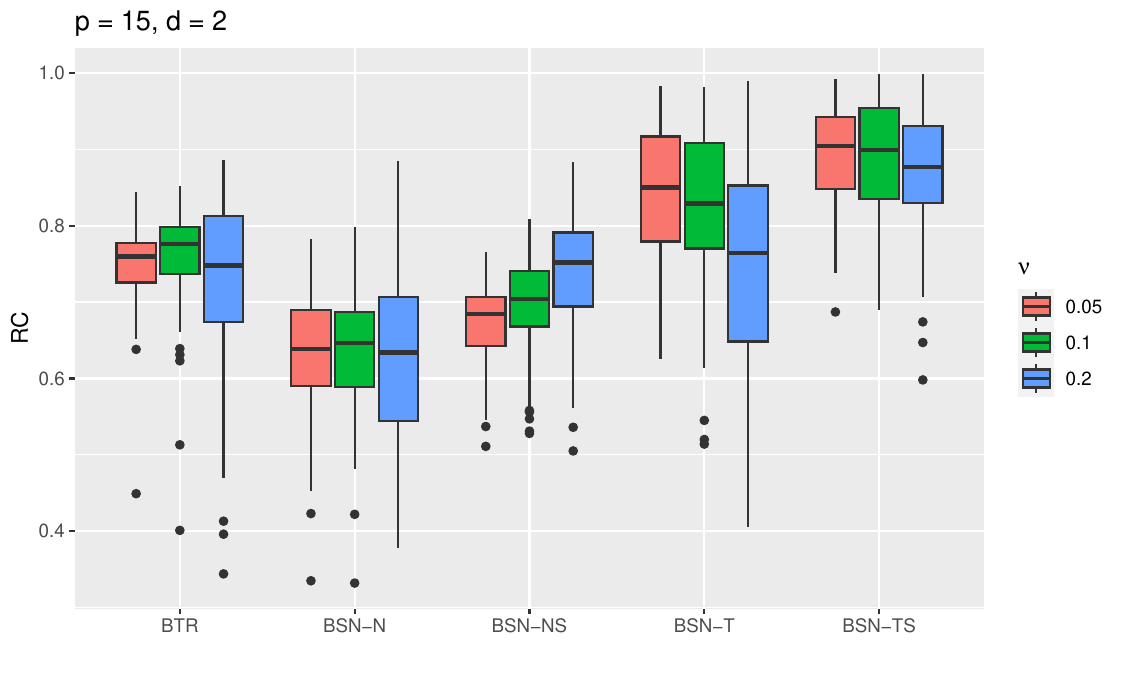}
\caption{Response coverage (RC) on test sets for settings with $p= 15$, $d = 2$, $n = 200$, and SNR = 1 from 100 runs in the ``misspecified'' case.}
\end{figure}

\begin{figure}[H]
  \centering
\includegraphics[width = 13cm, height = 8cm]{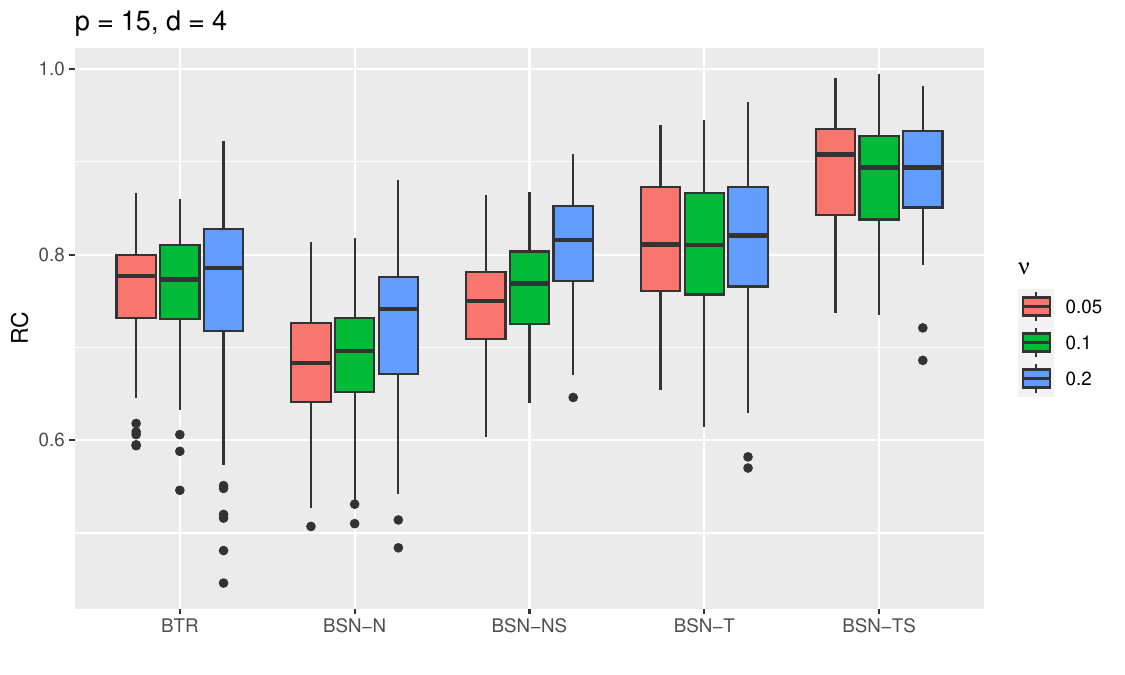}
\caption{Response coverage (RC) on test sets for settings with $p= 15$, $d = 4$, $n = 200$, and SNR = 1 from 100 runs in the ``misspecified'' case.}
\end{figure}

\subsection*{Length of credible intervals ($\text{Len}(\hat{Y}_{i})$)}

We show the density plots of the lengths of the 90\% credible intervals for different settings. For easier comparison with \textbf{BTR} and among the \textbf{BSN} methods, while avoiding many densities shown in the same figure, for each case we made two plots; both have the same \textbf{BTR} density, and for \textbf{BSN}, one plot has \textbf{BSN-N} and \textbf{BSN-NS},  and the other has \textbf{BSN-T} and \textbf{BSN-TS}. The x-axis in some figures is right-truncated to exclude the density estimated from a few very large intervals, allowing for easier comparison of the majority of the interval lengths on the same plot.

%% p = 5, d = 2

\begin{figure}[H]
\centering
\includegraphics[scale = 0.6]{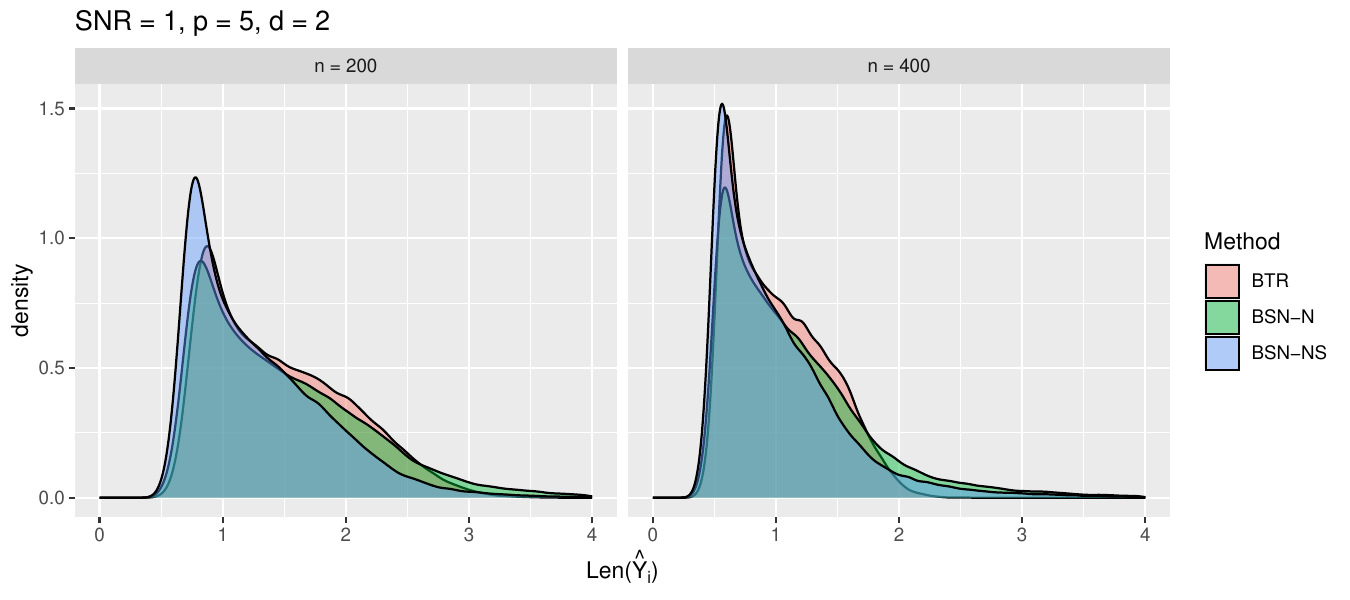}

\caption{Lengths of the 90\% credible intervals for settings with $p= 5$ and $d = 2$ under SNR = 1 from 100 runs in the ``correctly specified'' case.  The densities displayed are produced by \textbf{BTR}, \textbf{BSN-N} and \textbf{BSN-NS}.} 
\end{figure}

\begin{figure}[H]
\centering
\includegraphics[scale = 0.6]{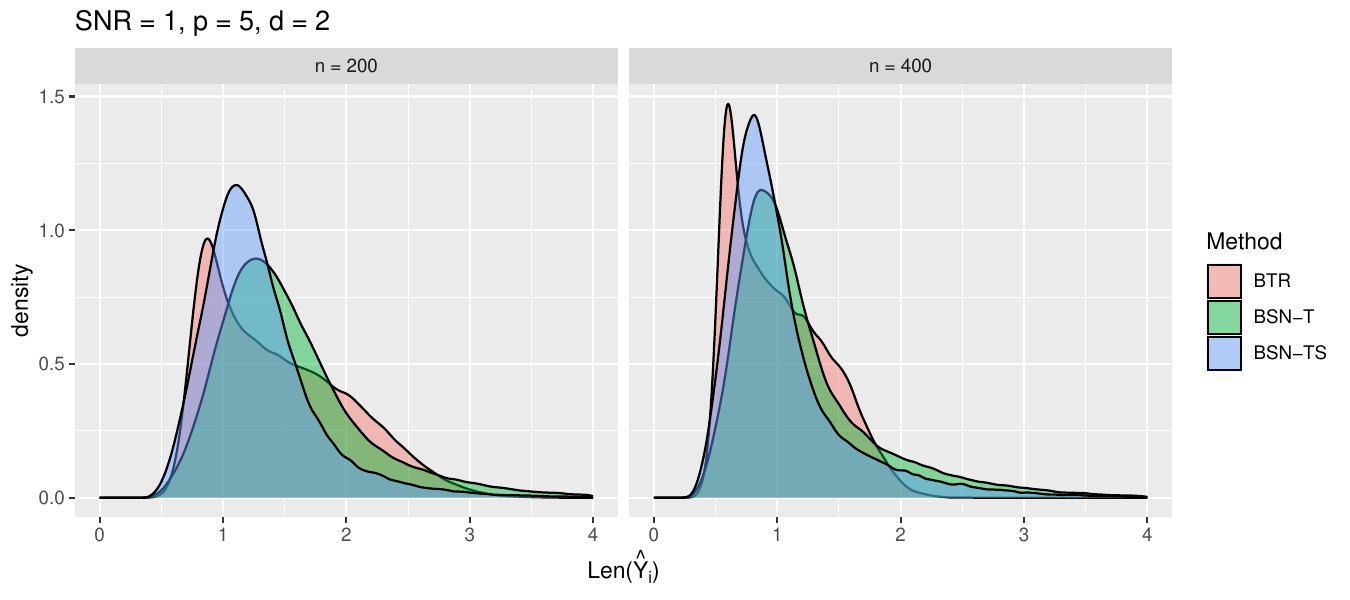}

\caption{Lengths of the 90\% credible intervals for settings with $p= 5$ and $d = 2$ under SNR = 1 from 100 runs in the ``correctly specified'' case.  The densities displayed are produced by \textbf{BTR}, \textbf{BSN-T} and \textbf{BSN-TS}.}
\end{figure}

\begin{figure}[H]
\centering
\includegraphics[scale = 0.6]{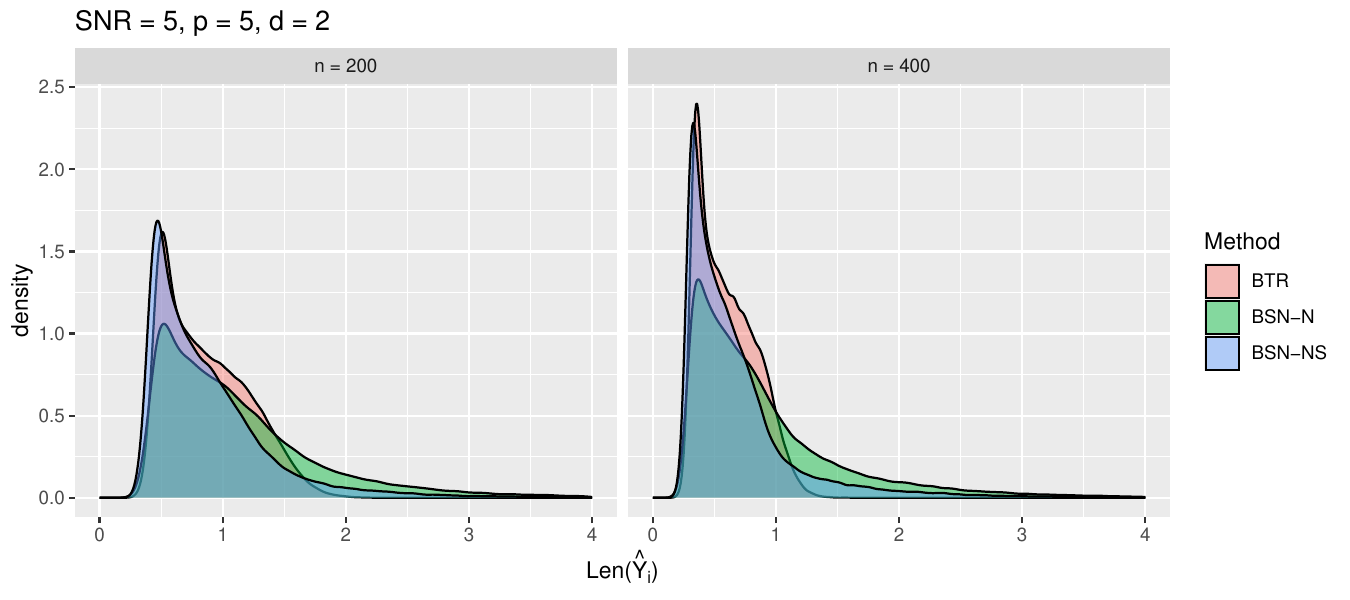}

\caption{Lengths of the 90\% credible intervals for settings with $p= 5$ and $d = 2$ under SNR = 5 from 100 runs in the ``correctly specified'' case.  The densities displayed are produced by \textbf{BTR}, \textbf{BSN-N} and \textbf{BSN-NS}.}
\end{figure}

\begin{figure}[H]
\centering
\includegraphics[scale = 0.6]{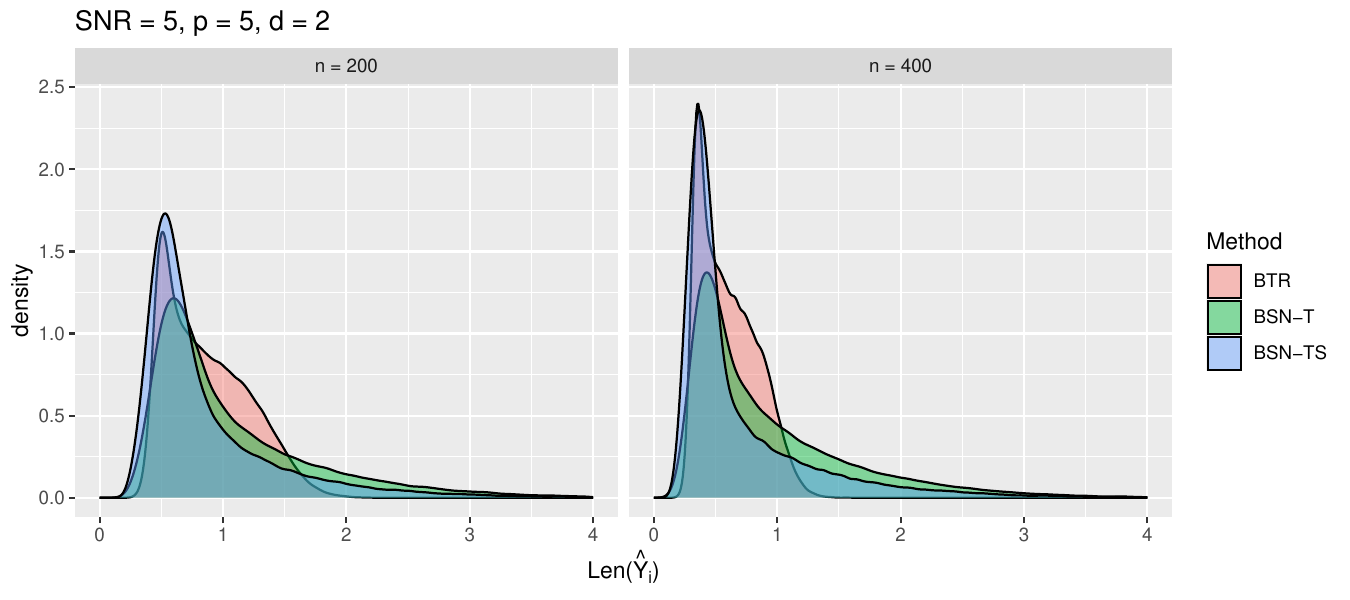}
\caption{Lengths of the 90\% credible intervals for settings with $p= 5$ and $d = 2$ under SNR = 5 from 100 runs in the ``correctly specified'' case.  The densities displayed are produced by \textbf{BTR}, \textbf{BSN-T} and \textbf{BSN-TS}.}
\end{figure}

%% p = 15, d = 2

\begin{figure}[H]
\centering
\includegraphics[scale = 0.6]{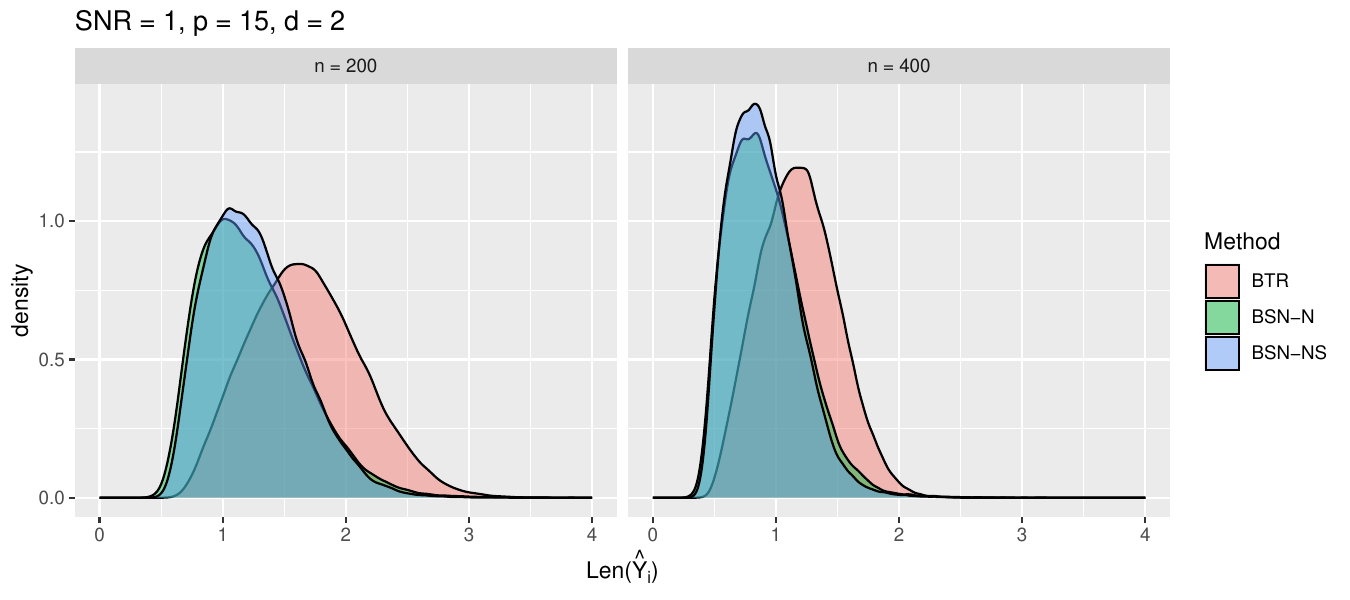}

\caption{Lengths of the 90\% credible intervals for settings with $p= 15$ and $d = 2$ under SNR = 1 from 100 runs in the ``correctly specified'' case.  The densities displayed are produced by \textbf{BTR}, \textbf{BSN-N} and \textbf{BSN-NS}.}
\end{figure}

\begin{figure}[H]
\centering
\includegraphics[scale = 0.6]{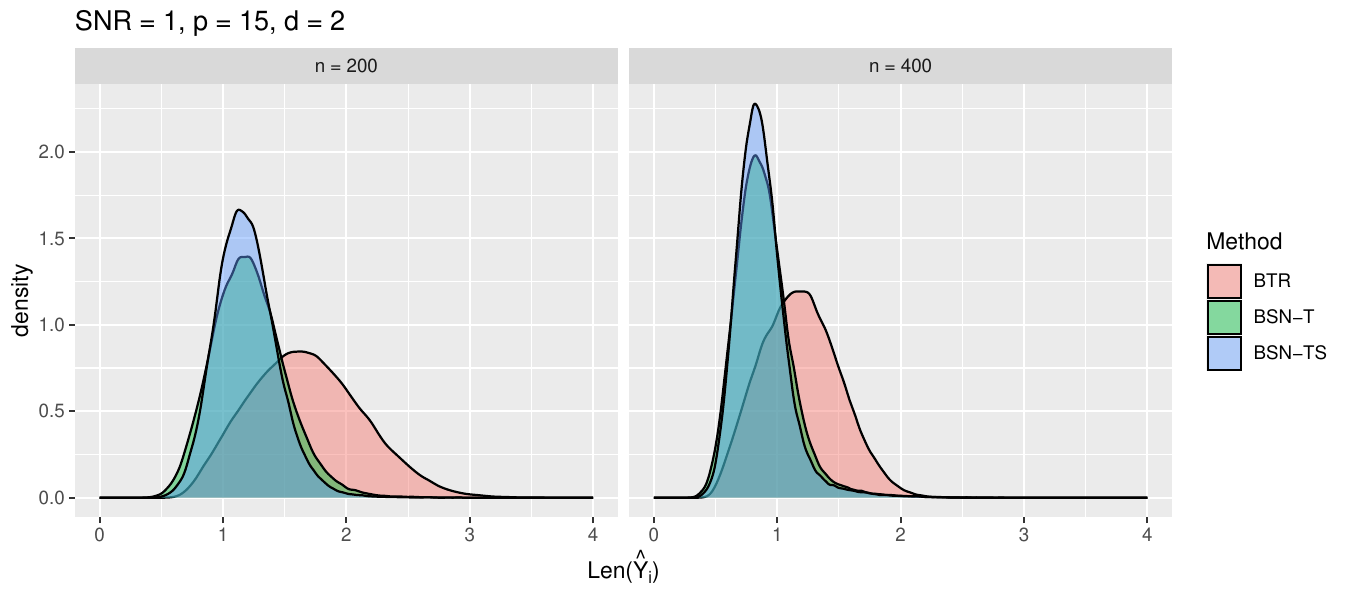}

\caption{Lengths of the 90\% credible intervals for settings with $p= 15$ and $d = 2$ under SNR = 1 from 100 runs in the ``correctly specified'' case.  The densities displayed are produced by \textbf{BTR}, \textbf{BSN-T} and \textbf{BSN-TS}.}
\end{figure}

\begin{figure}[H]
\centering
\includegraphics[scale = 0.6]{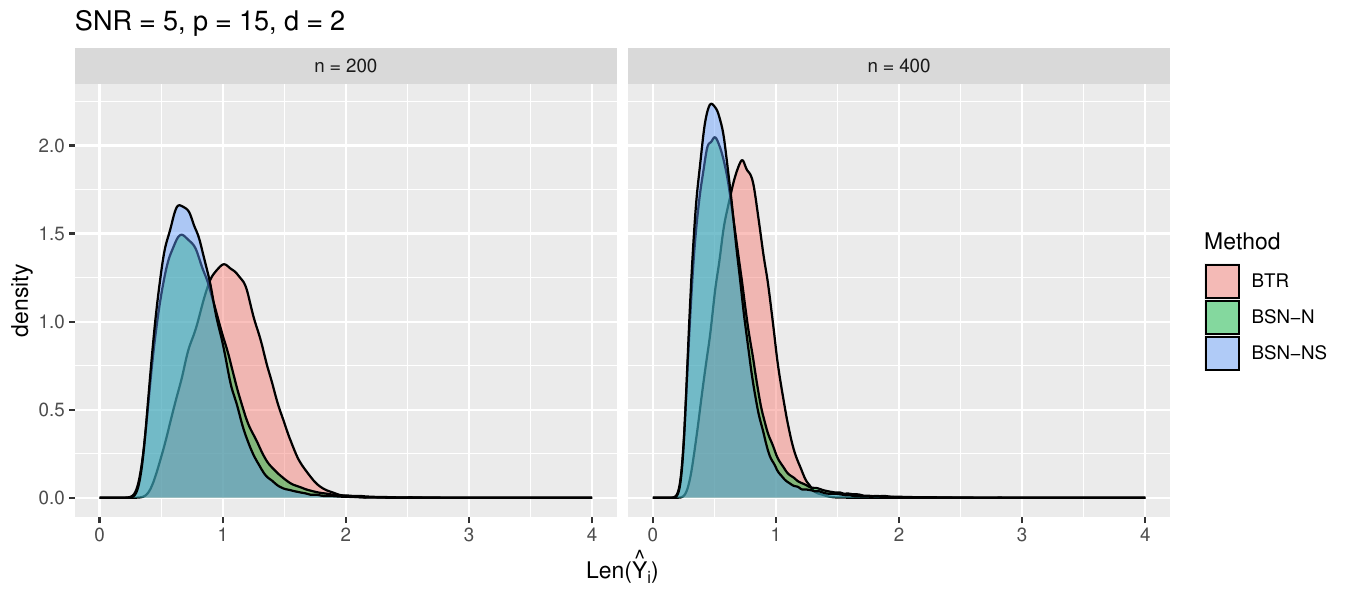}

\caption{Lengths of the 90\% credible intervals for settings with $p= 15$ and $d = 2$ under SNR = 5 from 100 runs in the ``correctly specified'' case.  The densities displayed are produced by \textbf{BTR}, \textbf{BSN-N} and \textbf{BSN-NS}.}
\end{figure}

\begin{figure}[H]
\centering
\includegraphics[scale = 0.6]{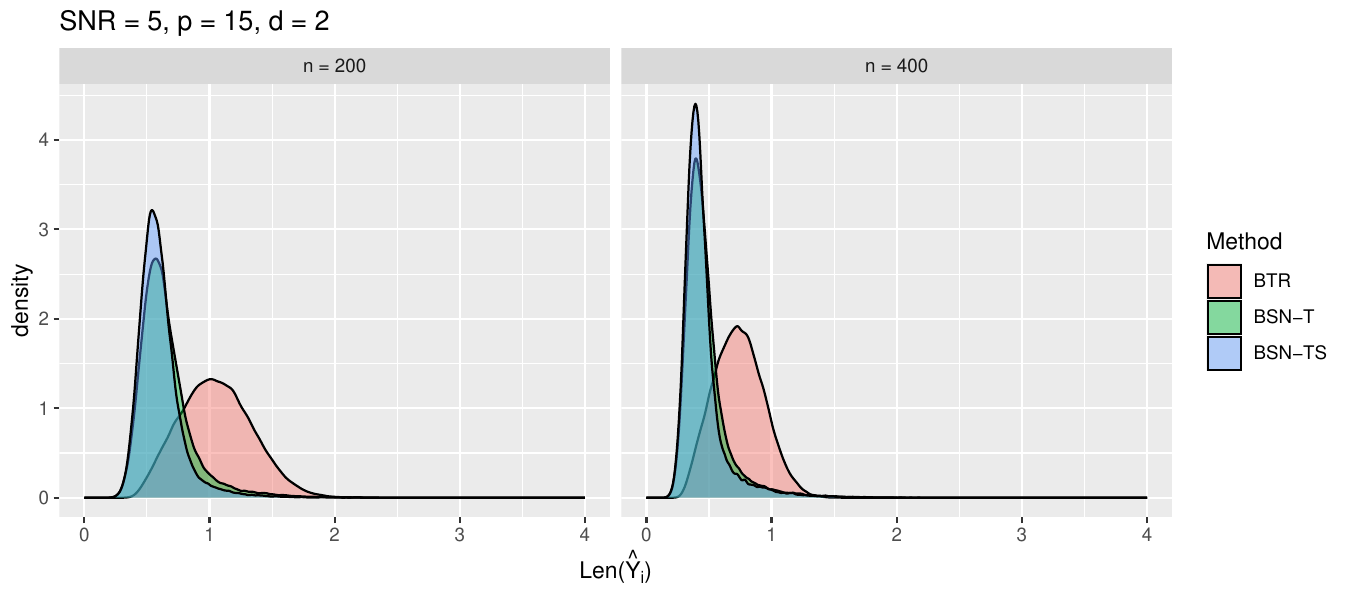}
\caption{Lengths of the 90\% credible intervals for settings with $p= 15$ and $d = 2$ under SNR = 5 from 100 runs in the ``correctly specified'' case.  The densities displayed are produced by \textbf{BTR}, \textbf{BSN-T} and \textbf{BSN-TS}.}
\end{figure}

%% p = 15, d = 4

\begin{figure}[H]
\centering
\includegraphics[scale = 0.6]{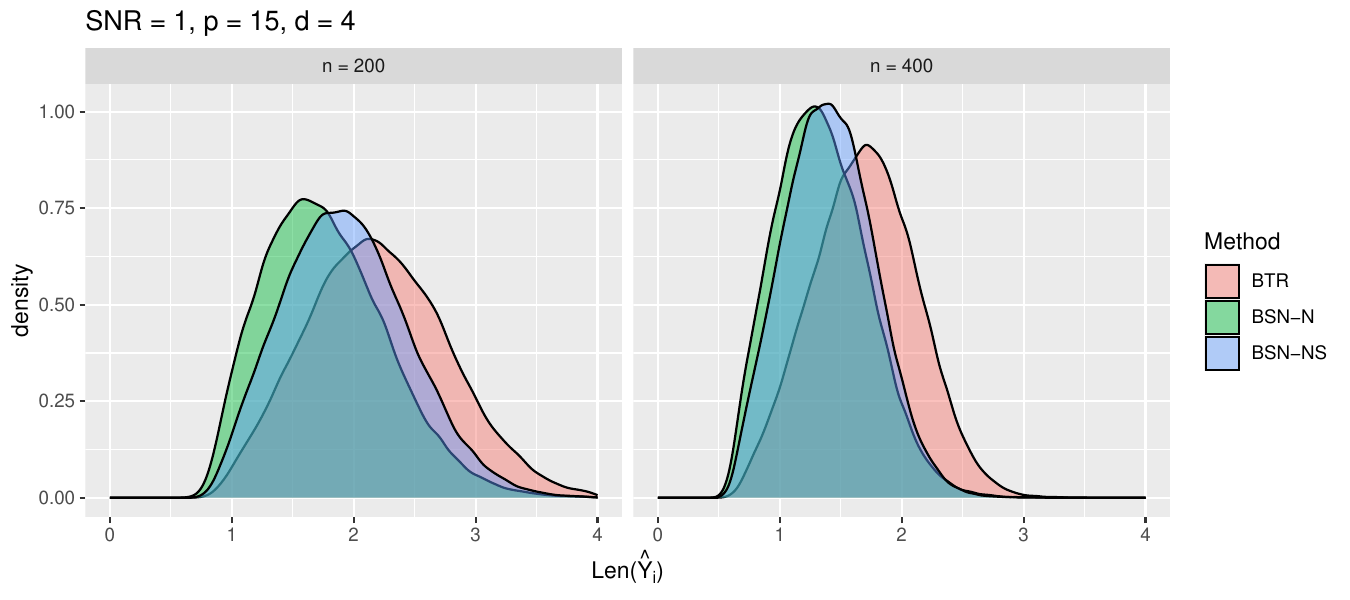}

\caption{Lengths of the 90\% credible intervals for settings with $p= 15$ and $d = 4$ under SNR = 1 from 100 runs in the ``correctly specified'' case.  The densities displayed are produced by \textbf{BTR}, \textbf{BSN-N} and \textbf{BSN-NS}.}
\end{figure}

\begin{figure}[H]
\centering
\includegraphics[scale = 0.6]{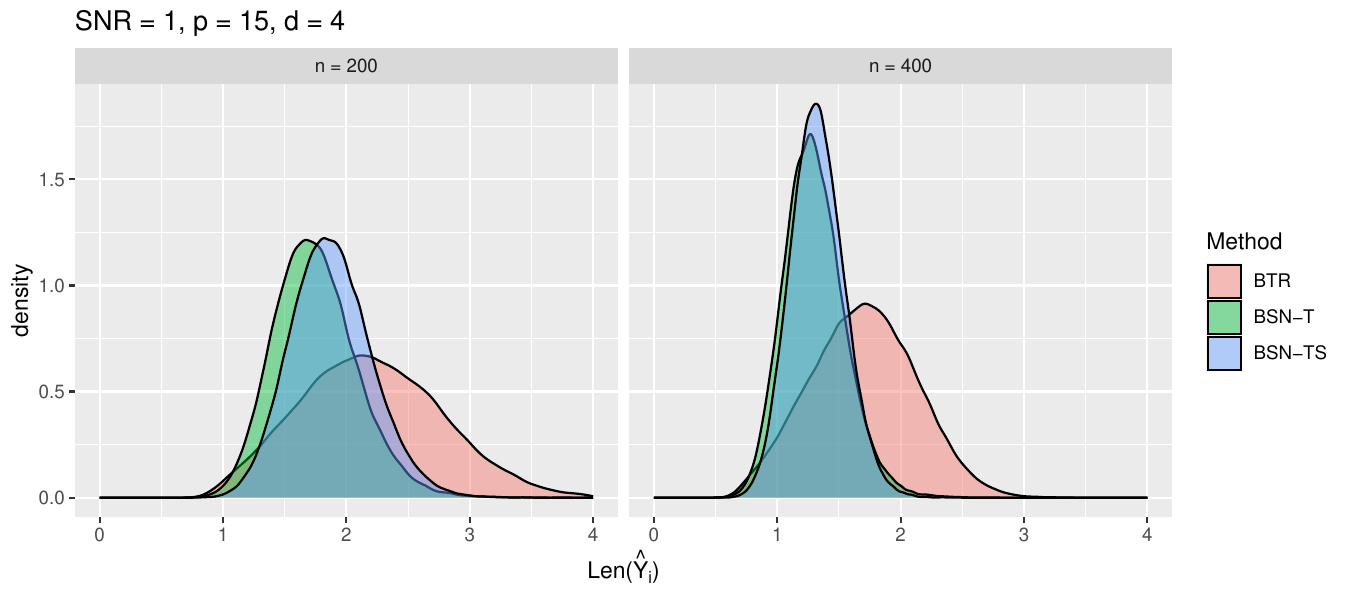}

\caption{Lengths of the 90\% credible intervals for settings with $p= 15$ and $d = 4$ under SNR = 1 from 100 runs in the ``correctly specified'' case.  The densities displayed are produced by \textbf{BTR}, \textbf{BSN-T} and \textbf{BSN-TS}.}
\end{figure}

\begin{figure}[H]
\centering
\includegraphics[scale = 0.6]{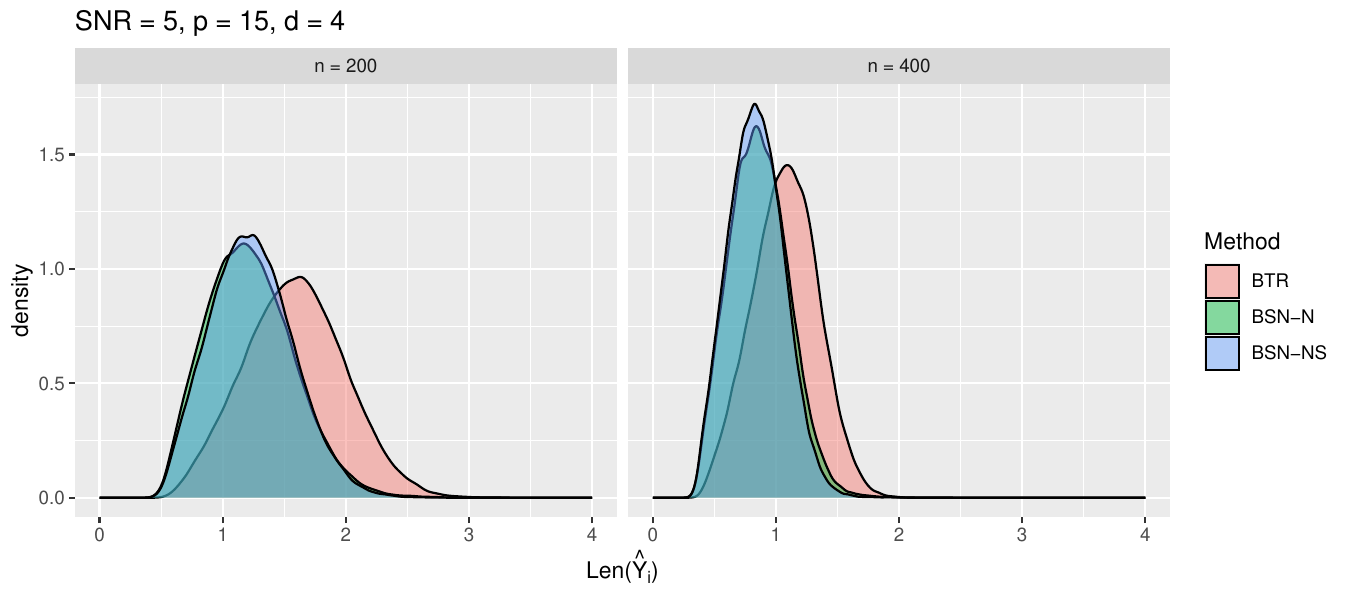}

\caption{Lengths of the 90\% credible intervals for settings with $p= 15$ and $d = 4$ under SNR = 5 from 100 runs in the ``correctly specified'' case.  The densities displayed are produced by \textbf{BTR}, \textbf{BSN-N} and \textbf{BSN-NS}.}
\end{figure}

\begin{figure}[H]
\centering
\includegraphics[scale = 0.6]{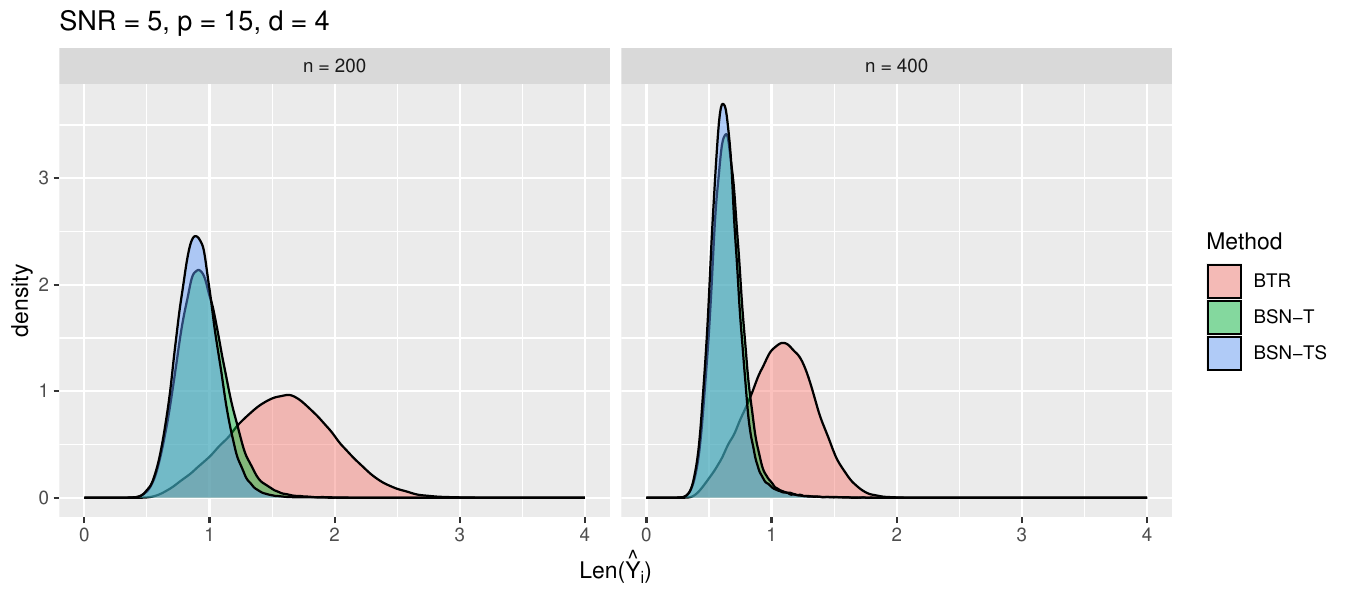}
\caption{Lengths of the 90\% credible intervals for settings with $p= 15$ and $d = 4$ under SNR = 5 from 100 runs in the ``correctly specified'' case.  The densities displayed are produced by \textbf{BTR}, \textbf{BSN-T} and \textbf{BSN-TS}.}
\end{figure}

%% Part (2)
\begin{figure}[H]
\centering
\includegraphics[scale = 0.6]{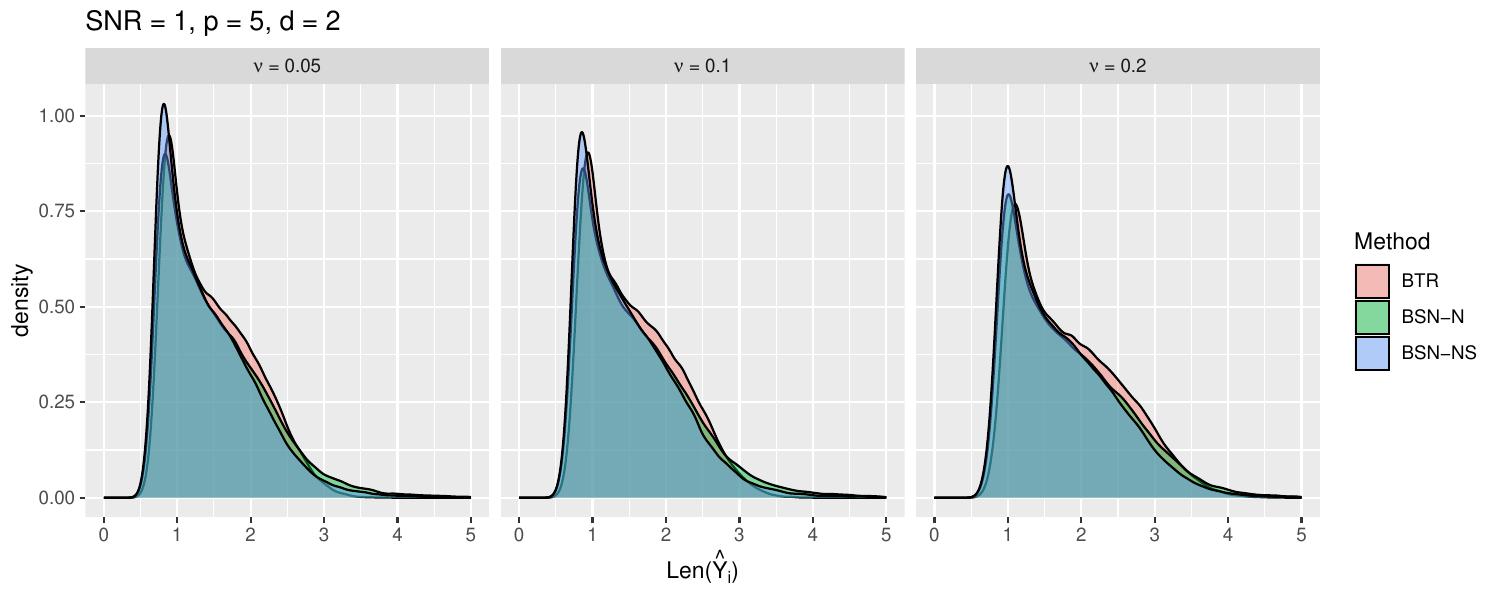}
\caption{Lengths of the 90\% credible intervals for settings with $p= 5$ and $d = 2$, $n = 200$, and SNR = 1 from 100 runs in the ``misspecified'' case.  The densities displayed are produced by \textbf{BTR}, \textbf{BSN-N} and \textbf{BSN-NS}.}
\end{figure}

\begin{figure}[H]
\centering
\includegraphics[scale = 0.6]{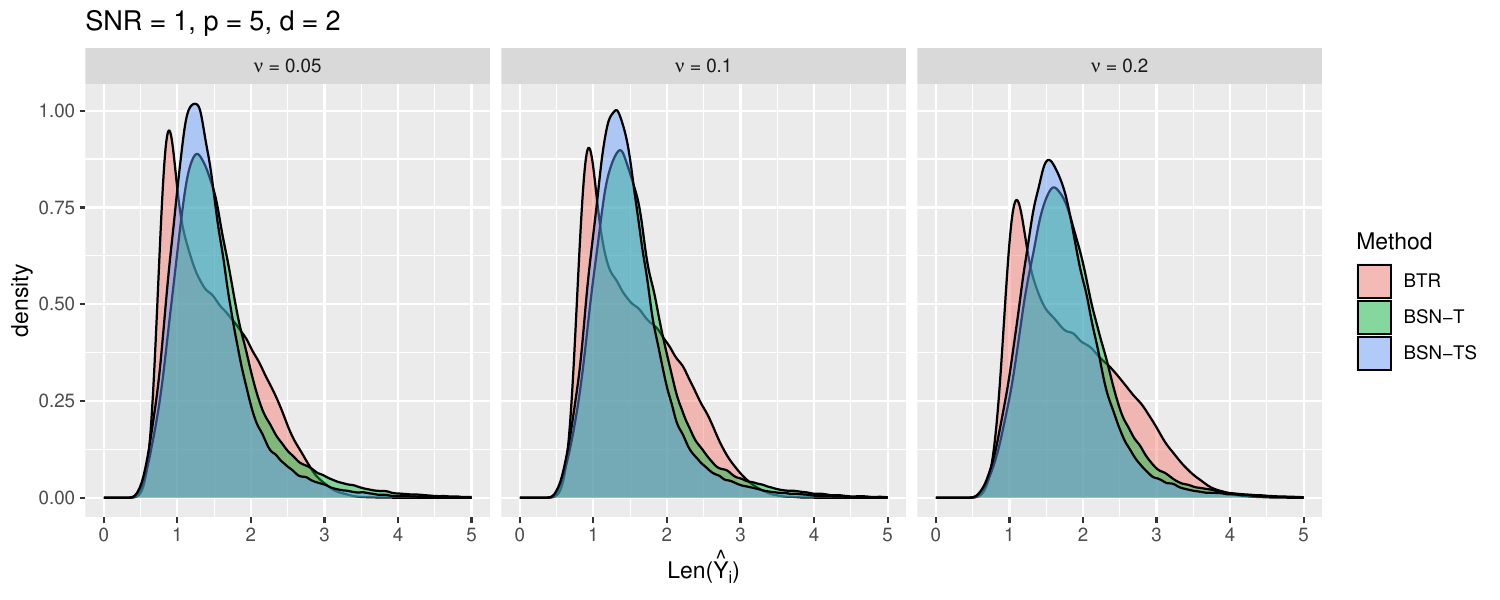}
\caption{Lengths of the 90\% credible intervals for settings with $p= 5$ and $d = 2$, $n = 200$, and SNR = 1 from 100 runs in the ``misspecified'' case.  The densities displayed are produced by \textbf{BTR}, \textbf{BSN-T} and \textbf{BSN-TS}.}
\end{figure}

\begin{figure}[H]
\centering
\includegraphics[scale = 0.6]{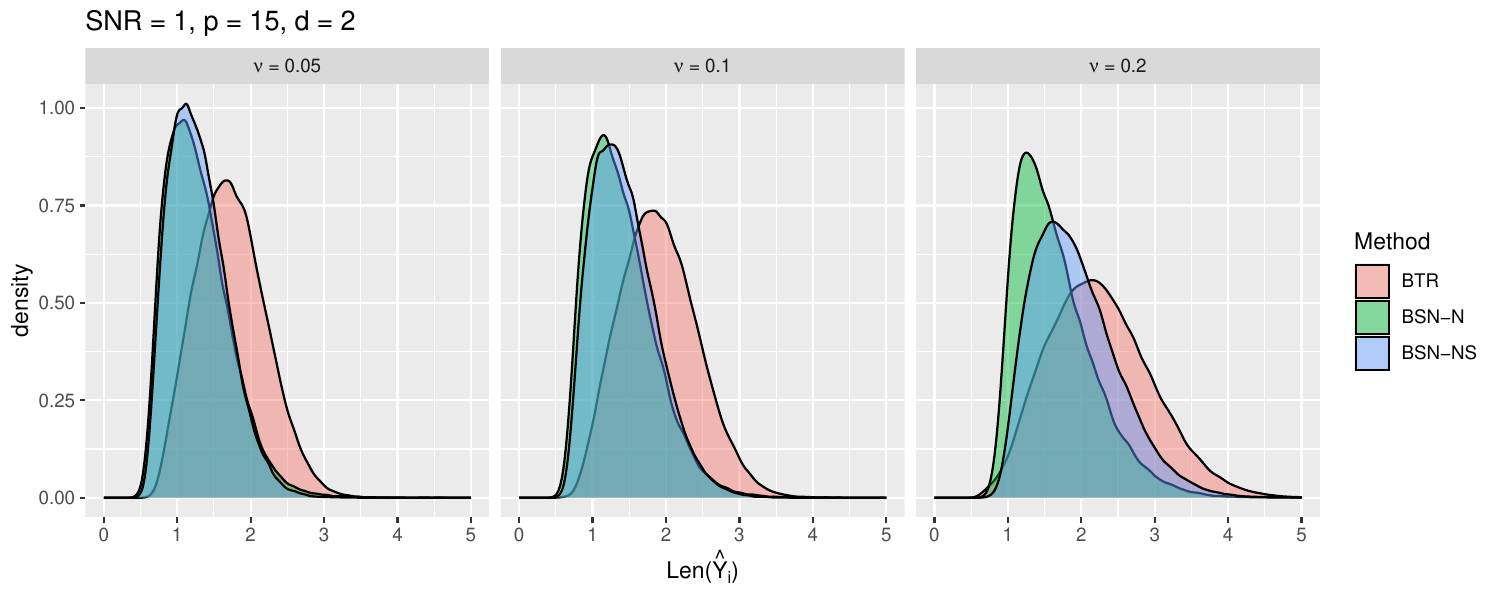}
\caption{Lengths of the 90\% credible intervals for settings with $p= 15$ and $d = 2$, $n = 200$, and SNR = 1 from 100 runs in the ``misspecified'' case.  The densities displayed are produced by \textbf{BTR}, \textbf{BSN-N} and \textbf{BSN-NS}.}
\end{figure}

\begin{figure}[H]
\centering
\includegraphics[scale = 0.6]{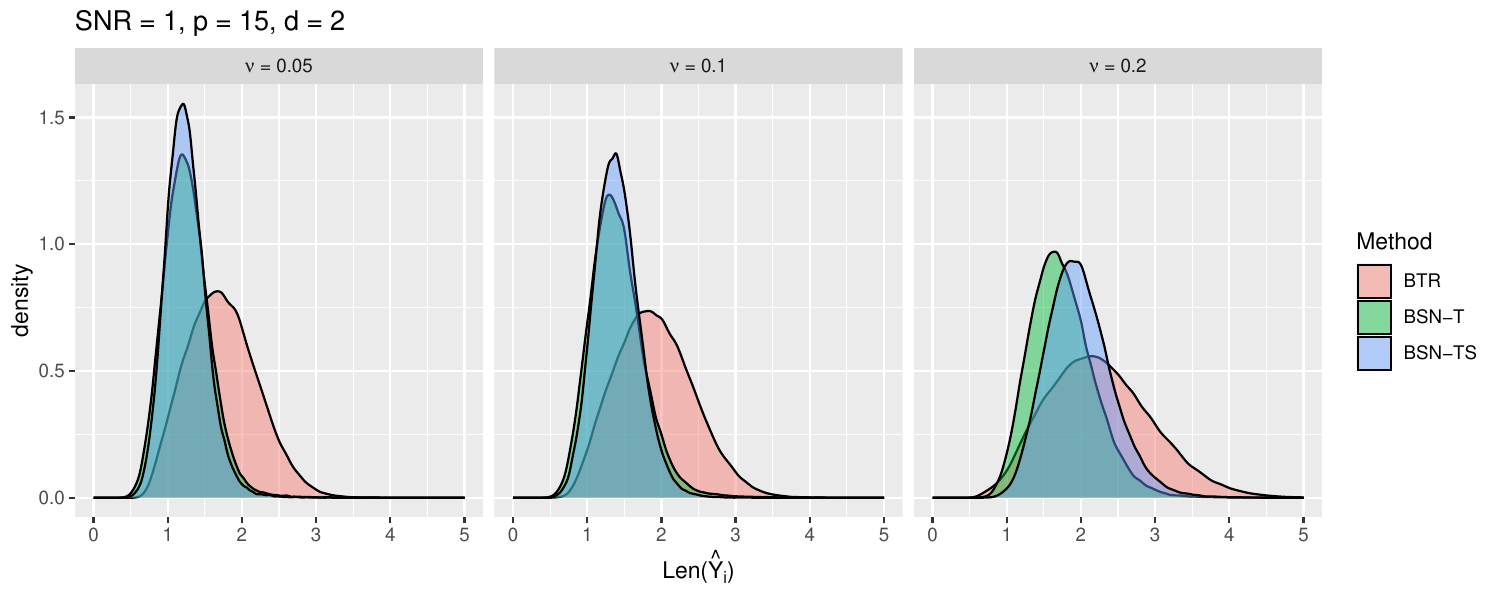}
\caption{Lengths of the 90\% credible intervals for settings with $p= 15$ and $d = 2$, $n = 200$, and SNR = 1 from 100 runs in the ``misspecified'' case.  The densities displayed are produced by \textbf{BTR}, \textbf{BSN-T} and \textbf{BSN-TS}.}
\end{figure}

\begin{figure}[H]
\centering
\includegraphics[scale = 0.6]{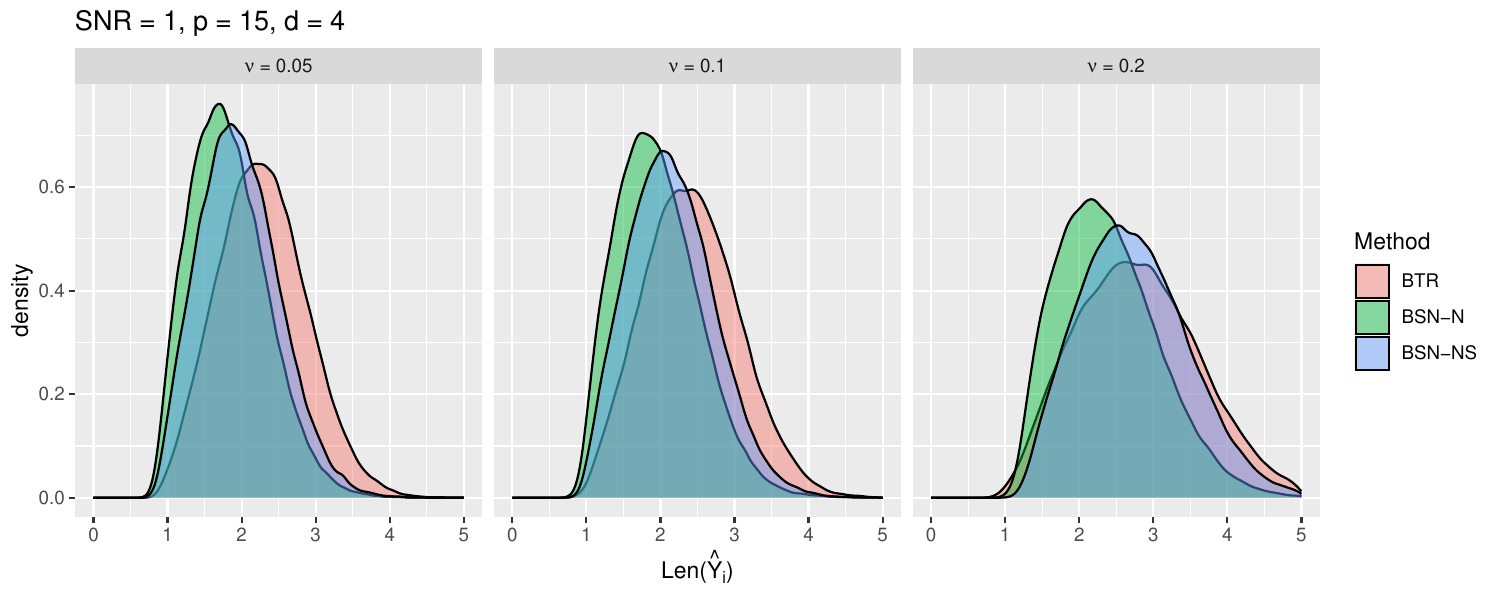}
\caption{Lengths of the 90\% credible intervals for settings with $p= 15$ and $d = 4$, $n = 200$, and SNR = 1 from 100 runs in the ``misspecified'' case.  The densities displayed are produced by \textbf{BTR}, \textbf{BSN-N} and \textbf{BSN-NS}.}
\end{figure}

\begin{figure}[H]
\centering
\includegraphics[scale = 0.6]{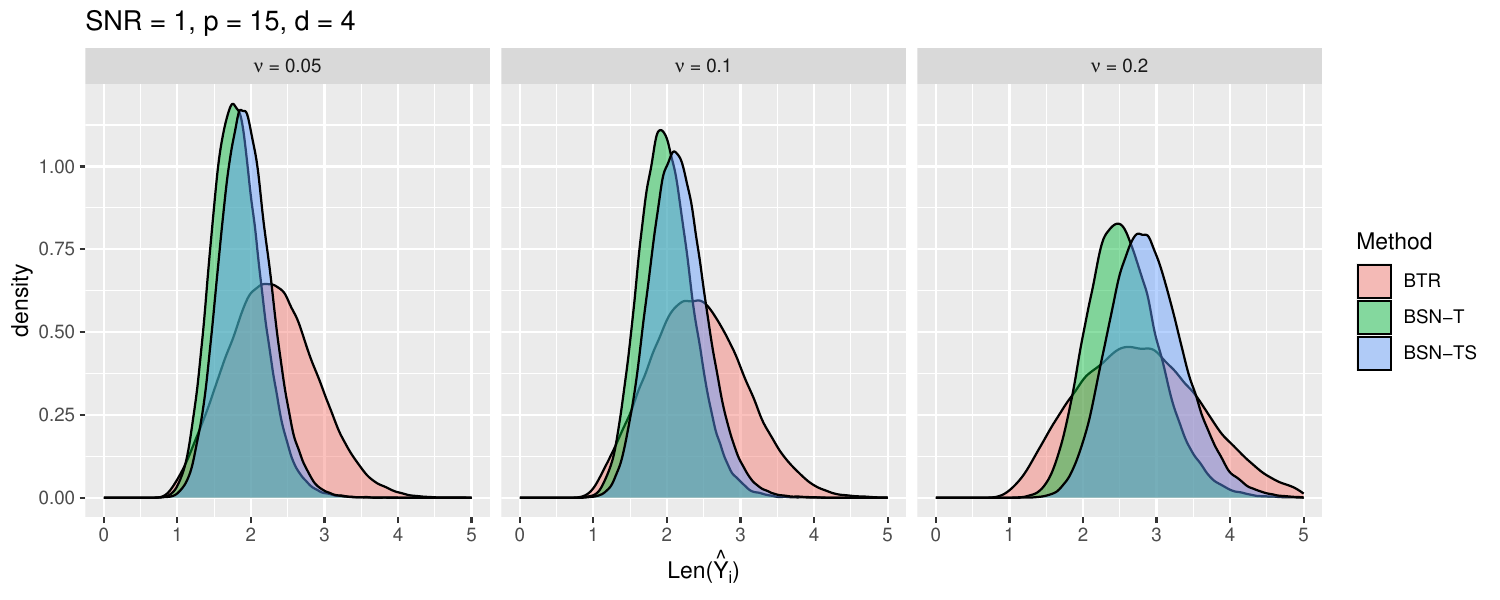}
\caption{Lengths of the 90\% credible intervals for settings with $p= 15$ and $d = 4$, $n = 200$, and SNR = 1 from 100 runs in the ``misspecified'' case.  The densities displayed are produced by \textbf{BTR}, \textbf{BSN-T} and \textbf{BSN-TS}.}
\end{figure}

\subsection*{Absolute cosine similarity (ACS)}
 %The x-axis represents the indices of the dimension $j = 1,..., d$ when computing $\textbf{ACS}(\bs{\gamma}_j, \bs{\gamma}_j^{(s)})$. 
\begin{figure}[H]
\begin{subfigure}{.49\textwidth}
  \centering
\includegraphics[scale = 0.45]{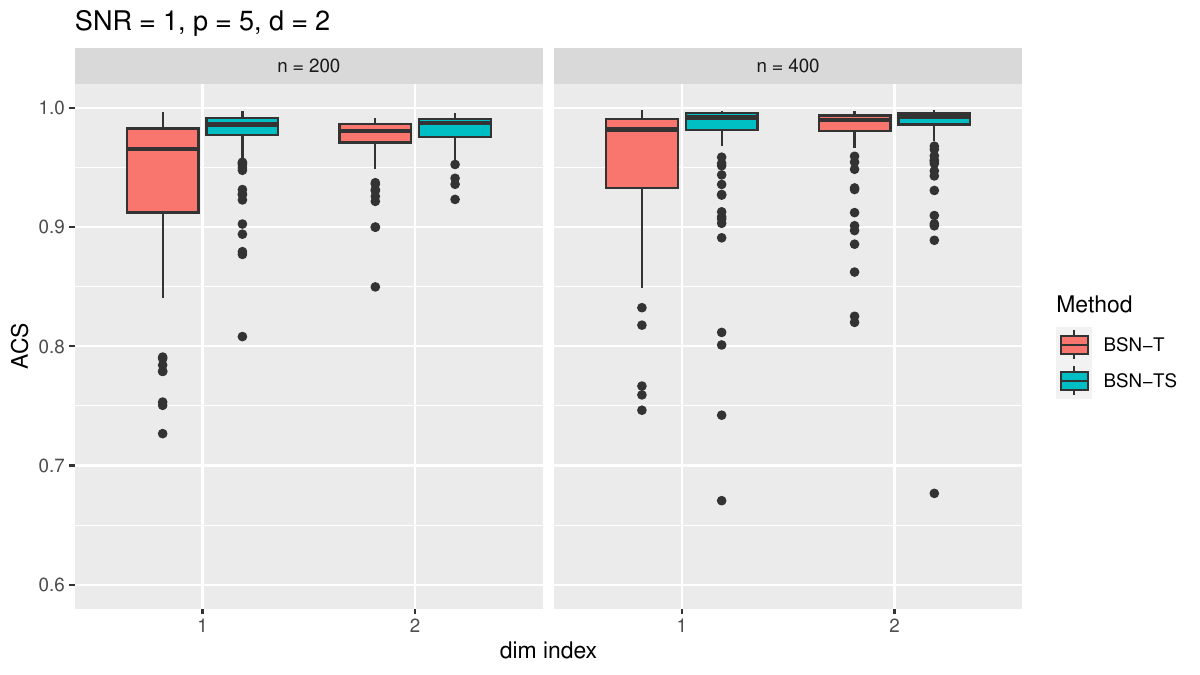}
\end{subfigure}
\begin{subfigure}{.49\textwidth}
\hspace{1cm}
\includegraphics[scale = 0.45]{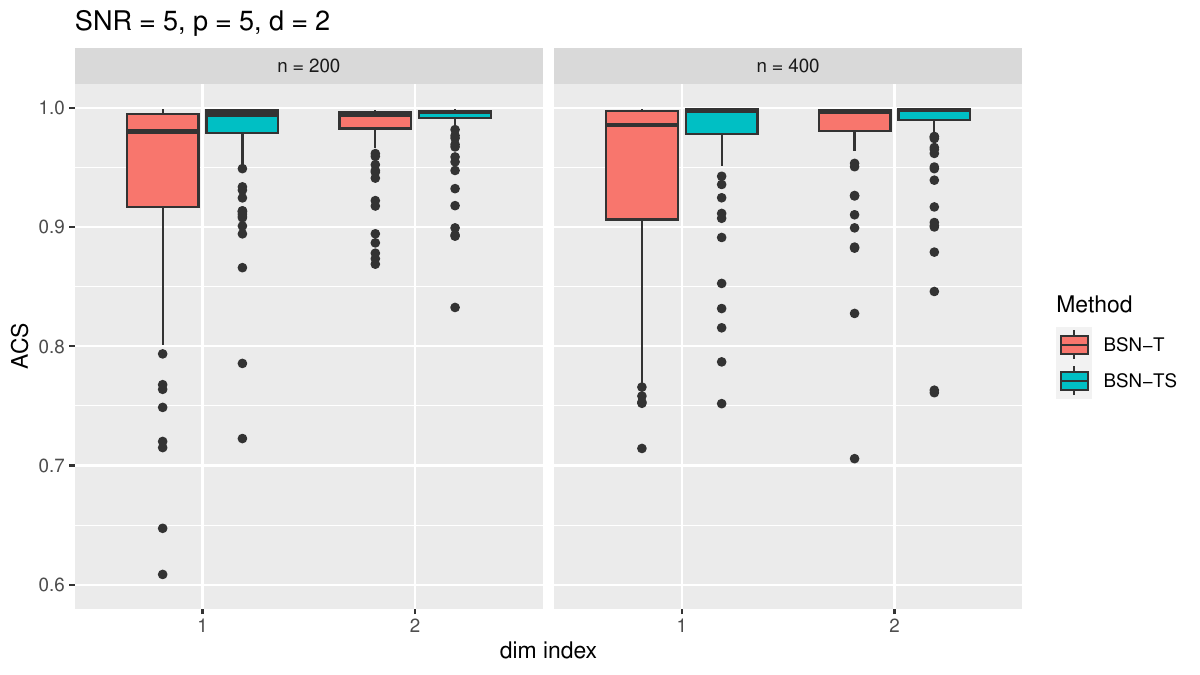}
\end{subfigure}
\caption{Absolute cosine similarity (ACS) for settings with $p= 5$ and $d = 2$ under SNR = 1 (left panel) and SNR = 5 (right panel), from 100 runs in the ``correctly specified'' case.}
\end{figure}

\begin{figure}[H]
\begin{subfigure}{.49\textwidth}
  \centering
\includegraphics[scale = 0.45]{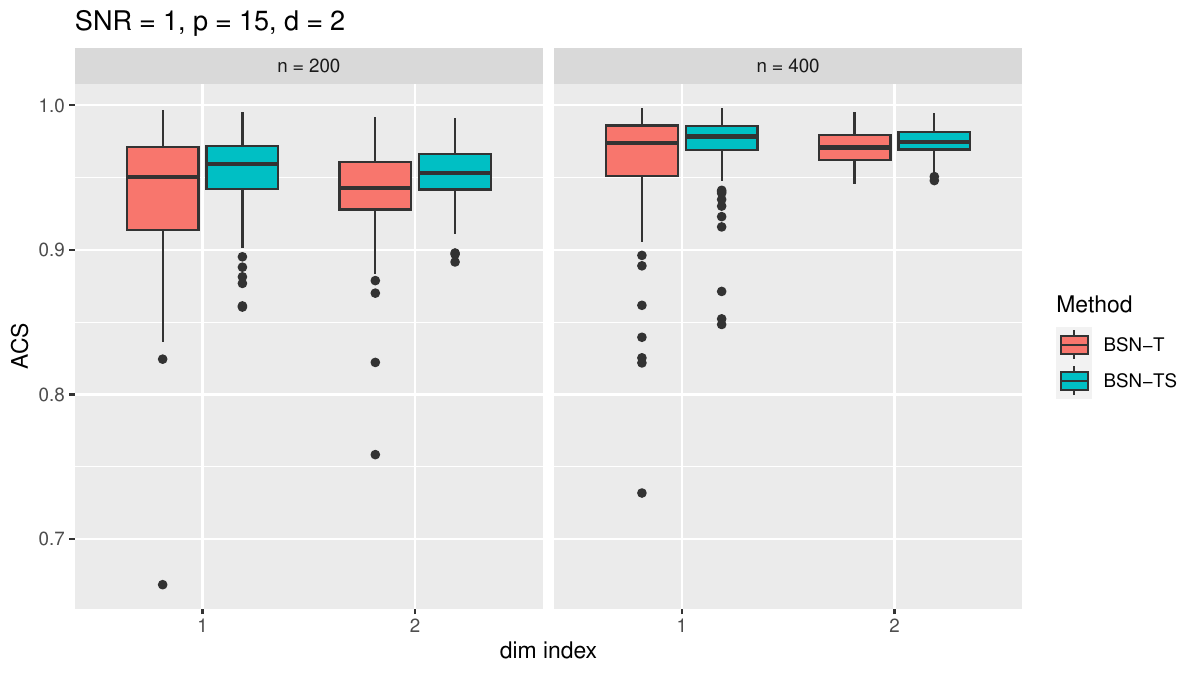}
\end{subfigure}
\begin{subfigure}{.49\textwidth}
\hspace{1cm}
\includegraphics[scale = 0.45]{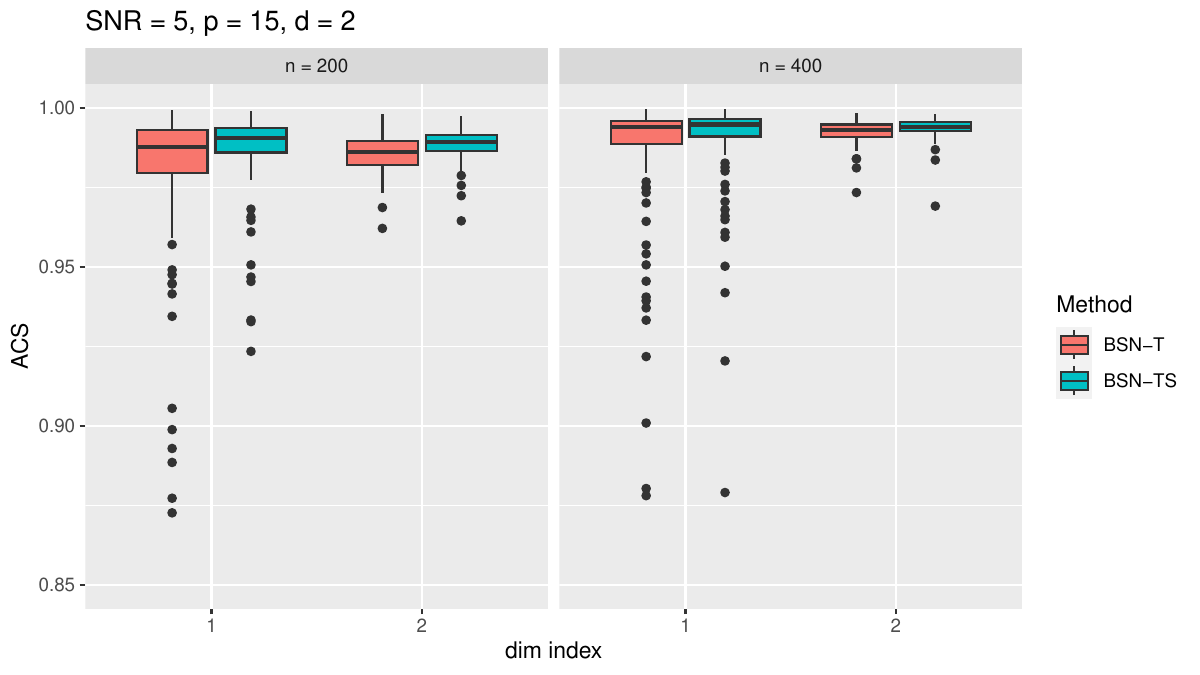}
\end{subfigure}
\caption{Absolute cosine similarity (ACS) for settings with $p= 15$ and $d = 2$ under SNR = 1 (left panel) and SNR = 5 (right panel), from 100 runs in the ``correctly specified'' case.}
\end{figure}

\begin{figure}[H]
\begin{subfigure}{.49\textwidth}
  \centering
\includegraphics[scale = 0.45]{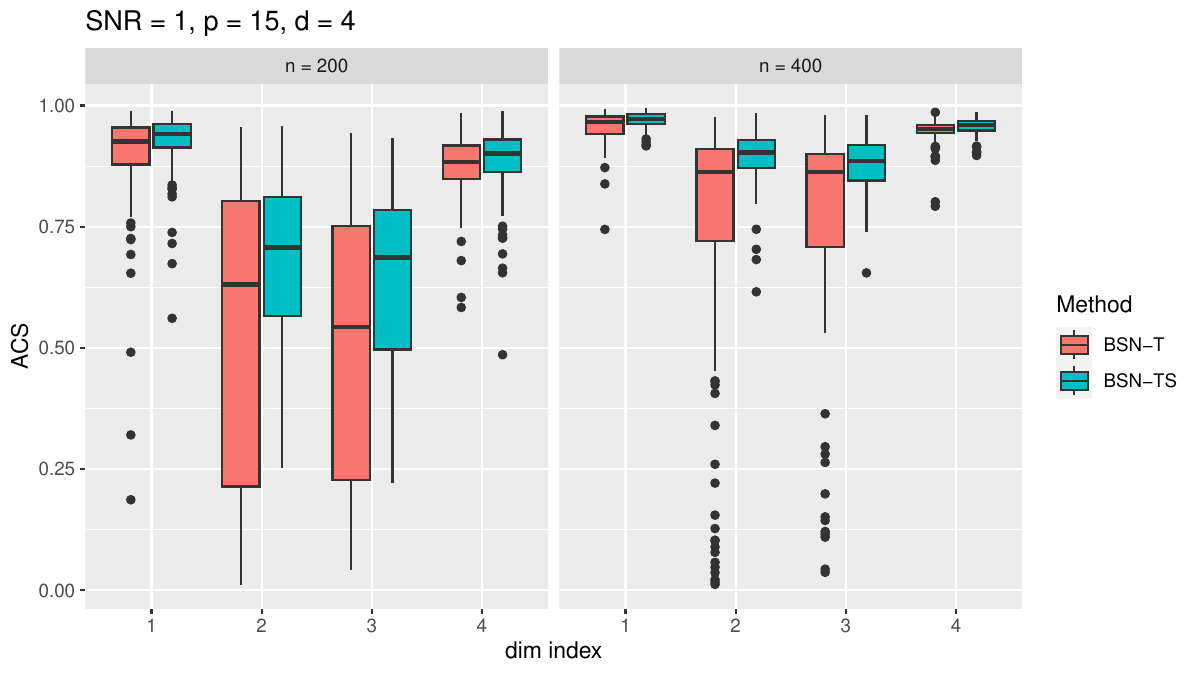}
\end{subfigure}
\begin{subfigure}{.49\textwidth}
\hspace{1cm}
\includegraphics[scale = 0.45]{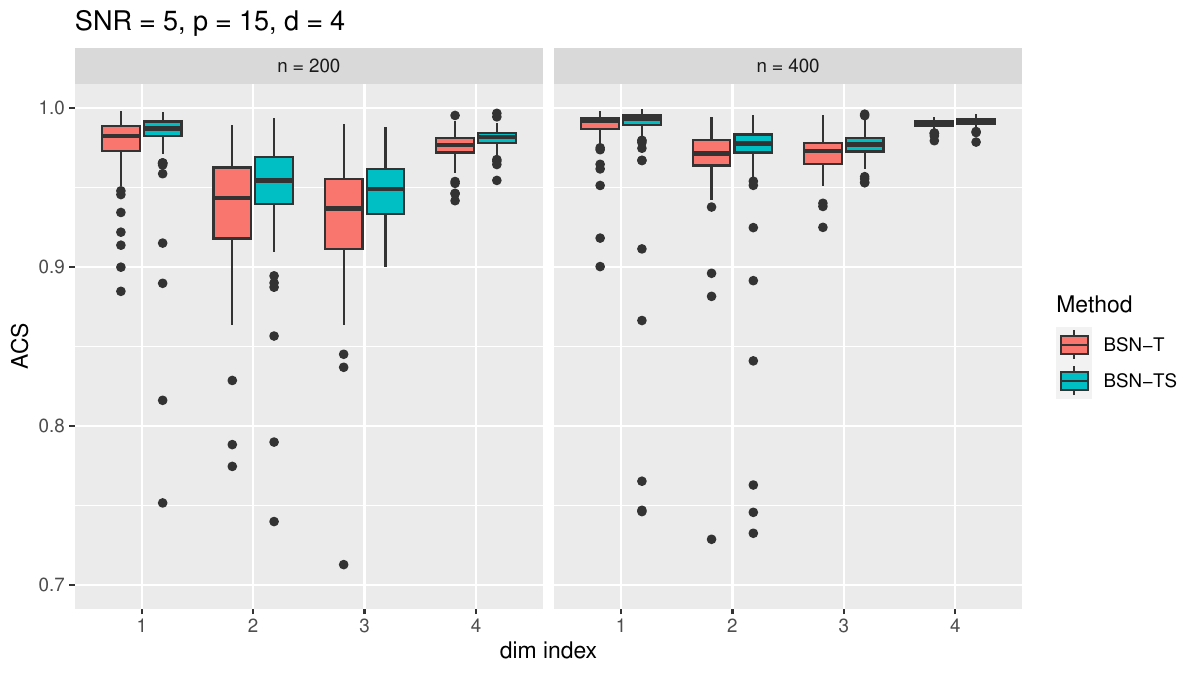}
\end{subfigure}
\caption{Absolute cosine similarity (ACS) for settings with $p= 15$ and $d = 4$ under SNR = 1 (left panel) and SNR = 5 (right panel), from 100 runs in the ``correctly specified'' case.}
\end{figure}

\begin{figure}[H]
  \centering
\includegraphics[scale = 0.7]{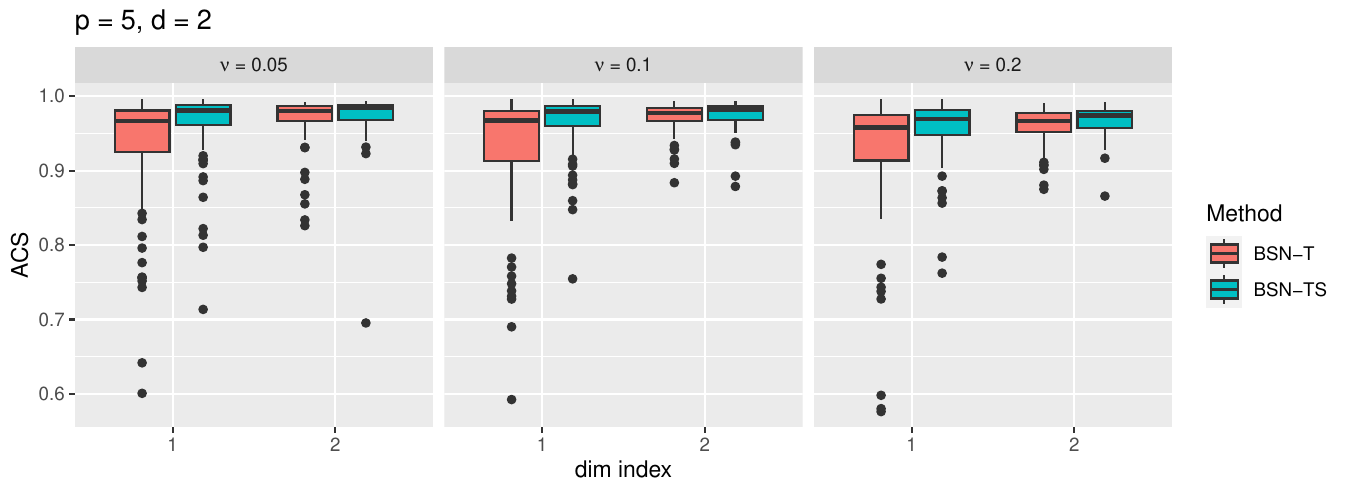}
\caption{Absolute cosine similarity (ACS) for settings with $p= 5$, $d = 2$, $n = 200$ and SNR = 1 from 100 runs in the ``misspecified'' case.}
\end{figure}

\begin{figure}[H]
  \centering
\includegraphics[scale = 0.7]{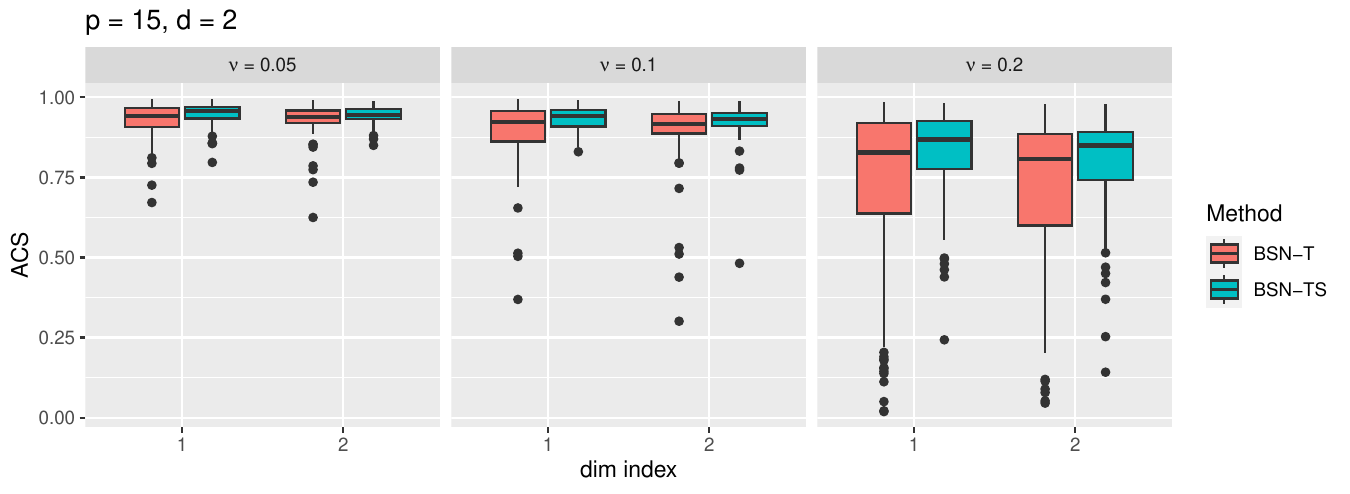}
\caption{Absolute cosine similarity (ACS) for settings with $p= 15$, $d = 2$, $n = 200$ and SNR = 1 from 100 runs in the ``misspecified'' case.}
\end{figure}

\begin{figure}[H]
  \centering
\includegraphics[scale = 0.7]{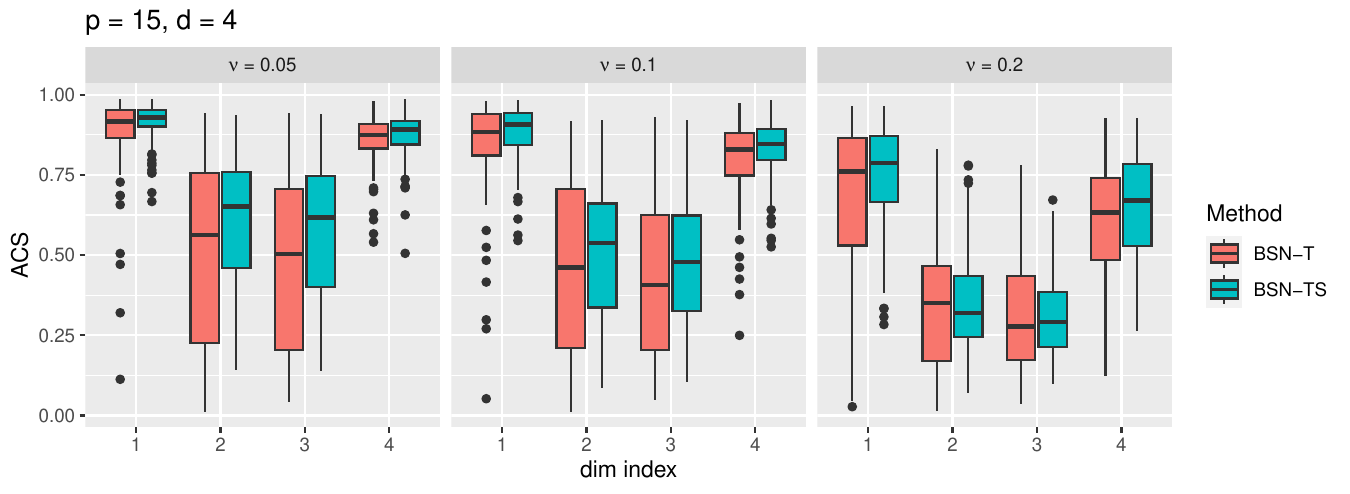}
\caption{Absolute cosine similarity (ACS) for settings with $p= 15$, $d = 4$, $n = 200$ and SNR = 1 from 100 runs in the ``misspecified'' case.}
\end{figure}

%% coverage of Gamma 

\subsection*{Coverage (Cover)}

The coverage (Cover) for each entry of  $\bs{\Gamma}$ and diagonal element of $\mb{B}$ was averaged  across 100 runs. For $\bs{\Gamma}$,   this produces an average coverage matrix of dimension $p \times  d$.  For easier comparison and visualization, we flattened these matrices into vectors column-wise and plotted the values. The x-axis represents the indices of the vectorized average coverage matrices.

\begin{figure}[H]
  \centering
\includegraphics[scale = 0.75]{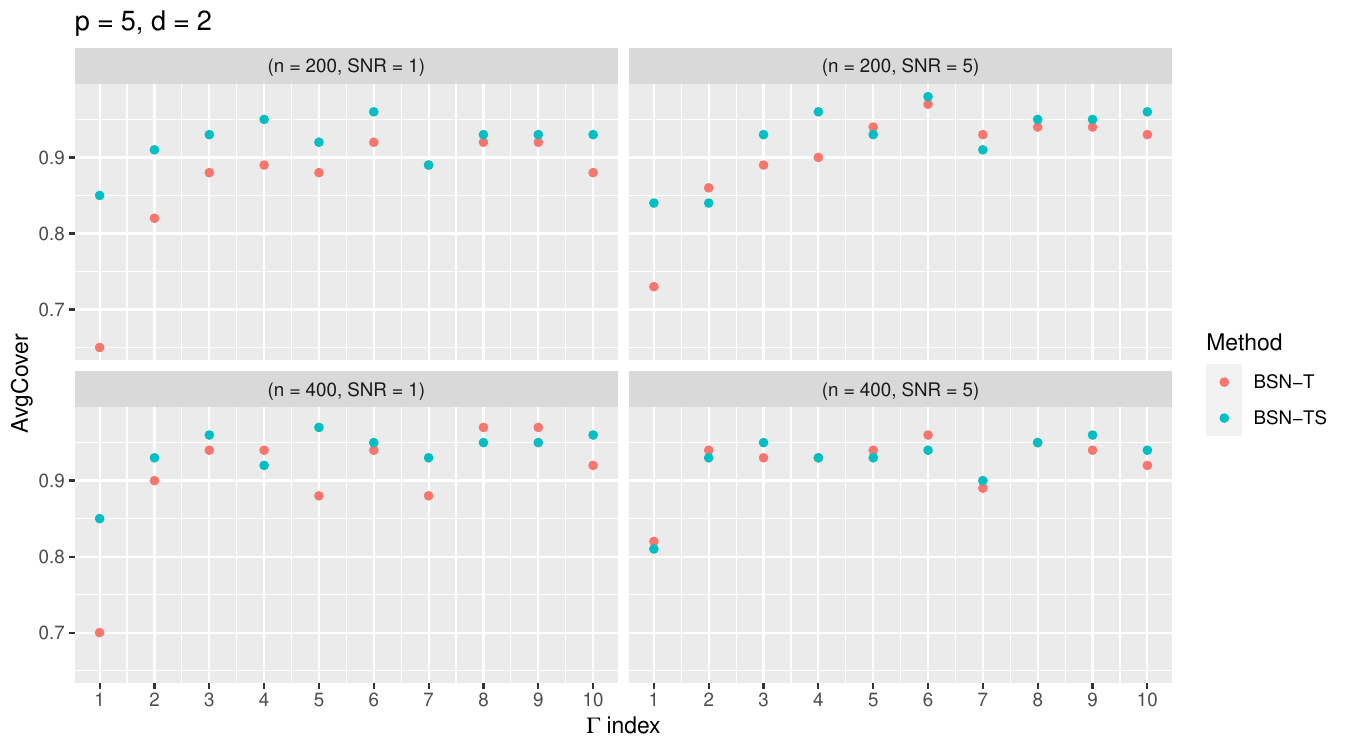}
\caption{Coverage (Cover) of the entries in $\bs{\Gamma}$  for settings with $p= 5$ and $d = 2$, averaged for 100 runs in the ``correctly specified'' case}
\end{figure}

\begin{figure}[H]
  \centering
\includegraphics[scale = 0.75]{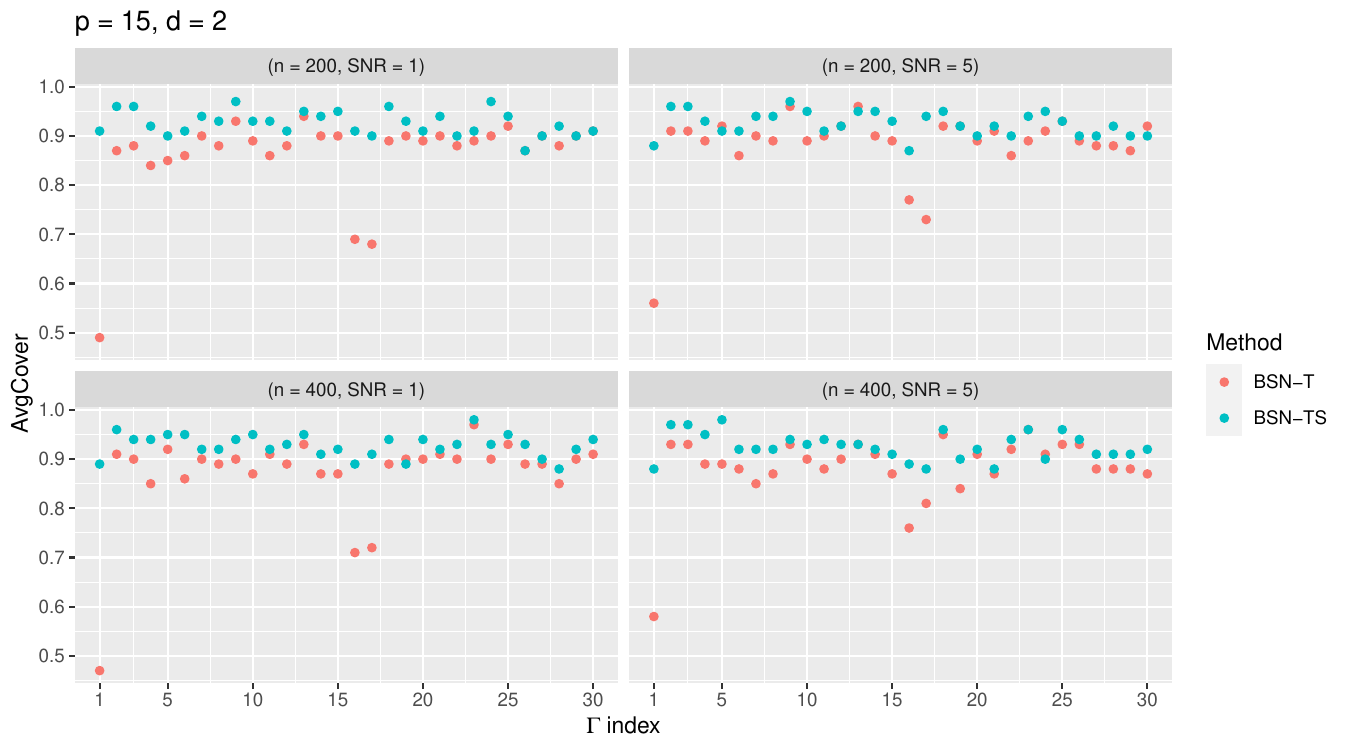}
\caption{Coverage (Cover) of the entries in $\bs{\Gamma}$  for settings with $p= 15$ and $d = 2$, averaged for 100 runs in the ``correctly specified'' case}
\end{figure}

\begin{figure}[H]
  \centering
\includegraphics[scale = 0.75]{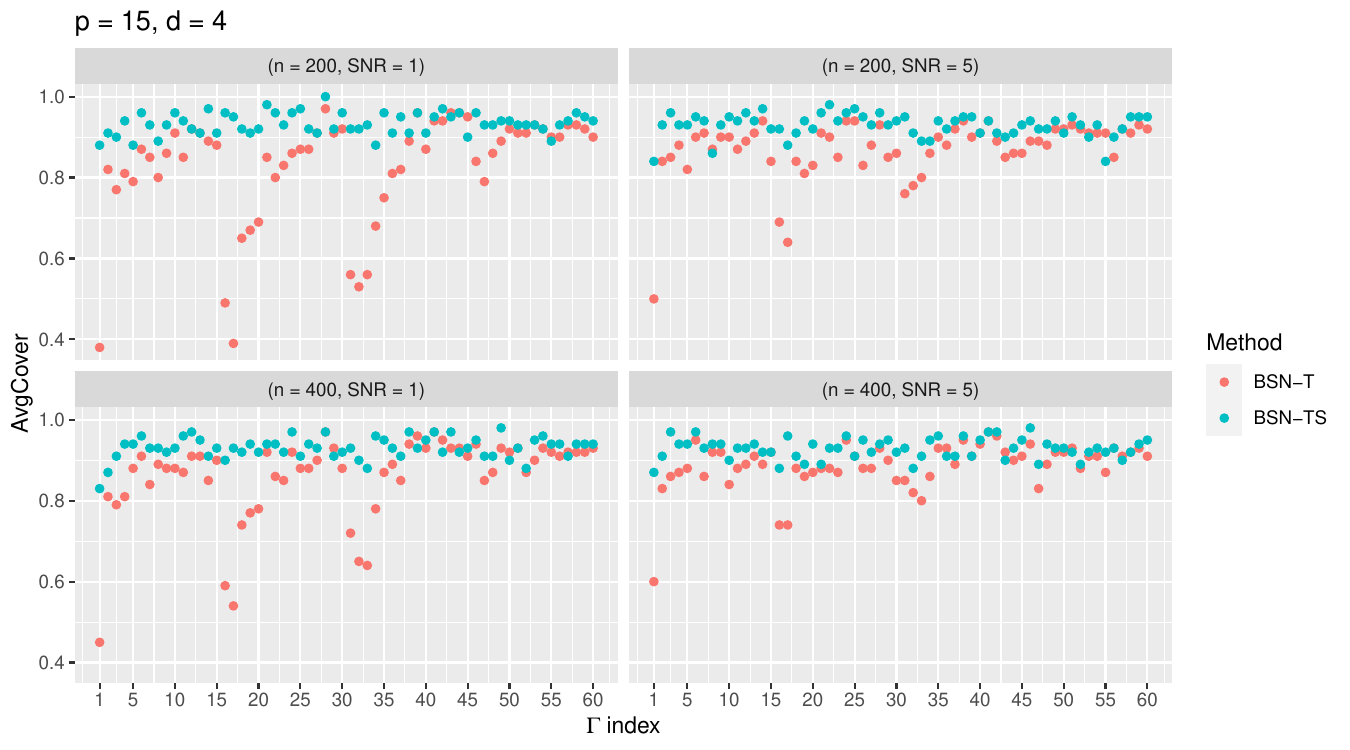}
\caption{Coverage (Cover) of the entries in $\bs{\Gamma}$ for settings with $p= 15$ and $d = 4$, averaged for 100 runs in the ``correctly specified'' case}
\end{figure}

\begin{figure}[H]
  \centering
\includegraphics[scale = 0.71]{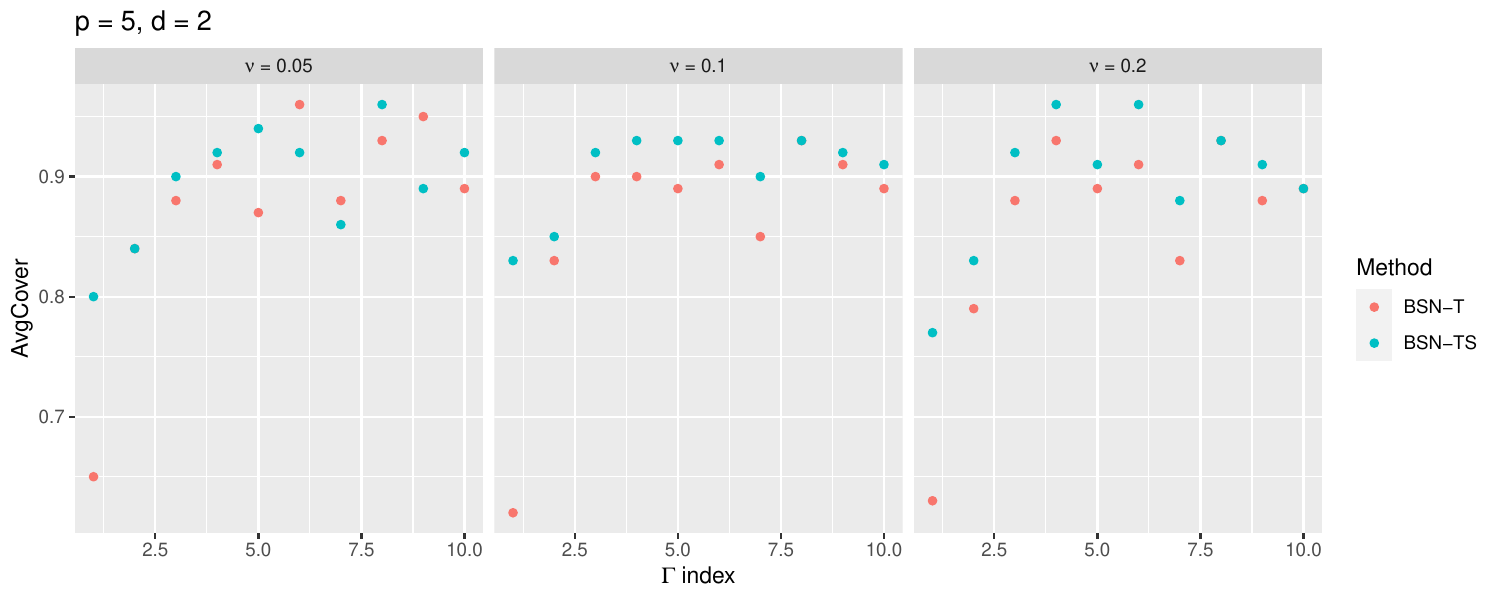}
\caption{Coverage (Cover) of the entries in $\bs{\Gamma}$ for settings with $p= 5$, $d = 2$, $n = 200$, and SNR = 1,  averaged for 100 runs in the ``misspecified'' case}
\end{figure}

\begin{figure}[H]
  \centering
\includegraphics[scale = 0.71]{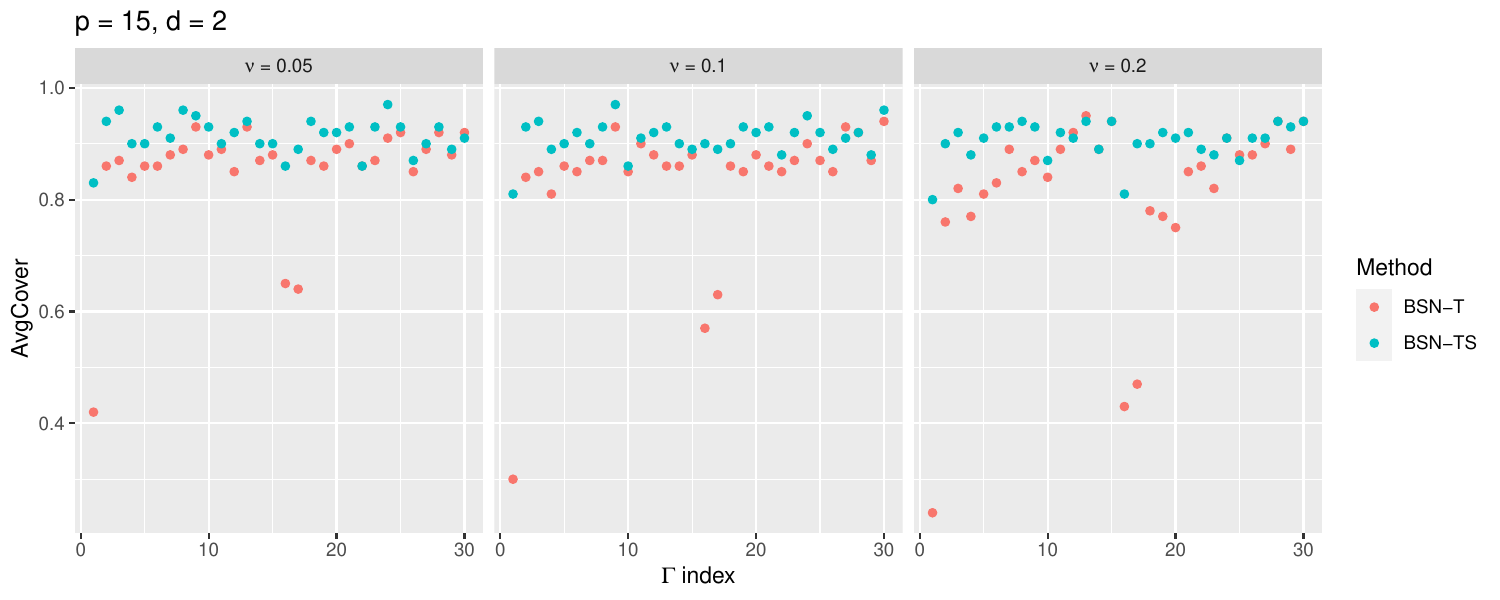}
\caption{Coverage (Cover) of the entries in $\bs{\Gamma}$ for settings with $p= 15$, $d = 2$, $n = 200$, and SNR = 1, , averaged for 100 runs in the ``misspecified'' case}
\end{figure}

\begin{figure}[H]
  \centering
\includegraphics[scale = 0.71]{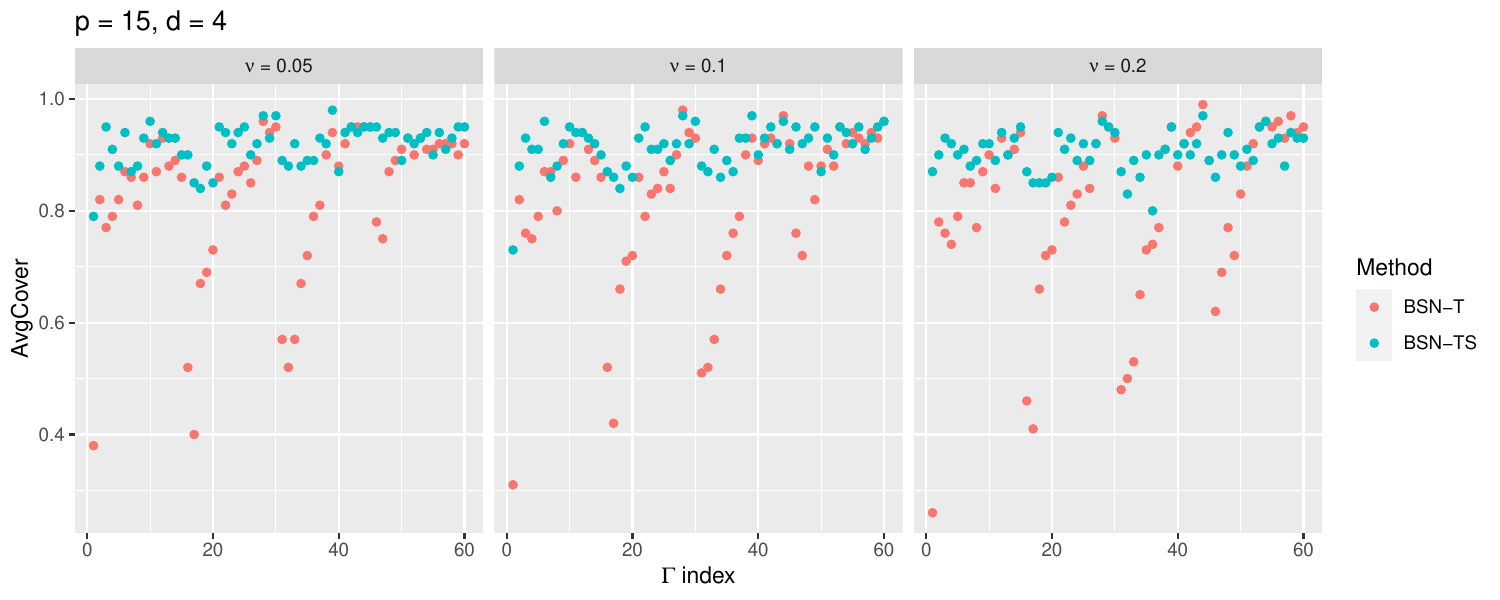}
\caption{Coverage (Cover) of the entries in $\bs{\Gamma}$ for settings with $p= 15$, $d = 4$, $n = 200$, and SNR = 1, averaged for 100 runs in the ``misspecified'' case}
\end{figure}

%%% coverage of b 

\begin{figure}[H]
  \centering
\includegraphics[scale = 0.75]{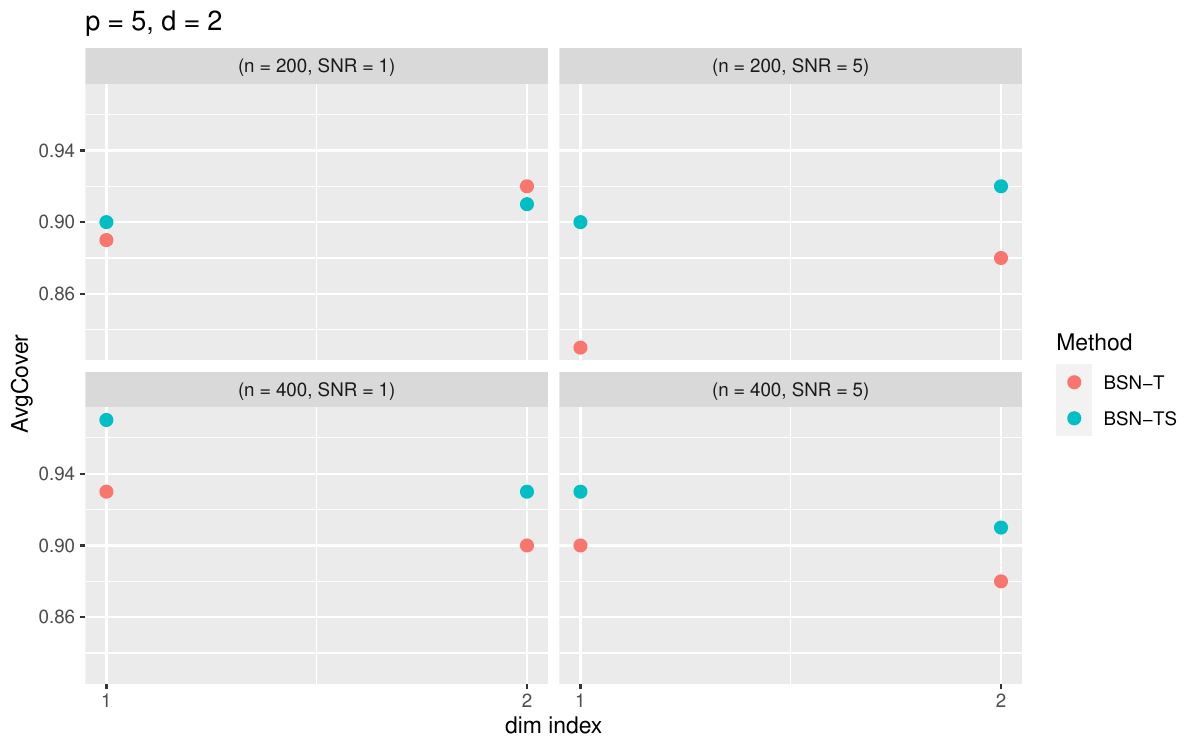}
\caption{Coverage (Cover) of the diagonal entries in $\mb{B}$  for settings with $p= 5$ and $d = 2$, averaged for 100 runs in the ``correctly specified'' case.}
\end{figure}

\begin{figure}[H]
  \centering
\includegraphics[scale = 0.75]{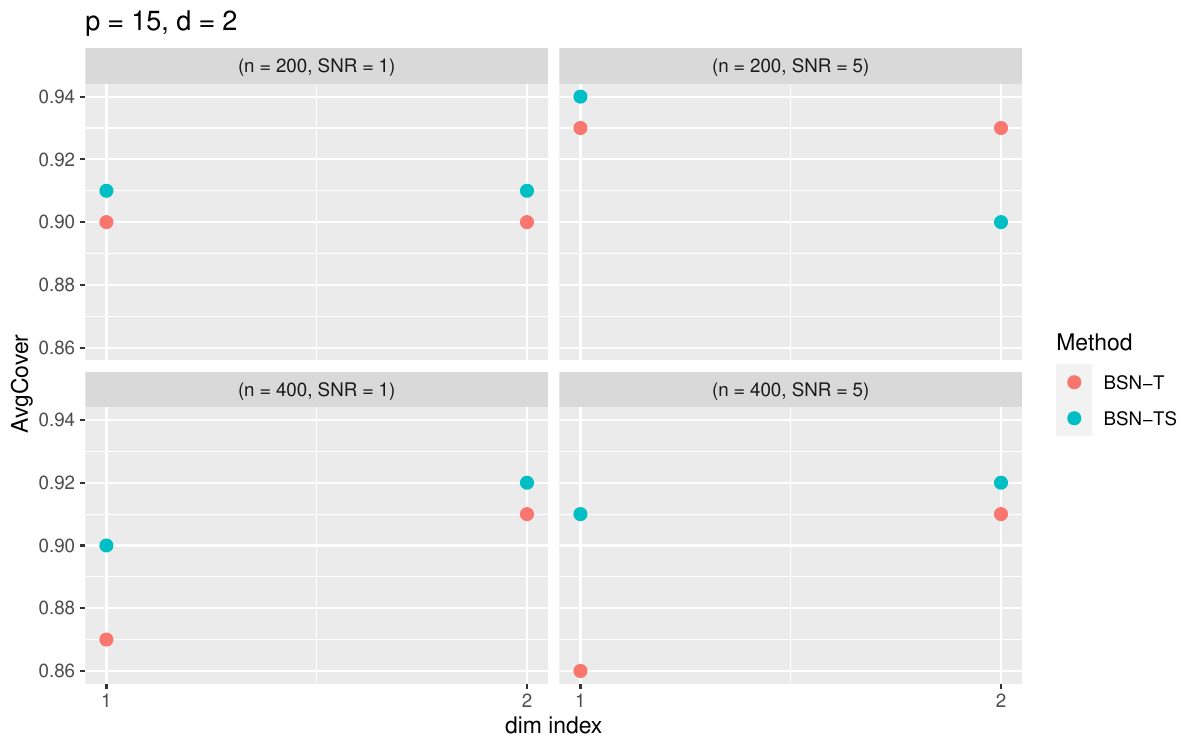}
\caption{Coverage (Cover) of the diagonal entries in $\mb{B}$  for settings with $p= 15$ and $d = 2$, averaged for 100 runs in the ``correctly specified'' case.}
\end{figure}

\begin{figure}[H]
  \centering
\includegraphics[scale = 0.75]{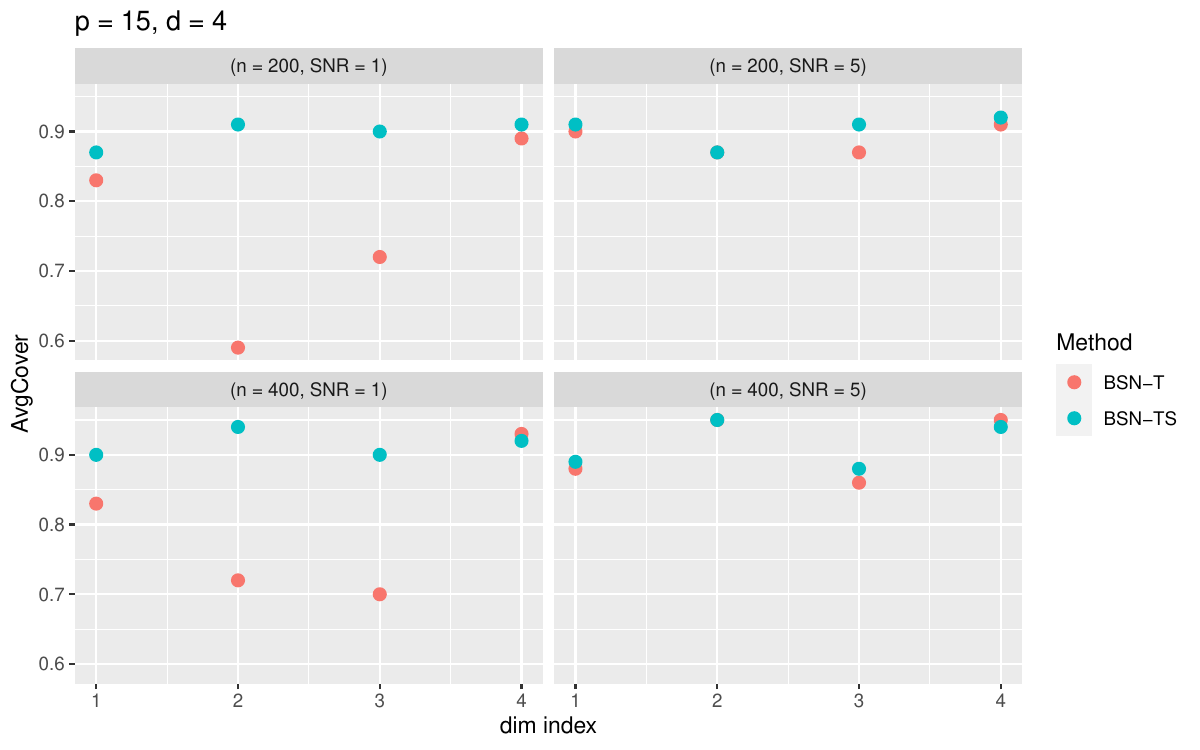}
\caption{Coverage (Cover) of the diagonal entries in $\mb{B}$ for settings with $p= 15$ and $d = 4$, averaged for 100 runs in the ``correctly specified'' case.}
\end{figure}

\begin{figure}[H]
  \centering
\includegraphics[scale = 0.71]{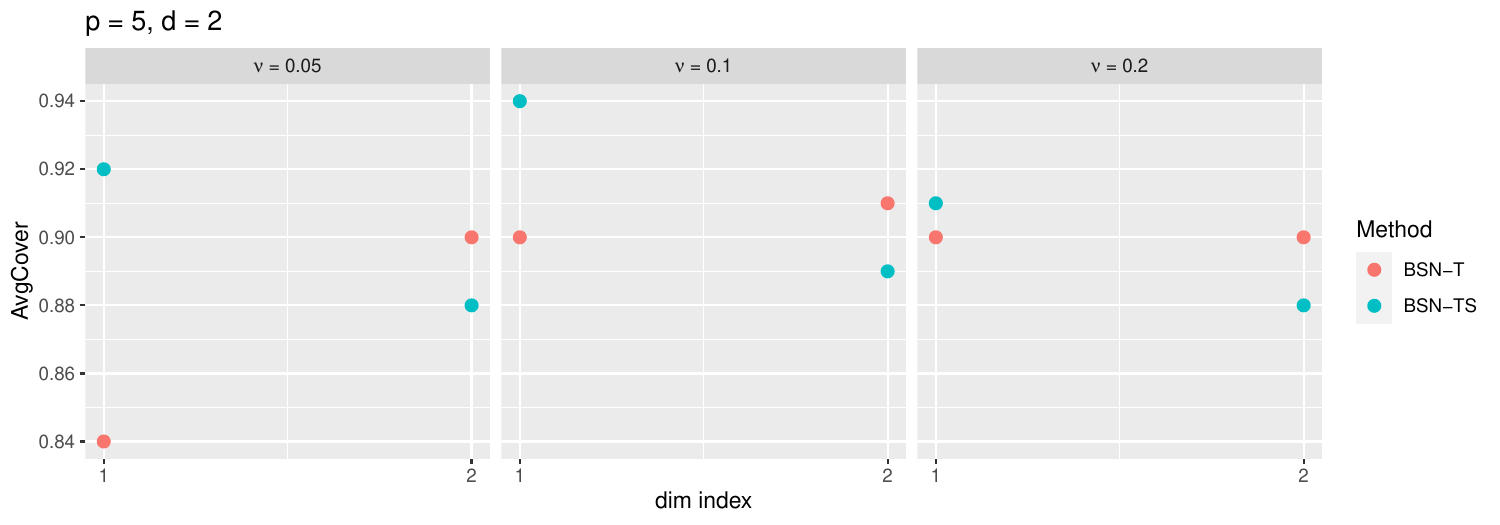}
\caption{Coverage (Cover) of the diagonal entries in $\mb{B}$ for settings with $p= 5$, $d = 2$, $n = 200$, and SNR = 1,  averaged for 100 runs in the ``misspecified'' case.}
\end{figure}

\begin{figure}[H]
  \centering
\includegraphics[scale = 0.71]{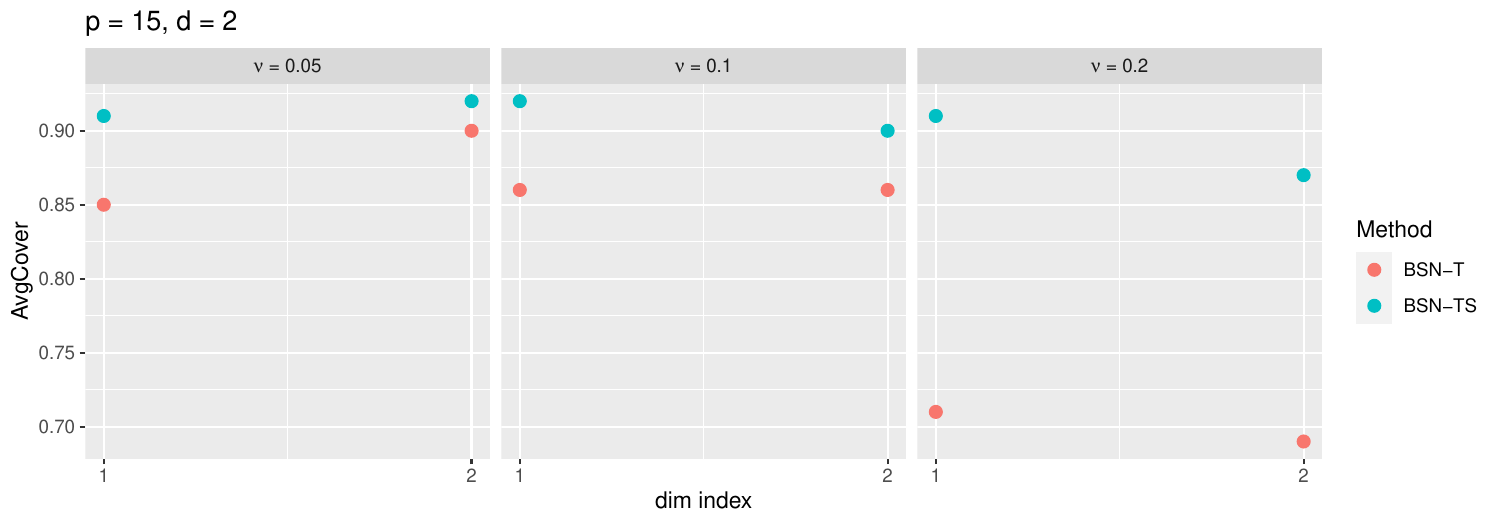}
\caption{Coverage (Cover) of the diagonal entries in $\mb{B}$ for settings with $p= 15$, $d = 2$, $n = 200$, and SNR = 1, , averaged for 100 runs in the ``misspecified'' case.}
\end{figure}

\begin{figure}[H]
  \centering
\includegraphics[scale = 0.71]{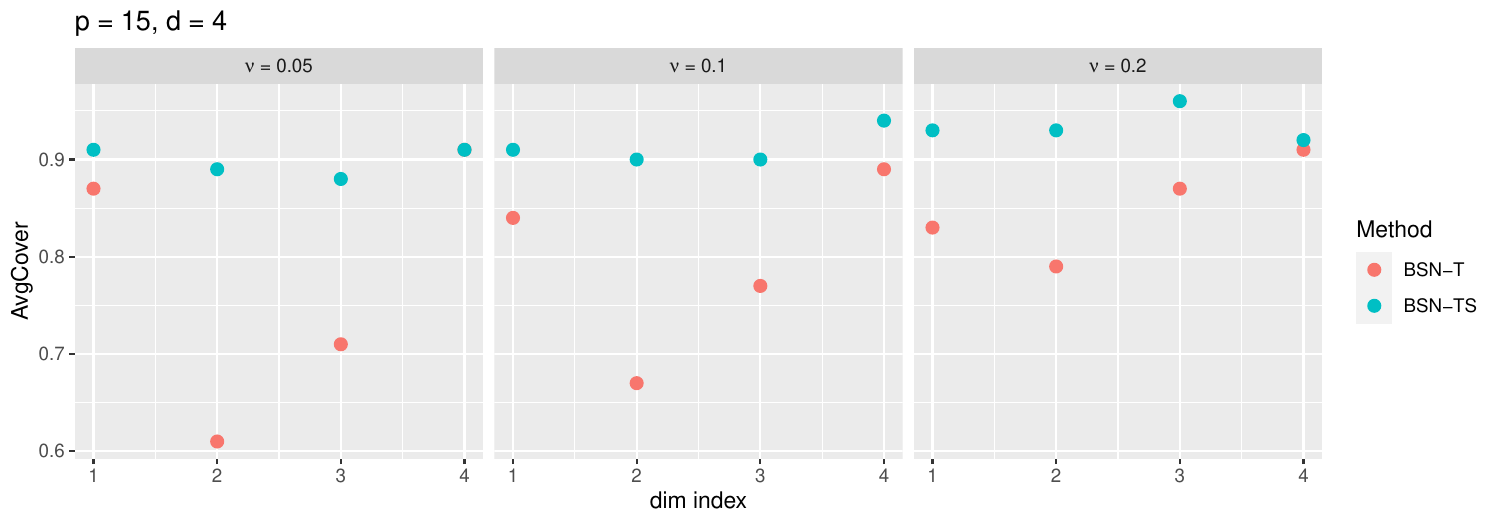}
\caption{Coverage (Cover) of the diagonal entries in $\mb{B}$ for settings with $p= 15$, $d = 4$, $n = 200$, and SNR = 1, averaged for 100 runs in the ``misspecified'' case.}
\end{figure}

\section*{Appendix E: Human Connectome Project}

\begin{figure}[H]
  \centering
\includegraphics[scale = 0.71]{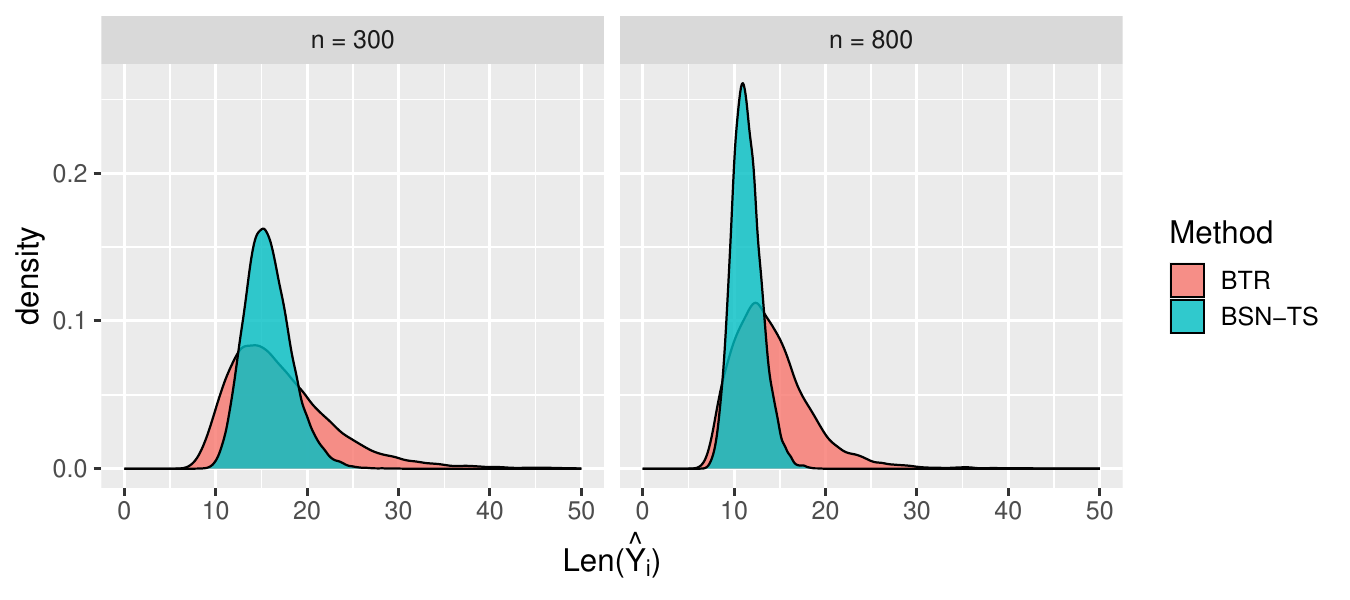}
\caption{Density plots showing distributions of 90\% credible interval lengths for \textbf{BTR} and \textbf{BSN-TS}. The plots were generated using aggregated results from the 50 runs, where each run used a random split of the HCP dataset. }
\label{eq:si:hcp1}
\end{figure}

Considering the temporal dependency in the time series of fMRI signals, we performed thinning of the observed data based on the effective sample size (ESS) for each subject $i$,
%, following the procedure described in \cite{park2023bayesian}. 
$$\text{ESS}(i) = \min_{ j \in \{1, ..., p\}} \frac{T_i}{1+ 2\sum_{t=1}^{T_i - 1} \text{cor}(x_{i,j}(t), x_{i,j}(1 + t))},$$
where  $x_{i,j}(t)$ denotes the signal at time $t$ in region $j$ for subject $i$.  The  $\text{ESS}(i)$ values range from 181 to 763, with an average of 349 across subjects.  We  subsampled $\text{ESS}(i)$ time points for each subject $i$.  \Cref{eq:si:hcp1} shows the distributions of the length of credible intervals for $n = 300$ and $n = 800$, respectively. \Cref{eq:si:hcp2}  shows the inference results of $\bs{\Gamma}^T{\mb{M}^{\ast}}^{-1/2}$for $n = 800$ to identify key brain regions. 

\begin{table}[H]
\centering
\begin{tabular}{lllll}
  \hline
 & Subnetwork 1 & Subnetwork 2 & Subnetwork 3 & Subnetwork 4 \\ 
  \hline
Node 1 & (-0.65, 0.68) & (-0.27, 0.85) & (-1.12, 0.35) & (-0.67, 1.19) \\ 
Node  2 & (-0.77, 0.68) &  \textbf{(0.07, 1.25)} & (-1.02, 0.32) & (-0.56, 1.01) \\ 
Node  3 & (-0.33, 0.27) & (-0.42, 0.17) & (-0.17, 0.24) & (-0.18, 0.19) \\ 
Node  4 & (-0.37, 0.49) & (-0.68, 0.25) & (-0.39, 0.72) & (-0.51, 0.70) \\ 
Node  5 & (-0.94, 0.53) & (-0.81, 0.57) & (-1.28, 1.12) & (-1.22, 0.71) \\ 
Node  6 & (-0.18, 0.95) & (-0.73, 0.62) & (-0.98, 0.44) & (-0.39, 1.16) \\ 
Node  7 & (-1.13, 0.21) & (-0.76, 0.96) & (-1.60, 1.21) & (-0.22, 1.72) \\ 
Node  8 & (-0.57, 0.96) & (-0.39, 0.99) & (-0.51, 0.33) & (-0.42, 0.38) \\ 
Node  9 & (-0.92, 1.19) & (-0.52, 1.15) & (-0.45, 0.37) & (-0.13, 0.53) \\ 
Node  10 &  \textbf{(0.06, 1.88)} & (-1.72, 1.35) & (-0.33, 0.98) & (-1.03, 0.35) \\ 
Node  11 & (-0.87, 0.60) & (-0.36, 1.15) & (-0.76, 0.65) & (-0.44, 0.99) \\ 
Node  12 & (-0.49, 0.40) & (-0.67, 0.28) & (-0.39, 0.34) & (-0.49, 0.15) \\ 
Node  13 & (-0.31, 1.17) & (-0.70, 0.85) & (-0.34, 0.50) & (-0.44, 0.41) \\ 
Node  14 & (-0.23, 1.51) & (-1.23, 0.91) & (-1.83, 1.37) & \textbf{(0.09, 1.93)} \\ 
Node  15 & (-0.47, 0.30) & (-0.13, 0.63) & \textbf{(0.13, 1.60)} & (-1.32, 1.10) \\ 
   \hline
\end{tabular}
\caption{The 90\% credible intervals of $\bs{\Gamma}^T {\mb{M}^{\ast}}^{-1/2} \in \mathbb{R}^{d \times p} (=\mathbb{R}^{4 \times 15}$) with the intervals of its $j$-th component  $\bs{\gamma}_j^T {\mb{M}^{\ast}}^{-1/2}~\in~\mathbb{R}^p$ ($=\mathbb{R}^{15}$), representing a ``subnetwork'' of the 15 network nodes, displayed in the corresponding $j$-th column of the table for $j \in  \{1, 2, 3, 4\}$. The 90\% credible intervals were computed from aggregated MCMC posterior samples across the 50 runs, each with a random data split of the HCP data.  The intervals highlighted in boldface are the ones that do not contain zero. }
   \label{eq:si:hcp2}
\end{table}

\end{document}